\documentclass[a4paper,preprintnumbers,showpacs,onecolumn,superscriptaddress,nofootinbib,amsmath,amssymb,notitlepage]{revtex4-1}

\usepackage{amssymb,amsthm}

\usepackage{enumitem}
\usepackage{times}
\usepackage{dcolumn}
\usepackage{mathrsfs}
\usepackage{comment}
\usepackage{hyperref}
\usepackage{amsmath}
\usepackage{mathtools}
\usepackage{bbm}
\usepackage{bm}
\usepackage{amsfonts}
\usepackage{amssymb}
\usepackage{mathrsfs}
\usepackage{wasysym}
\usepackage{graphicx}
\usepackage[]{subfigure}
\usepackage{filecontents}
\usepackage[dvipsnames]{xcolor}

\hypersetup{
    bookmarks=true,         
    pdftoolbar=true,        
    pdfmenubar=true,        
    pdffitwindow=true,     
    pdfstartview={FitH},     
    pdftitle={My title},     
    pdfauthor={author},      
    pdfsubject={Subject},    
    pdfcreator={Creator},    
    pdfproducer={Producer},  
    pdfkeywords={keyword1} 
    pdfnewwindow=true,       
    colorlinks=true ,        
    linkcolor=Blue  ,        
    citecolor=Blue,      
    filecolor=green,       
    urlcolor=Blue            
}

\newcommand{\arccosh}{\operatorname{arccosh}}

\newcommand{\ed}{\mathrm{d}}

\newcommand{\st}{\mathrm{st}}
\newcommand{\bg}{\bar{\gamma}}
\newcommand{\sQ}{\mathscr{Q}}
\newcommand{\fu}{\mathfrak{u}}
\newcommand{\fv}{\mathfrak{v}}
\newcommand{\fw}{\mathfrak{w}}
\newcommand{\fp}{\mathfrak{p}}
\newcommand{\fq}{\mathfrak{q}}
\newcommand{\fr}{\mathfrak{r}}
\newcommand{\fs}{\mathfrak{s}}
\newcommand{\ft}{\mathfrak{t}}

\newcommand{\ff}{\mathfrak{f}}
\newcommand{\im}{\mathrm{i}}
\newcommand{\mZ}{\mathcal{Z}}

\begin{document}

\title{Spherical photon orbits around a rotating black hole with quintessence and cloud of strings}

\author{Mohsen Fathi}
\email{mohsen.fathi@postgrado.uv.cl}
\affiliation{Instituto de F\'{i}sica y Astronom\'{i}a, Universidad de Valpara\'{i}so,
Avenida Gran Breta\~{n}a 1111, Valpara\'{i}so, Chile}

\author{Marco Olivares}
\email{marco.olivaresr@mail.udp.cl}
\affiliation{Facultad de
              Ingenier\'{i}a y Ciencias, Universidad Diego Portales,  Avenida Ej\'{e}rcito
             Libertador 441, Santiago, Chile}

\author{J.R. Villanueva}
\email{jose.villanueva@uv.cl}
\affiliation{Instituto de F\'{i}sica y Astronom\'{i}a, Universidad de Valpara\'{i}so,
Avenida Gran Breta\~{n}a 1111, Valpara\'{i}so, Chile}


\begin{abstract}
In this paper we calculate the analytical solutions for the radii of planar and polar spherical photon orbits around a rotating black hole that is associated with quintessential field and cloud of strings. This includes a full analytical treatment of a quintic that describes orbits on the equatorial plane. Furthermore, The radial profile of the impact parameters is studied and the radii corresponding to the extreme cases are derived. For the more general cases, we also discuss the photon regions that form around this black hole. To simulate the orbits that appear in different inclinations, we analytically solve the latitudinal and azimuth equations of motion in terms of the Weierstra{\ss}ian elliptic functions, by considering the radii of spherical orbits, in their general form, as the initial conditions. The period and the stability conditions of the orbits are also obtained analytically.

\bigskip

{\noindent{\textit{keywords}}: Black holes, photon orbits, photon regions, quintessence, cloud of strings}\\

\end{abstract}

\maketitle

\section{Introduction and outlook}\label{sec:intro}

Black holes are indeed among the most mysterious astrophysical objects. From their first observational evidences, which were due to Cygnus X-1 in the early 70's \cite{webster_cygnus_1972,bolton_identification_1972}, to the latest observations of the shadows of M87* and SgrA* \cite{Akiyama:2019,Akiyama:2022}, the quest for obtaining astrophysical data from black holes have been on a constant course. While the former was indicating an extreme X-ray source, the latter were indeed optical resemblances of what general relativity had been formulated for the exterior geometry of such objects. According to general relativity, the near-horizon spacetime around black holes is so wrapped, that photons travel on spherical orbits. Such photon orbits are essentially unstable, and together, they form a photon ring that confines the black hole shadow. From the theoretical viewpoint, the determination of spherical orbits for static spacetimes (like Schwarzschild), is easy, since they are planar and are, essentially, circles on the equatorial plane. This situation, however, changes for stationary (rotating) black holes, since the frame-dragging effect produces a region filled with photons on constant-radius orbits; the so-called photon regions. For the case of Kerr black holes and applying the geodesic equations for light rays, the radii of planar and polar orbits were first given in Refs.~\cite{Bardeen:1972a,Bardeen:1973b}, and discussed further in the context of Kerr geodesics in Ref.~\cite{Chandrasekhar:1998}. Ever since, numerous publications have been devoted to the study of spherical photon orbits, photon regions and photon rings in Kerr and Kerr-like black hole spacetimes (see for example Refs.~\cite{stoghianidis_polar_1987,cramer_using_1997,Teo:2003,Johannsen:2013,Grenzebach:2014,Perlick:2017,charbulak_spherical_2018,Johnson_universal_2020,Himwich:2020,Gelles:2021,Ayzenberg:2022,Das:2022}). It is important to highlight that, the photon regions are spatially bounded by the aforementioned radii of planar orbits, which have been given analytical expressions for Kerr black holes. On the other hand, the analytical determination of the radii of non-planar photon orbits in their most general form in Kerr-like spacetimes is a formidable task, because it leads to solving polynomials of the orders of six and above. Nevertheless, some rigorous, by approximated studies have been done so far in order to provide some semi-analytical expressions for the radii of photon orbits on Kerr black holes (see Refs.~\cite{Hod:2013,Tavlayan:2020}). 

In this paper, we also deal with the spherical photon orbits on a Kerr-like black hole, whose exterior geometry is a rotating counterpart of a static spherically symmetric spacetime associated with quintessence. In fact, it is plausible that the black hole evolution in the current cosmic era, could be affected by the dark side of the universe \cite{JimenezMadrid:2005,Jamil:2009,Li:2019,Roy:2020}. Technically, one can add specific cosmological components to the black hole's spacetime geometry, so that the dark features could also contribute. Such components may be contributed, for example, by including a dark fluid energy-momentum tensor in the Einstein field equations, in the form of a halo \cite{Xu:2018,Das:2021}, or a quintessential field \cite{Kiselev:2003,Saadati:2019,AliKhan:2020}. In fact, quintessential fields are supposed to explain, dynamically, the late time accelerated expansion of the universe. Additionally, we assume that the black hole is also associated with cloud of strings, which means that instead of point particles, the cosmic fluid in which the black hole resides, consists of one-dimensional strings. Based on the same criteria, a generalization to the Schwazschild black hole spacetime has been done in  Refs.~\cite{Stachel:1977,Letelier:1979}. On the other hand, if both of the quintessence and cloud of strings are present for a Schwarzschild black hole, the spacetime is endowed with extra gravitational potentials, that can be regarded similarly as those in the Mannheim-Kazanas spherically symmetric solution to the fourth order Weyl conformal gravity, which was claimed to recover the flat galactic rotation curves \cite{Mannheim:1989}. In such spacetime, the source of gravity is extended, and hence, the black hole resides in a stringy universe. A static spherically symmetric spacetime metric for this black hole, has been derived and analyzed in Refs.~\cite{Toledo:2018,Dias:2019,Toledo:2019}. Applying a modified Newman-Janis algorithm, this spacetime was then assigned a rotating counterpart in Ref.~\cite{Toledo:2020}, which reduces correctly to that of Kerr, in the absence of the parameters of quintessence and cloud of strings.

In this work, we pursue two main objectives. First, we go deep into the analytical derivation of the radii of the planar orbits around the black hole. This requires a precise treatment of a quintic that governs such orbits. Secondly, we obtain the exact analytical solutions for the evolution of the polar and azimuth coordinates, that happen on constant radii. To obviate these aims, we organize the paper as follows: In Sect. \ref{sec:theBH}, we briefly introduce the static spacetime, its components and casual structure. This is followed by discussing the rotating counterpart, in terms of the horizons, the ergoregion and the properties of the extremal case. In Sect.~\ref{sec:sphericalGeneral}, we begin our study of the spherical photon orbits on the black hole, by means of the geodesic equations. There, we calculate the critical impact parameters, and this way, a general octic equation is generated than governs the radii of spherical orbits. We continue this section by confining ourselves to the equatorial plane, so the aforementioned octic reduces to a quintic. This quintic will be treated analytically and its solutions are  expressed in terms of the generalized hypergeometric functions. The detailed mathematical methods are then described in the appendices. Further in this section, we also calculate the radii of polar orbits that cross the axis of symmetry. With the help of these information, we discuss some examples of the photon regions that form in the exterior of the black hole, for various spin parameters. In Sect.~\ref{sec:analyticalSpherical}, we give a rigorous study of the latitudinal and azimuth motion, by solving analytically their first order equations of motion. Accordingly, the integrals of motion are given solutions in terms of the three Weierstra{\ss}ian elliptic functions. In this section, the period of the latitudinal oscillations is calculated separately, and is compared to that inferred from the profile of the polar coordinate, for some specific examples. In Sect.~\ref{sec:examples}, we apply the above solutions for some radii determined by solving numerically the octic for a variety of initial inclinations, in order to simulate several categories of spherical orbits, for sub-extremal, extremal and super-extremal spacetimes. In this section, we also discuss the stability of the orbits.  We conclude in Sect.~\ref{sec:conclusion}. Throughout this work, we apply a geometrized system of units, in which $G=c=1$.

\section{The black hole solution in the dark background 
}\label{sec:theBH}

The static, spherically symmetric black hole solution in the quintessential background, which is surrounded by cloud of strings, is described by the following metric in the $x^\mu = (t,r,\theta,\phi)$ coordinates:
\begin{equation}\label{eq:metric}
    \ed s^2 = -B(r) \ed t^2 + \frac{\ed r^2}{B(r)}+r^2 \ed\theta^2+r^2\sin^2\theta \ed\phi^2
\end{equation}
with the lapse function defined as \cite{Toledo:2018,Dias:2019}
\begin{equation}\label{eq:lapse}
    B(r) = 1-\alpha - \frac{2M}{r}-\frac{\gamma}{r^{3w_q+1}},
\end{equation}
in which, $\alpha$, $M$, $\gamma$ and $w_q$, represent, respectively, the dimensionless string cloud parameter ($0<\alpha<1$), the black hole mass, the quintessence parameter and the equation of state (EoS) parameter. For a perfect fluid distribution of matter/energy, this latter is defined by $P_q = w_q \rho_q$, with $P_q$ and $\rho_q$ as the quintessential energy pressure and density, and lies within the range $-1<w_q<-\frac{1}{3}$. This parameter is set to be responsible for the cosmological acceleration and the special case of $w_q=-1$ recovers the cosmological constant. 

To proceed further with our study, we will consider the case of $w_q = -\frac{2}{3}$ which corresponds to the black hole spacetime with the lapse function
\begin{equation}\label{eq:lapse_1}
    B(r) = 1-\alpha - \frac{2M}{r}-\gamma r,
\end{equation}
located in a matter dominated universe \cite{Wei:2008}. This spacetime is not asymptotically flat, however, its three-dimensional subspace has an asymptotic deficit of angle \cite{Macias:2002}. Such effect is also intensified by the presence of the cloud of strings. Note that, for this particular choice for the $w_q$, the dimension of $\gamma$ is $\mathrm{m}^{-1}$. 

Let us define the {mass function} \cite{Toshmatov:2017,Toledo:2020}
\begin{equation}\label{eq:newParameter}
    \rho(r) = M+\frac{\alpha r}{2}+\frac{\gamma r^2}{2},
\end{equation}
that vanishes at $r_0 = \frac{-\alpha+\sqrt{\alpha^2-8M\gamma}}{2\gamma}$. {This way, the lapse function \eqref{eq:lapse_1} can be recast as $B(r)=\frac{1}{r}[r-2\rho(r)]$. Hence, at $r_0$ we have $B(r_0)=1$, which corresponds to the Minkowski spacetime. On the other hand, the condition $\rho(r_0)=0$ implies that the mass parameter can also opt negative values. Based on the fact that $\alpha, \gamma>0$, negative values for $\rho(r)$ correspond to negative radial distances. This, however, is not allowed physically, since the geometry of the static black hole has a real singularity at $r=0$.}

For a quintessential energy tensor $T_{\mu\nu} = (\varepsilon,P_r,P_\theta,P_\phi)$ with a constituent of cloud of strings, one can confirm that \cite{Toledo:2020}
\begin{subequations}\label{eq:TmunuComp}
\begin{align}
    & \varepsilon = \frac{2\rho' }{8\pi }=-P_r,\\
    & P_\theta = P_r-\frac{\rho'' r + 2\rho'}{8\pi r} = P_\phi,
\end{align}
\end{subequations}
where primes denote differentiations with respect to the $r$-coordinate. The above relations hold in the context of general relativity $G_{\mu\nu} = 8\pi T_{\mu\nu}$, where $G_{\mu\nu}$ is the Einstein tensor. Hence, the solution \eqref{eq:lapse_1} can be regarded as a static black hole spacetime surrounded by cloud of strings, that is located in a universe filled with quintessential dark energy. Note that, for a comoving time-like observer with a velocity four-vector field $u^\mu = (1,0,0,0)$, the values in Eq. \eqref{eq:TmunuComp} provide
\begin{equation}\label{eq:WEC}
   T_{\mu\nu} u^\mu u^\nu = \frac{\alpha +2 \gamma r}{8 \pi  r^2}.
\end{equation}
Hence, $T_{\mu\nu} u^\mu u^\nu>0$ is guaranteed for all $\gamma>0$ and therefore, we can infer that the weak energy condition (WEC) is respected. 
Note that $\gamma_c \rightarrow 0$ for $\alpha \rightarrow 1$, and $\gamma_c = \frac{1}{8M}$ for $\alpha \rightarrow 0$. The black hole admits the two horizons \cite{Cardenas:2021}
\begin{eqnarray}
	&& r_{++} =\frac{1-\alpha}{\gamma}\cos^2\left(\frac{1}{2}\arcsin\left(\frac{2\sqrt{2M\gamma}}{1-\alpha}\right)\right),\label{eq:rg_sB}\\
	&& r_+ = \frac{1-\alpha}{\gamma}\sin^2\left(\frac{1}{2}\arcsin\left(\frac{2\sqrt{2M\gamma}}{1-\alpha}\right)\right),\label{eq:rH_sB}
\end{eqnarray}
that correspond, respectively, to the (quintessential) cosmological, and the event horizons (note that, these horizons are only valid for the case of $\gamma\neq0$). This way, the extremal black hole has a unique horizon $r_+=r_{++}=r_e=\frac{4M}{1-\alpha}$ for $\gamma=\gamma_c\equiv\frac{(1-\alpha)^2}{8M}$, and a naked singularity is obtained for $\gamma>\gamma_c$ (therefore the static black hole is valid for the range $0<\gamma<\gamma_c$). This black hole has been studied in Ref.~\cite{Mustafa:2021}, regarding the radial and circular orbits of mass-less and massive particles. This study has been completed in Ref.~\cite{Fathi:2022a}, by investigating all types of possible orbits for these particles. Furthermore, in Ref.~\cite{He:2022}, the shadow and the photon sphere of this black hole has been studied.\\

To obtain the rotating counterpart of this black hole spacetime, in Ref. \cite{Toledo:2020}, a modified version of the Newman-Janis algorithm \cite{Newman:1965}, proposed by Azreg-A\"{\i}nou \cite{Azreg:2014} was applied. This algorithm generates the stationary spacetime
\begin{multline}\label{eq:metric_rotating}
    \ed s^2 = -\frac{\Delta-a^2\sin^2\theta}{\Sigma} \ed t^2 + \frac{\Sigma}{\Delta} \ed r^2 - 2 a \sin^2\theta\left(1-\frac{\Delta-a^2\sin^2\theta}{\Sigma}\right)\ed t\ed \phi + \Sigma \ed\theta^2\\
    + \sin^2\theta\left[\Sigma+a^2\sin^2\theta\left(2-\frac{\Delta-a^2\sin^2\theta}{\Sigma}\right)\right]\ed\phi^2,
\end{multline}
in which $a$ is the black hole's spin parameter which is directly related to its angular momentum through the relation $J = a M$,  and referring to the lapse function \eqref{eq:lapse_1}, we have defined
\begin{subequations}
\begin{align}
    & \Delta(r) = a^2 + r^2 B(r) = (1-\alpha) r^2+a^2 - 2Mr - \gamma r^3,\label{eq:Delta}\\
    & \Sigma(r,\theta) = r^2+a^2 \cos^2\theta.
\end{align}
\end{subequations}
The metric \eqref{eq:metric_rotating} can resemble the Kerr-like form
\begin{equation}\label{eq:metric_rotating_KerrLike}
    \ed s^2 = - \left(1-\frac{2\rho r}{\Sigma}\right)\ed t^2 + \frac{\Sigma}{\Delta}\ed r^2 -\frac{4 \rho r a \sin^2\theta}{\Sigma}\ed t\ed\phi
    +\Sigma \ed\theta^2 + \sin^2\theta\left(
    r^2+ a^2+\frac{2  \rho r a^2\sin^2\theta}{\Sigma}
    \right)\ed\phi^2,
\end{equation}
by means of the definition \eqref{eq:newParameter},
according to which, $\Delta = r^2+a^2-2 \rho r$. {Unlike the static case, the mass function $\rho(r)$ may encounter its zero value at $r_0$ as well as becoming negative. As mentioned before, for positive $\alpha$ and $\gamma$, this latter corresponds to negative $r$. This case has been included in the study of particle geodesics, for example in Refs. \cite{calvani_complete_1981,Hackmann_KerrAds:2010, hackmann_analytical_2010,PhysRevD.87.124030}. Accordingly, since $r=0$ is not a singularity for the spacetime \eqref{eq:metric_rotating_KerrLike}, the test particles can enter the negative sub-manifold of the spacetime, by avoiding the ring singularity. For large negative values of $r$, this sub-manifold corresponds to a \textit{negative universe}, where the black hole possesses a negative mass. In this study, however, we are not concerned about this case and only positive values of $r$ are taken into account.}  

Now the energy tensor components of Eqs.~\eqref{eq:TmunuComp} change to \cite{Toledo:2020}
\begin{subequations}\label{eq:TmunuComp-1}
\begin{align}
    & \varepsilon = \frac{2\rho' r^2}{8\pi \Sigma^2}=-P_r,\\
    & P_\theta = P_r-\frac{\rho'' r + 2\rho'}{8\pi \Sigma} = P_\phi.
\end{align}
\end{subequations}
Furthermore, for a comoving time-like observer  the values in Eq. \eqref{eq:TmunuComp-1} provide
\begin{equation}\label{eq:WEC}
   T_{\mu\nu} u^\mu u^\nu = \frac{r^2 (\alpha +2 \gamma r)}{8 \pi  \left(a^2 \cos ^2\theta+r^2\right)^2}, 
\end{equation}
which for all $\gamma>0$ implies $T_{\mu\nu} u^\mu u^\nu>0$. Hence, the solution \eqref{eq:metric_rotating_KerrLike} can be regarded as a stationary black hole spacetime associated with cloud of strings, and located in a universe filled with quintessential dark energy.
The rotating black hole defined in Eq. \eqref{eq:metric_rotating_KerrLike} admits three horizons located at the real roots of the equation $\Delta(r) = 0$, which are 
\begin{eqnarray}
&& r_{++} =R_{*}+ 4\sqrt{\frac{\chi_2}{3}} \cos\left(\frac{1}{3} \arccos \left(\sqrt{\frac{27 \chi_3^2}{\chi_2^3}}\right)\right),\label{eq:rQ}\\
&& r_+ = R_{*}+ 4\sqrt{\frac{\chi_2}{3}} \cos\left(\frac{1}{3} \arccos \left(\sqrt{\frac{27 \chi_3^2}{\chi_2^3}}\right)+\frac{4\pi}{3}\right),\label{eq:rH}\\
&& r_{-} = R_{*}+ 4\sqrt{\frac{\chi_2}{3}} \cos\left(\frac{1}{3} \arccos\left(\sqrt{\frac{27 \chi_3^2}{\chi_2^3}}\right)+\frac{2\pi}{3}\right),\label{eq:rC}
\end{eqnarray}
denoting, respectively, the (quintessential) cosmological, event and Cauchy horizons, where
\begin{subequations}
\begin{align}
& \chi_2 = \frac{3R_*^2}{4}\left(1-\frac{3\gamma}{4\gamma_c}\right),\label{eq:chi2}\\
& \chi_3 = \frac{a^2}{16 \gamma}+\frac{R_*^{3}}{8} \left(1-\frac{9 \gamma}{8 \gamma_c}\right),\label{eq:chi3}
\end{align}
\end{subequations}
and
\begin{equation}\label{rstar}
    R_* =\frac{1-\alpha}{3\gamma}=\frac{\sqrt{8 M \gamma_c}}{3\gamma}.
\end{equation}
Note that, unlike the static case, the extremal rotating black hole corresponds to $\gamma=\bg_c$, where the discriminant of the cubic $\Delta(r)=0$ vanishes, with 
\begin{equation}
\bg_c=\frac{2}{27a^4}\left[\sqrt{\left[4M^2-3a^2(1-\alpha)\right]^3}+9Ma^2(1-\alpha)-8M^3\right],
    \label{eq:bgammac}
\end{equation}
and hence, $\bg_c$ is only well-defined in the context of the stationary black hole with $a\neq0$. {For an extremal black hole with $\gamma=\bg_c$, the exterior horizons of the black hole merge, resulting in $r_+=r_{++}$. } The naked singularity then corresponds to $\gamma>\bg_c$ (this means that the stationary black hole solution is valid for $0<\gamma<\bg_c$). One can also obtain the value for the spin parameter that corresponds to the extremal black hole, which reads
\begin{equation}
\bar{a}_c=\frac{1}{3\gamma}\sqrt{\frac{2}{3}}\left[
9M\gamma(1-\alpha)-(1-\alpha)^3+\sqrt{\left[(1-\alpha)^2-6M\gamma\right]^3}
\right]^\frac{1}{2},
    \label{eq:bac}
\end{equation}
and is well-defined only for $\gamma\neq0$ (accordingly, the black hole exists for $a<\bar{a}_c$). {Note that, the extremality in this case results in $r_-=r_+$, which will be considered further in this paper as a particular case, in the demonstration of the photon regions.}

\begin{figure}[t]
    \centering
    \includegraphics[width=8cm]{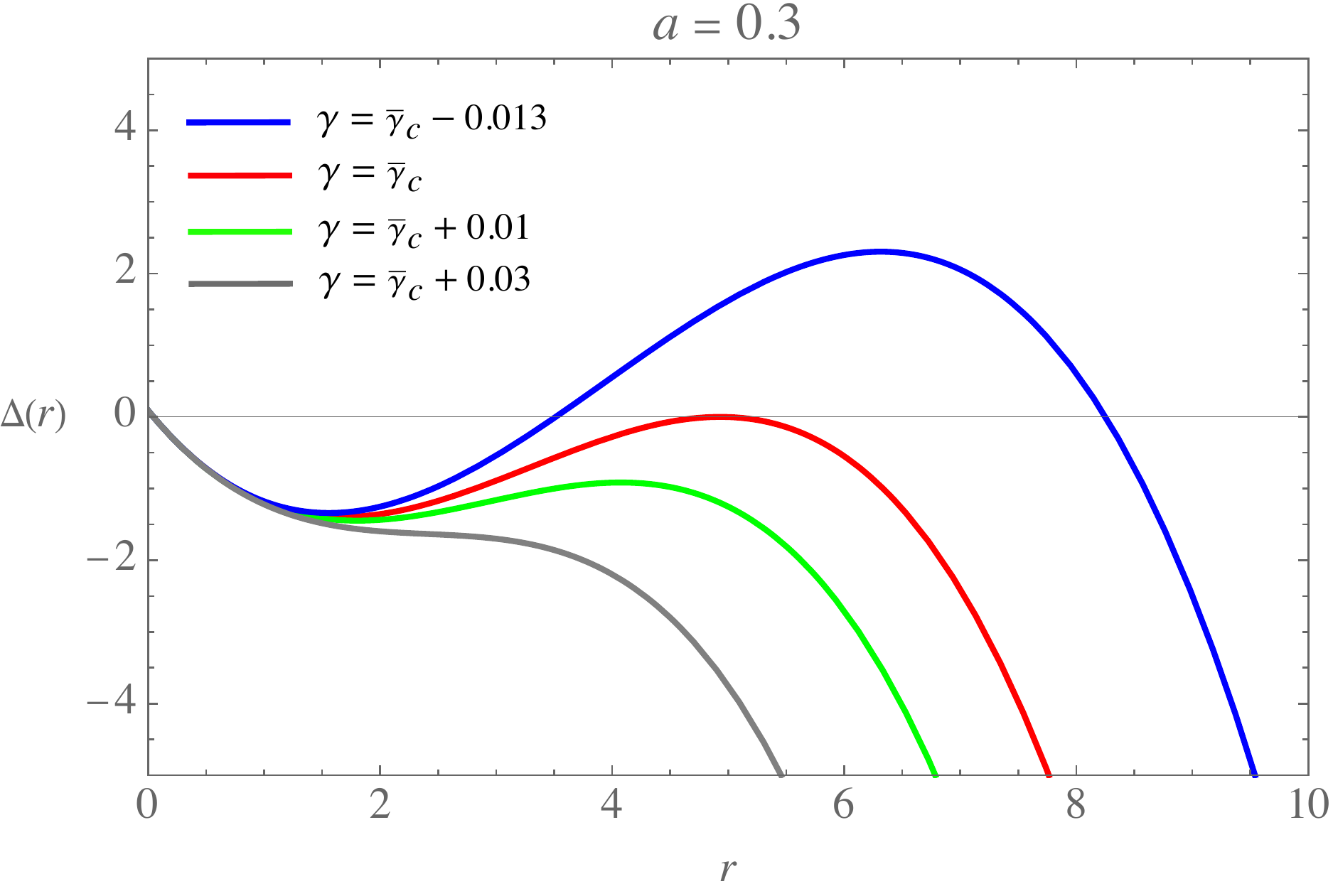}\qquad\qquad
    \includegraphics[width=8cm]{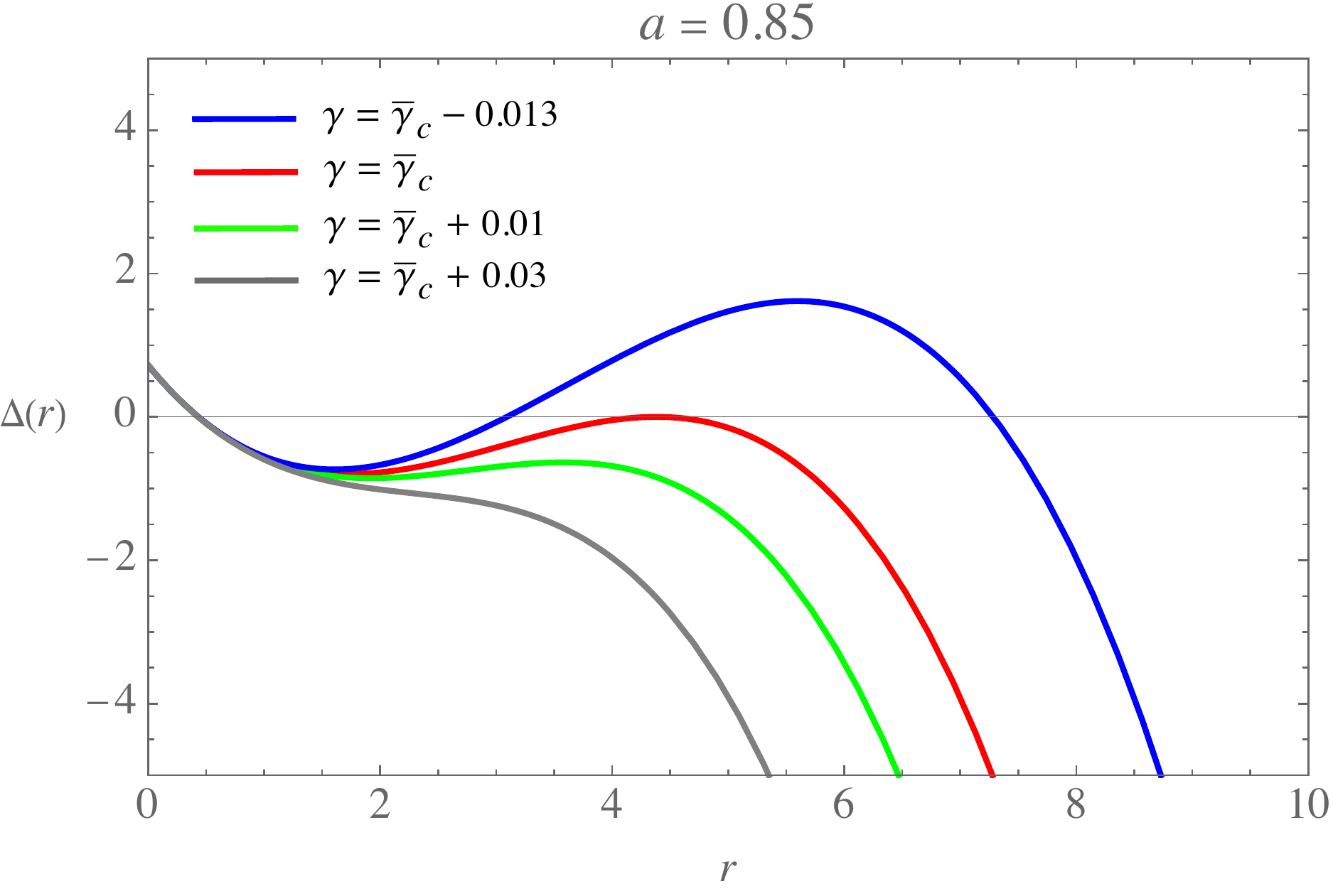}
    \caption{The behavior of $\Delta(r)$ and its zeros, for two values of the spin parameter, corresponding to a slow and a fast rotating black hole. Different values of $\gamma$ have been used around the extremal value $\bar{\gamma}_c$, and we have considered $\alpha=0.2$. This way, $\bg_c=0.0807$ for $a=0.3$, and $\bg_c=0.0870$ for $a=0.85$. The unit of length along the axis is taken as $M$.}
    \label{fig:rootsofDelta}
\end{figure}

\begin{figure}[t]
    \centering
    \includegraphics[width=8cm]{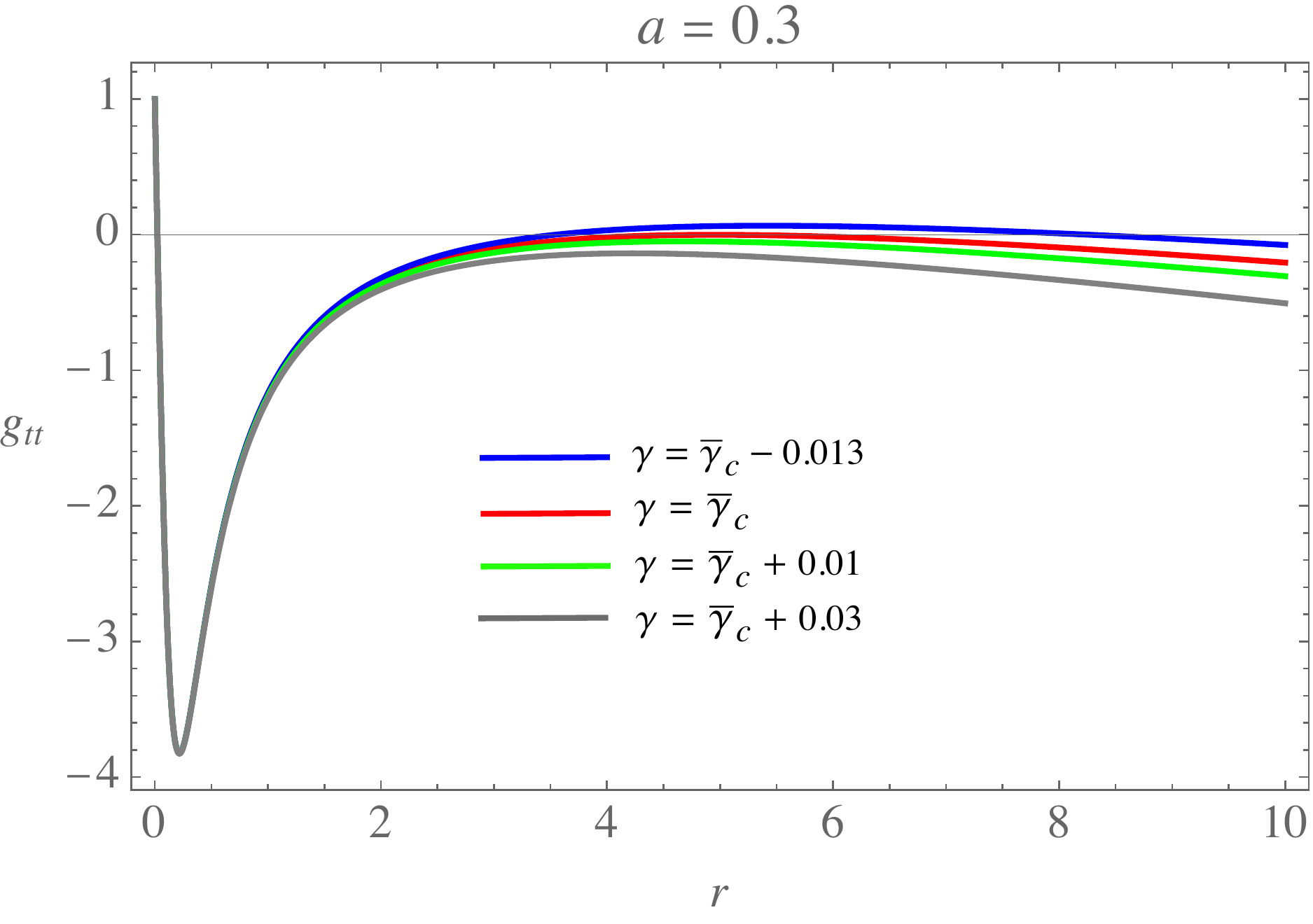}\qquad\qquad
    \includegraphics[width=8cm]{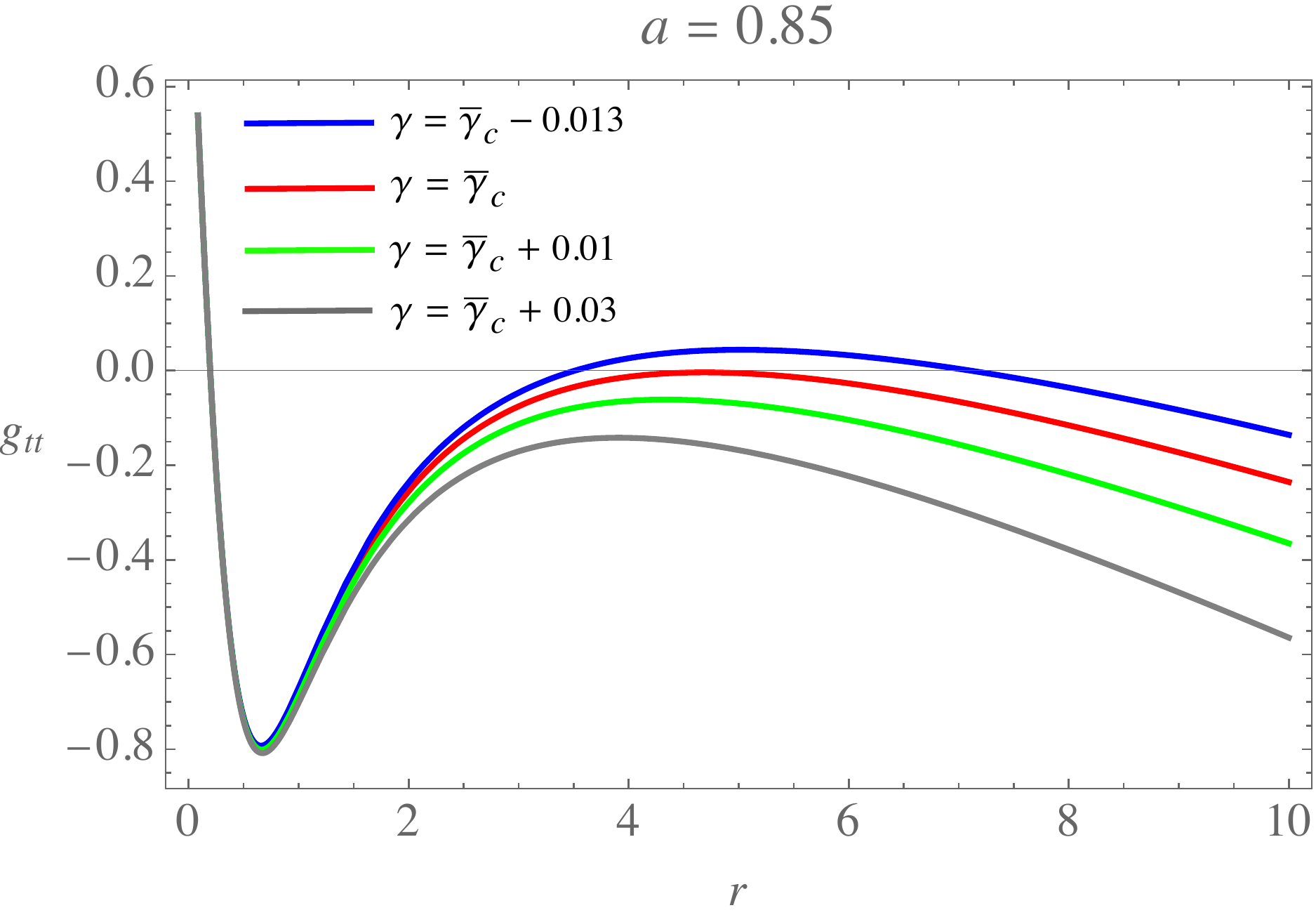}
    \caption{The behaviors of $g_{tt}$ and its real roots, plotted for $\theta=\frac{\pi}{4}$ and the same parameters as considered for Fig.~\ref{fig:rootsofDelta}.}
    \label{fig:rsl}
\end{figure}

In Fig. \ref{fig:rootsofDelta}, the behavior of the third order polynomial of $\Delta(r)$ has been plotted by indicating its three roots.
Furthermore, as it is well-known, the static limits correspond to hypersurfaces at the radii obtained from solving the equation $g_{tt}=0$. These hypersurfaces, together with those formed by the horizons, constitute the ergoregions. Inside the ergoregions, no static observer can exist and all the observers are in the state of corotation with the black hole. The equation $g_{tt}=0$ results in the solutions
\begin{eqnarray}
&& r_{\st_{++}} =R_{*}+ 4\sqrt{\frac{\chi_2}{3}} \cos\left(\frac{1}{3} \arccos \left(\sqrt{\frac{27 \bar{\chi}_3^2}{\chi_2^3}}\right)\right),\label{eq:rstpp}\\
&& r_{\st_{+}} = R_{*}+ 4\sqrt{\frac{\chi_2}{3}} \cos\left(\frac{1}{3} \arccos \left(\sqrt{\frac{27 \bar{\chi}_3^2}{\bar{\chi}_2^3}}\right)+\frac{4\pi}{3}\right),\label{eq:rstp}\\
&& r_{\st_{-}} = R_{*}+ 4\sqrt{\frac{\chi_2}{3}} \cos\left(\frac{1}{3} \arccos\left(\sqrt{\frac{27 \bar{\chi}_3^2}{\chi_2^3}}\right)+\frac{2\pi}{3}\right),\label{eq:rstn}
\end{eqnarray}
that satisfy the conditions $0<r_{\st_-}<r_-$, and $r_+<r_{\st_+}<r_{\st_{++}}<r_{++}$, where $\bar{\chi}_3=\chi_3-\frac{a^2\sin^2\theta}{16\gamma}$ (so it is verified that the static limits and the horizons coincide for the case of $\theta=0$, or as viewed from the axis of symmetry). In Fig.~\ref{fig:rsl}, the radial profiles of $g_{tt}$ have been shown for the same parameters exploited in Fig.~\ref{fig:rootsofDelta}.
In Ref. \cite{Toledo:2020}, the ergoregion structure of this black hole has been discussed, qualitatively, for different values of $\alpha$ and $\gamma$, and accordingly, it has been shown that the presence of the cloud of strings causes the ergoregion to shrink.


\section{The spherical photon orbits and the photon regions}\label{sec:sphericalGeneral}

In the case of $a\neq0$
there are several spherical
light rays around the black hole. Each of them stays on a sphere $r =
\mathrm{const}.$\footnote{{Note that, the $r$-constant photonic surfaces, instead of being spherical, are indeed \textit{spheroidal}. This has been discussed for example in Ref. \cite{Ferraro_untangling_2014}. On the other hand, the term \textit{spherical orbits} refers directly to every single photon orbit with constant radial distance from the black hole, and is commonly used in the literature.}}, with the $\theta$-coordinate varying between
two turning points (explicit derivations and discussions on this subject for the case of Kerr black holes can be found in standard texts such as those in Refs.~\cite{Chandrasekhar:1998,Bardeen:1972a,Bardeen:1973b}). These spherical light rays exist
for radius values in a certain interval; only the
innermost and the outermost ones are circular, and
all the other ones are non-planar (see below). The photon region
is the region of all points, through which, such
spherical light rays exist. In what follows, we follow the standard Carter's equations of geodesic motion for the light rays (obtained from the method of the separation of Hamilton-Jacobi equations), in order to obtain the aforementioned radii of planar light orbits. Further in this section, the photon regions are demonstrated and discussed.  

We use the Hamilton-Jacobi equation and the Carter’s separation method to determine the geodesic equations in the exterior geometry of the black hole given by the metric \eqref{eq:metric_rotating}. These equations take the first-order differential forms \cite{Carter:1968,Chandrasekhar:1998} 
\begin{eqnarray}
M\frac{\ed t}{\ed\lambda} &=& \frac{r^2+a^2}{\Delta}\left[E\left(r^2+a^2\right)-aL\right]-a\left(aE\sin^2\theta-L\right),\label{eq:dt}\\
M\frac{\ed r}{\ed\lambda} &=&\pm\sqrt{\mathcal{R}(r)},\label{eq:dr}\\
M\frac{\ed \theta}{\ed\lambda} &=&\pm\sqrt{\Theta(\theta)},\label{eq:dtheta}\\
M\frac{\ed \phi}{\ed\lambda} &=&\frac{a}{\Delta}\left[E\left(r^2+a^2\right)-aL\right]-\left(aE-\frac{L}{\sin^2\theta}\right),\label{eq:dphi}
\end{eqnarray}
by making use of the dimensionless \textit{Mino time}, $\lambda$, as $\Sigma\ed\lambda=M\ed\tau$ \cite{Mino:2003}, with $\tau$ as the trajectory affine parameter. In the above first order differential equations, $E$ and $L$ are the constants of motion related to the temporal and axisymmetrical symmetries of the spacetime. Note that, here $E$ cannot be regarded as the conserved energy of the photons because the spacetime is not asymptotically flat. On the other hand, $L$ is the component of the angular momentum associated with the photons which is stretched along the axis of symmetry, and $\sQ$ is the Carter's constant. Furthermore,
\begin{subequations}\label{eq:R,Theta}
\begin{align}
    & \mathcal{R}(r) =\left[E\left(r^2+a^2\right)-aL\right]^2 -\Delta\left[\left(aE-L\right)^2+\sQ\right]\label{eq:R},\\
    &\Theta(\theta) = \sQ-\left(\frac{L^2}{\sin^2\theta}-a^2E^2\right)\cos^2\theta\label{eq:Theta}.
\end{align}
\end{subequations}
For convenience, we choose the positive segments of Eqs.~\eqref{eq:dr} and \eqref{eq:dtheta}, and define the two impact parameters
\begin{eqnarray}
&& \xi=\frac{L}{E},\label{eq:xi}\\
&& \eta=\frac{\sQ}{E^2}.\label{eq:eta}
\end{eqnarray}
Note that, $\sQ$ has a crucial role in the determination of the particles' orbits in the sense that trajectories confined to the equatorial plane (i.e. $\theta=\frac{\pi}{2}$) correspond to $\sQ=0$ (or $\eta=0$), which also constitutes the boundary of the orbits with constant radii that satisfy $\sQ\geq0$.

The necessary conditions for unstable photon orbits 
are characterized by the equations $R(r)=0=R'(r)$, which by means of Eq.~\eqref{eq:R} provide
\cite{Kumar:2020}
\begin{equation}
 \xi_p = \frac{\left(r^2+a^2\right)\Delta'(r)-4r\Delta(r)}{a\Delta'(r)},\label{eq:xip}
\end{equation}
\begin{equation}
 \eta_p = \frac{r^2}{a^2\Delta'(r)^2}\left[
 8\left(2a^2+r\Delta'(r)\right)\Delta(r)-16\Delta(r)^2-r^2\Delta'(r)^2
 \right].\label{eq:etap}
\end{equation}
One can define the effective inclination angle \cite{Ryan:1995}
\begin{equation}
\cos i=\frac{L}{\sqrt{L^2+\sQ}},
    \label{eq:cosi}
\end{equation}
given in terms of the two conserved quantities $L$ and $\sQ$. Now applying the above definition and by eliminating $E$ from the Eqs.~\eqref{eq:xip} and \eqref{eq:etap}, we can derive the following octic equation:
\begin{equation}
p_8(x) = \sum_{j=0}^{8} m_j x^j=0,
    \label{eq:p8=0}
\end{equation}
where
\begin{subequations}\label{eq:mj}
\begin{align}
    & m_0 = 4 u^4\nu,\label{eq:m0}\\
    & m_1 = 8 u^4 \nu(1+\alpha),\label{eq:m1}\\
    & m_2 = 4u^2\nu\left[u^2\left(3b+(1+\alpha)^2\right)-6\right],\label{eq:m2}\\
    & m_3 = 4u^2\left[3bu^2\nu(1+\alpha)-4(1+2\alpha\nu)\right],\label{eq:m3}\\
    & m_4 = 9\left(4+b^2u^4\nu\right)+8u^2\nu\left(1-\alpha^2-5b\right),\label{eq:m4}\\
    & m_5 = 8\left[bu^2(1-2\alpha\nu)-3(1-\alpha)\right],\label{eq:m5}\\
    & m_6 = 4\left[3b+(1-\alpha)^2-6b^2u^2\nu\right],\label{eq:m6}\\
    & m_7 = -4b(1-\alpha),\label{eq:m7}\\
    & m_8 = b^2,\label{eq:m8}
\end{align}
\end{subequations}
given the dimensionless parameters
\begin{subequations}\label{eq:notations}
\begin{align}
    & x = \frac{r}{M},\label{eq:x}\\
    & u = \frac{a}{M},\label{eq:u}\\
    & b = \gamma M,\label{eq:b}\\
    & \nu = \sin^2 i.\label{eq:nu}
\end{align}
\end{subequations}
This way, the geometrical parameters of the spacetime can be recast as
$\Delta(x) = 
(1-\alpha) x^2+u^2-2x-bx^3
$ and $\Sigma(x,\theta)= 
x^2+u^2\cos^2\theta
$, and he critical impact parameters become
\begin{equation}\label{eq:xip(x)}
    \xi_p(x) = \frac{
    \left[\left(x^2+u^2\right)\Delta'(x)-4x\Delta(x)\right]}{u\Delta'(x)},
\end{equation}
\begin{equation}\label{eq:etap(x)}
    \eta_p(x) = \frac{
    x^2}{u^2\Delta'(x)^2}\left[8\left(2
    u^2+x\Delta'(x)\right)\Delta(x)-16\Delta(x)^2-x^2\Delta'(x)^2\right].
\end{equation}
Accordingly, the solution to the Eq.~\eqref{eq:p8=0} will be given in terms of $x(u,\alpha,b,\nu)$ for which $x>0, 0\leq |u|< \bar{u}_c, 0\leq\alpha<1, 0\leq b< \bar{b}_c$, and $0\leq\nu\leq1$, with $\bar{u}_c=\frac{\bar{a}_c}{M}$ and $\bar{b}_c=\bg_c M$. We continue by studying the planar and polar photon orbits in the spacetime.


\subsection{Radii of planar and polar orbits in the Kerr limit}\label{subsec:planarKerr}

The case of planar orbits on the equatorial plane corresponds to $\sQ=0$ (i.e. $i=0, \pi$ or $\nu=0$). When a Kerr black hole is concerned with (i.e. $\alpha=b=0$), the octic equation \eqref{eq:p8=0} reduces to the cubic
\begin{equation}\label{eq:p3=0}
    x^3-6x^2+9x-4u^2=0.
\end{equation}
Clearly, for the case of the Schwarzschild black hole (i.e. $u=\alpha=b=0$), the cubic \eqref{eq:p3=0} reduces to $x-3=0$, which gives the radius of a circular photon orbit (or the photon ring) on the equatorial plane. The general form of  Eq.~\eqref{eq:p3=0} has been solved in Refs.~\cite{Bardeen:1972a,Chandrasekhar:1998}, which based on our notations in Eqs.~\eqref{eq:notations}, can be expressed as \cite{Tavlayan:2020}
\begin{equation}
x_{p_\pm}=2\left[1+\cos\left(\frac{2}{3}\arccos\left(\pm u\right)\right)\right],
    \label{eq:xpmKerr}
\end{equation}
that puts the retrograde (counter-rotating) photon orbits in the domain $3\leq x_{p_+}\leq 4$, and the prograde (corotating) ones in the domain $1\leq x_{p_-}\leq 3$.

It is also possible to apply the same method to calculate the radius of polar orbits, for which $L=0$ (i.e. $i=\pm\frac{\pi}{2}$ or $\nu=1$). This way, we obtain another cubic
\begin{equation}\label{eq:p3pol=0}
    x^3-3x^2+u^2 x+ u^2=0,
\end{equation}
with the unique solution \cite{Tavlayan:2020}
\begin{equation}
x_{\mathrm{pol}}=1+2\sqrt{1-\frac{u}{3}}\cos\left(\frac{1}{3}\arccos\left(
\frac{1-u}{\sqrt{\left(1-\frac{u}{3}\right)^3}}
\right)\right),
    \label{eq:xpKerr}
\end{equation}
which is valid in the domain $1\leq x_{\mathrm{pol}}\leq 3$. Photons on these orbits are in the sate of corotation with the black hole as a consequence of the dragging of their inertial frames.

\subsection{Radii of planar and polar orbits in the general case}\label{subsec:planarGeneral}

For the case of planar orbits, the octic equation \eqref{eq:p8=0} reduces to the quintic
\begin{equation}
p_5(x)=\sum_{j=0}^5 \bar{m}_j x^j = 0,
    \label{eq:p5=0}
\end{equation}
where
\begin{subequations}\label{eq:bmj}
\begin{align}
    & \bar{m}_0 =  -4u^2 <0,\label{eq:bm0}\\
    & \bar{m}_1 = 9>0,\label{eq:bm1}\\
    & \bar{m}_2 = 2\left[bu^2-3(1-\alpha)\right],\label{eq:bm2}\\
    & \bar{m}_3 = 3b+(1-\alpha)^2>0,\label{eq:bm3}\\
    & \bar{m}_4 = -b(1-\alpha)<0,\label{eq:bm4}\\
    & \bar{m}_5 = \frac{b^2}{4}>0.\label{eq:bm5}
\end{align}
\end{subequations}
 According to the coefficients given in Eqs.~\eqref{eq:bmj}, the quintic \eqref{eq:p5=0} has either three or five sign variations, depending on the sign of $\bar{m}_2$. Hence, the Descartes' rule of sign changes implies that
\begin{equation}\label{eq:signVar}
  \mathrm{number~ of~positive~ roots~ of} ~ p_5(x) = \left\{
             \begin{array}{ll}
             0<u<\sqrt{\frac{3}{b}(1-\alpha)},&\,~~~~ 3, 1, \mathrm{or}~0, \\
             \medskip\\
            u>\sqrt{\frac{3}{b}(1-\alpha)},&\,~~~~ 5, 3, 1, \mathrm{or}~0, \\
             \end{array} \right.
\end{equation}
according to whether $\bar{m}_2>0$ or $\bar{m}_2<0$. According to the celeberated Abel–Ruffini theorem, it is impossible to express exact solutions to polynomials beyond quartic, in terms of finite radicals. There are, however, several algebraic methods at hand, that can propose analytical solutions to quintic equations. Among those, the Mellin hypergeometric representation \cite{Mellin:1921} and the Hermite–Kronecker–Brioschi characterization in terms of elliptic integrals \cite{Hermite:1858,Brioschi:1858,Kronecker:1858} can be named. 
It is common to reduce the general quintic to its Bring-Jerrard form \cite{Bring:1786}, by means of a Tschirnhausen transformation. The solutions to the quintic can be then given in terms of generalized hypergeometric functions and finally, in the form of Bring radicals. Applying a proper Tschirnhausen transformation, reduces the quintic \eqref{eq:p5=0} into the Bring-Jerrard form (see appendix \ref{app:A}) 
\begin{equation}
p_{5_\mathrm{BJ}}(\ft) = \ft^5- \ft + K= 0, 
\label{eq:p5=0Bring}    
\end{equation}
where $K = d_5\ff^5$ and $\ff=d_4^{-\frac{1}{4}}$ (see Eqs.~\eqref{eq:A13d} and \eqref{eq:A13e} in appendix \ref{app:A}, for the definitions of $d_4$ and $d_5$, and also the further derivations therein). The simplified Bring-Jerrard quintic \eqref{eq:p5=0Bring} has the five solutions \cite{weisstein_crc_2002}
\begin{eqnarray}
  \ft_1&=&K\mathcal{F}_1(K),\label{eq:t1sol}\\
  \ft_2&=&-\mathcal{F}_1(K)-\frac{1}{4}K\mathcal{F}_2(K)+\frac{5}{32}K^2\mathcal{F}_3(K)-\frac{5}{32}K^3\mathcal{F}_4(K),\label{eq:t2sol}\\
  \ft_3&=&-\mathcal{F}_1(K)-\frac{1}{4}K\mathcal{F}_2(K)-\frac{5}{32}K^2\mathcal{F}_3(K)-\frac{5}{32}K^3\mathcal{F}_4(K),\label{eq:t3sol}\\
  \ft_4&=&-\im\mathcal{F}_1(K)-\frac{1}{4}K\mathcal{F}_2(K)-\frac{5}{32}\im K^2\mathcal{F}_3(K)+\frac{5}{32}K^3\mathcal{F}_4(K),\label{eq:t4sol}\\
  \ft_5&=&-\im\mathcal{F}_1(K)-\frac{1}{4}K\mathcal{F}_2(K)+\frac{5}{32}\im K^2\mathcal{F}_3(K)+\frac{5}{32}K^3\mathcal{F}_4(K)\label{eq:t4sol},\label{eq:t5sol}
\end{eqnarray}
where $\im=\sqrt{-1}$, and we have used the definitions \cite{Slater:2008}
\begin{subequations}\label{eq:F1234}
\begin{align}
    & \mathcal{F}_1(K)=\mathcal{F}_2(K),\label{eq:F1234a}\\
    & \mathcal{F}_2(K) = {}_4F_3\left(\left\{\frac{1}{5},\frac{2}{5},\frac{3}{5},\frac{4}{5}\right\};\left\{\frac{1}{2},\frac{3}{4},\frac{5}{4}\right\};\frac{3125}{256}K^4\right),\label{eq:F1234b}\\
    & \mathcal{F}_3(K) = {}_4F_3\left(\left\{\frac{9}{20},\frac{13}{20},\frac{17}{20},\frac{21}{20}\right\};\left\{\frac{3}{4},\frac{5}{4},\frac{3}{2}\right\};\frac{3125}{256}K^4\right),\label{eq:F1234c}\\
    & \mathcal{F}_4(K) = {}_4F_3\left(\left\{\frac{7}{10},\frac{9}{10},\frac{11}{10},\frac{13}{10}\right\};\left\{\frac{5}{4},\frac{3}{2},\frac{7}{4}\right\};\frac{3125}{256}K^4\right),\label{eq:F1234d}\\
\end{align}
\end{subequations}
with 
\begin{equation}
{}_kF_{k-1}\left(\left\{\frac{1}{k+1},\cdots,\frac{k}{k+1}\right\};\left\{\frac{2}{k},\frac{3}{k},\cdots,\frac{k-1}{k},\frac{k+1}{k} \right\};\mathcal{D}_k\left[s\left(1-s^k\right)\right]^k\right)
    \label{eq:kFk-1def}
\end{equation}
for $k=2,3,\cdots$ and $0\leq s\leq(k+1)^{-\frac{1}{k}}$, as the generalized hypergeometric function, in which we have defined $\mathcal{D}_k=k^{-k}(k+1)^{k+1}$. In particular, 
\begin{equation}
{}_4F_3\left(\{\beta_1,\beta_2,\beta_3,\beta_4\};\{\zeta_1,\zeta_2,\zeta_3\};s\right) = \sum_{k\geq0}\frac{(\beta_1)_k(\beta_2)_k(\beta_3)_k(\beta_4)_k}{k!(\zeta_1)_k(\zeta_2)_k(\zeta_3)_k}s^k,
    \label{eq:4F3def}
\end{equation}
with
\begin{equation}
(\beta_j)_k=\frac{\Gamma(\beta_j+k)}{\Gamma(\beta_j)},
    \label{eq:(beta)k}
\end{equation}
where $\Gamma(s)$ is the gamma function. Note that, the solutions \eqref{eq:t1sol}--\eqref{eq:t5sol} are merely analytical and their nature is only revealed by observing their numerical values. 
On the other hand, in order to obtain the relevant values of $x$ in the context of the original quintic \eqref{eq:p5=0}, we need to go though a quartic and a quadratic equation. Pursuing this, one obtains eight sets of  the solutions in the form of $x_{jil}$ with $j=\overline{1,5}$, $i=\overline{1,4}$ and $l=1,2$ (see appendix \ref{app:B} and in particular, Eq.~\eqref{eq:B12}). In Table \ref{table:1}, a variety of values for the black hole parameters have been taken into account together with their corresponding value of $K$ in the Bring-Jerrard form \eqref{eq:p5=0Bring}. Note that, the quintic provides, at least, a real value and a complex conjugate pair for $\ft$, in accordance with Eqs.~\eqref{eq:t1sol}--\eqref{eq:t5sol}. The rest of the roots can be either two distinct real values or another complex conjugate pair. These roots have been calculated by means of the aforementioned equations and have been put in their appropriate positions within Table \ref{table:1}.
\begin{table}[t]
\centering
\begin{tabular}{c | c | c | c | c } 
 \hline
 $u$ & $\alpha$ & $b$ & $K$ & $\ft$\\ [0.3ex] 
 \hline\hline
 & $ 10^{-1}$ & $10^{-2}$ & $-0.4057$ &  $-0.8504, -0.4186, 1.0828, 0.0931\pm1.0220 \im$ \\
  & $2\times 10^{-1}$ & $10^{-2}$ & $0.0378$ &  $-1.009, 0.0378,  0.9903, -0.0094 \pm 1.0002 \im$ \\
 0.85  & $ 10^{-2}$ & $10^{-4}$ & $-0.5971$ &  ${1.1133, -0.6838\pm0.1417 \im, 0.1272\pm1.0410 \im}$ \\
   & $ 10^{-4}$ & $10^{-6}$ & $-0.6470$ &  ${1.1207, -0.6954\pm0.1877 \im,  0.1351\pm1.0460 \im}$ \\
   & $ 10^{-6}$ & $10^{-8}$ & $4.3165$ &  ${-1.418 , -0.351\pm1.326 \im, 1.060\pm0.7033 \im}$ \\
   \hline
   & $ 10^{-1}$ & $10^{-2}$ & $0.3999$ &  $0.4117, 0.85394, -1.082, -0.0919\pm1.021 \im$ \\
   1 & $ 10^{-2}$ & $10^{-4}$ & $0.2499$ &  $0.9242, 0.2509, -1.055, -0.06026\pm1.009 \im$ \\
   & $ 10^{-6}$ & $10^{-8}$ & $-0.3872$ &  ${0.0000, -0.5578\pm0.5578 \im,  0.5578\pm0.5578 \im}$ \\
   \hline
   & $ 10^{-1}$ & $10^{-2}$ & $0.0653$ &  ${-1.016, 0.0653, 0.9830, -0.01628\pm1.001 \im}$ \\
   $\bar{u}_c$ & $ 10^{-4}$ & $10^{-6}$ & $0.1948$ &  ${-1.044, 0.1951, 0.94383,   -0.04762\pm1.006 \im}$ \\
   & $ 10^{-6}$ & $10^{-8}$ & $0.5652$ &  ${-1.108, -0.1219\pm1.038 \im, 0.6762\pm0.0996 \im}$ \\
   \hline
  $\bar{u}_c+0.2$ & $ 10^{-1}$ & $10^{-2}$ & $0.0820$ &  ${-1.02, 0.0820, 0.97836, -0.02041\pm1.001 \im}$ \\
 [1ex] 
 \hline
\end{tabular}
\caption{The values of the constant $K$ obtained from the method pursued in appendix \ref{app:A}, given for specific cases of the black hole parameters. Their corresponding solutions to the $\ft$-parameter in the Bring-Jerrard form \eqref{eq:p5=0Bring}, have been then calculated by means of Eqs.~\eqref{eq:t1sol}--\eqref{eq:t5sol}.}
\label{table:1}
\end{table}

\begin{figure}[t]
\centering
    \includegraphics[width=7cm]{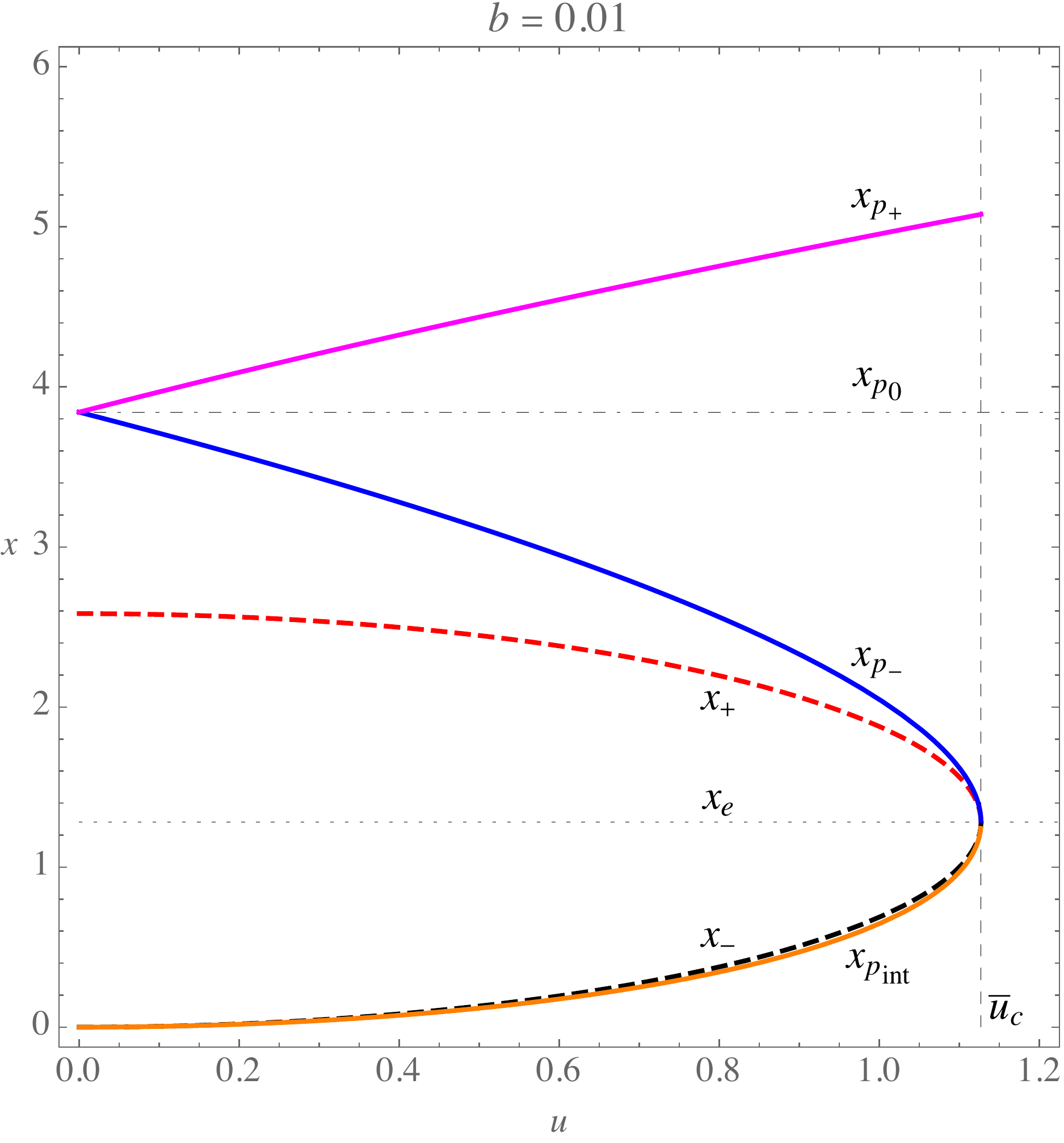}\qquad\qquad
    \includegraphics[width=7cm]{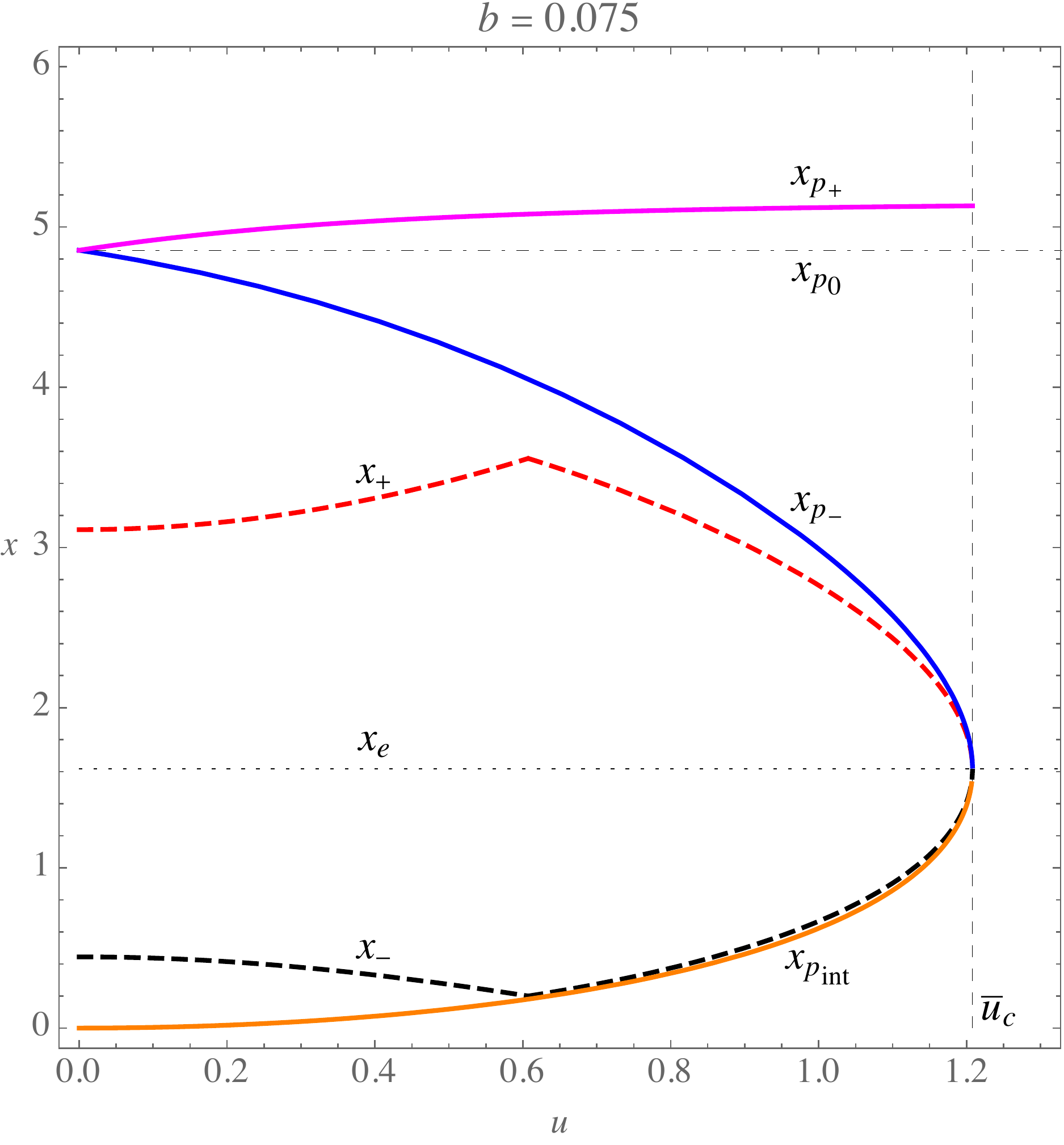}
    \caption{The behaviors of $x_{p}$ for $\alpha=0.2$ and two values for the quintessential parameter $b$, with respect to the changes in $u$, where $x_+$, $x_-$ and $x_{e}$, correspond respectively to the dimensionless radii of the event, Cauchy, and the extremal black hole horizons. The latter is obtained for $u=\bar{u}_c$. In the diagrams, the radius of prograde (retrograde) photon orbits has been indicated by $x_{p_-}$ ($x_{p_+}$), and $x_{p_{\mathrm{int}}}$ is the interior radius of photon orbits. Furthermore, $x_{p_0}$ is where the solutions $x_{p_\pm}$ connect to each other, and corresponds to $u=0$. This radius corresponds to the circular photon orbit (ring) for a quintessential Schwarzschild black hole associated with cloud of strings. Note that, since the corresponding values of the cosmological horizons are too large, they are not shown in the figures.}
    \label{fig:solutions}
\end{figure}

On the other hand, one can directly solve, numerically, the quintic \eqref{eq:p5=0} for definite values of the black hole parameters. This way, one finds that there are three real and two complex conjugate solutions for the quintic. {This is indeed expected, since the condition $\bar{m}_2>0$ is always satisfied for the chosen values of $\alpha$ and $b$ given in Table \ref{table:1}. Hence according to Eq. \eqref{eq:signVar}, the quintic can possess, at most, three positive roots.} The behavior of these solutions (termed as $x_p$) with respect to the changes in the spin parameter $u$, have been plotted in Fig.~\ref{fig:solutions} for two definite values of $b$. As it is observed in the diagrams, the solutions are simply connected by passing definite extremal cases, where the black hole characteristic hypersurfaces unite. 

Note that, since the polar orbits correspond to $i=\frac{\pi}{2}$ (or $\nu=1$), the octic \eqref{eq:p8=0} does not reduce. There is, however, another way of obtaining these radii, which passes through the definition of the impact parameter $\xi_p$. In fact, since $\xi_p$ corresponds to the angular momentum of the photons around the $\phi$-axis, a vanishing $\xi_p$ means that the photon trajectories do not have any changes in their $\phi$-coordinate; hence, they are completely polar. Applying the expression in Eq.~\eqref{eq:xip(x)}, we can observe that $\xi_p(x)=0$ is an equation of fourth order, reading as
\begin{equation}
p_4(x)=\sum_{j=0}^4\bar{\bar{m}}_jx^j=0,
    \label{eq:p4=0}
\end{equation}
in which
\begin{subequations}\label{eq:barbarm}
\begin{align}
& \bar{\bar{m}}_0=-2u^2,\\
& \bar{\bar{m}}_1 = 2(1-\alpha)u^2-4u^2,\\
& \bar{\bar{m}}_2 = 6-3bu^2,\\
& \bar{\bar{m}}_3 = -2(1-\alpha),\\
& \bar{\bar{m}}_4 = b.
\end{align}
\end{subequations}
The above quartic has the solutions (see appendix \ref{app:B})
\begin{equation}
 x_j=\bar{x}_j-\frac{\bar{\bar{m}}_3}{4},
 \label{eq:x_j_polar}
\end{equation}
in which
\begin{eqnarray}
 && \bar{x}_1 = \bar{\mathrm{A}}+\sqrt{\bar{\mathrm{A}}^2-\bar{\mathrm{B}}},\label{eq:bar_x_1}\\
 && \bar{x}_2 = \bar{\mathrm{A}}-\sqrt{\bar{\mathrm{A}}^2-\bar{\mathrm{B}}},\label{eq:bar_x_2}\\
 && \bar{x}_3 = -\bar{\mathrm{A}}+\sqrt{\bar{\mathrm{A}}^2-\bar{\mathrm{C}}},\label{eq:bar_x_3}\\
 && \bar{x}_4 = -\bar{\mathrm{A}}-\sqrt{\bar{\mathrm{A}}^2-\bar{\mathrm{C}}},\label{eq:bar_x_4}
\end{eqnarray}
where
\begin{subequations}
\begin{align}
    & \bar{\mathrm{A}} = \sqrt{\bar{\mathrm{U}}-\frac{\mathrm{A}}{6}},\\
    & \bar{\mathrm{B}} = 2\bar{\mathrm{A}}^2+\frac{\mathrm{A}}{2}+\frac{\mathrm{B}}{4\bar{\mathrm{A}}},\\
    & \bar{\mathrm{C}} = 2\bar{\mathrm{A}}^2+\frac{\mathrm{A}}{2}-\frac{\mathrm{B}}{4\bar{\mathrm{A}}},
\end{align}
\end{subequations}
with
\begin{subequations}
\begin{align}
    & {\mathrm{A}} = \bar{\bar{m}}_2-\frac{3\bar{\bar{m}}_3^2}{8},\\
    & {\mathrm{B}} = \bar{\bar{m}}_1+\frac{\bar{\bar{m}}_3^3}{8}-\frac{\bar{\bar{m}}_2\bar{\bar{m}}_3}{2},\\
    & {\mathrm{C}} = \bar{\bar{m}}_0+\frac{\bar{\bar{m}}_2\bar{\bar{m}}_3^2}{16}-\frac{3\bar{\bar{m}}_3^4}{256}-\frac{\bar{\bar{m}}_1\bar{\bar{m}}_3}{4},
\end{align}
\end{subequations}
and
\begin{equation}
\bar{\mathrm{U}} = \sqrt{\frac{\varphi_2}{3}}\cosh\left(\frac{1}{3}\arccosh\left(3\varphi_3\sqrt{\frac{3}{\varphi_2^3}}\right)\right),
    \label{eq:barU}
\end{equation}
where
\begin{subequations}
\begin{align}
    & \varphi_2 = \frac{\mathrm{A}^2}{12}+\mathrm{C},\\
    & \varphi_3 = \frac{\mathrm{A}^3}{216}-\frac{\mathrm{A}\mathrm{C}}{6}+\frac{\mathrm{B}^2}{16}.
\end{align}
\end{subequations}
The values of $x_j$ in Eq.~\eqref{eq:x_j_polar}, are therefore, the radii of the polar spherical orbits around the black hole. Note that, the solutions $x_j$ adopt one negative and three positive real values. To facilitate the comparison with the other radii in what follows in this study, we choose the particular positive value that is comparable to the radial size of the event horizon.

\subsection{Photon regions}\label{subsec:photonregions}

\begin{figure}[t]
\centering
    \includegraphics[width=4.5cm]{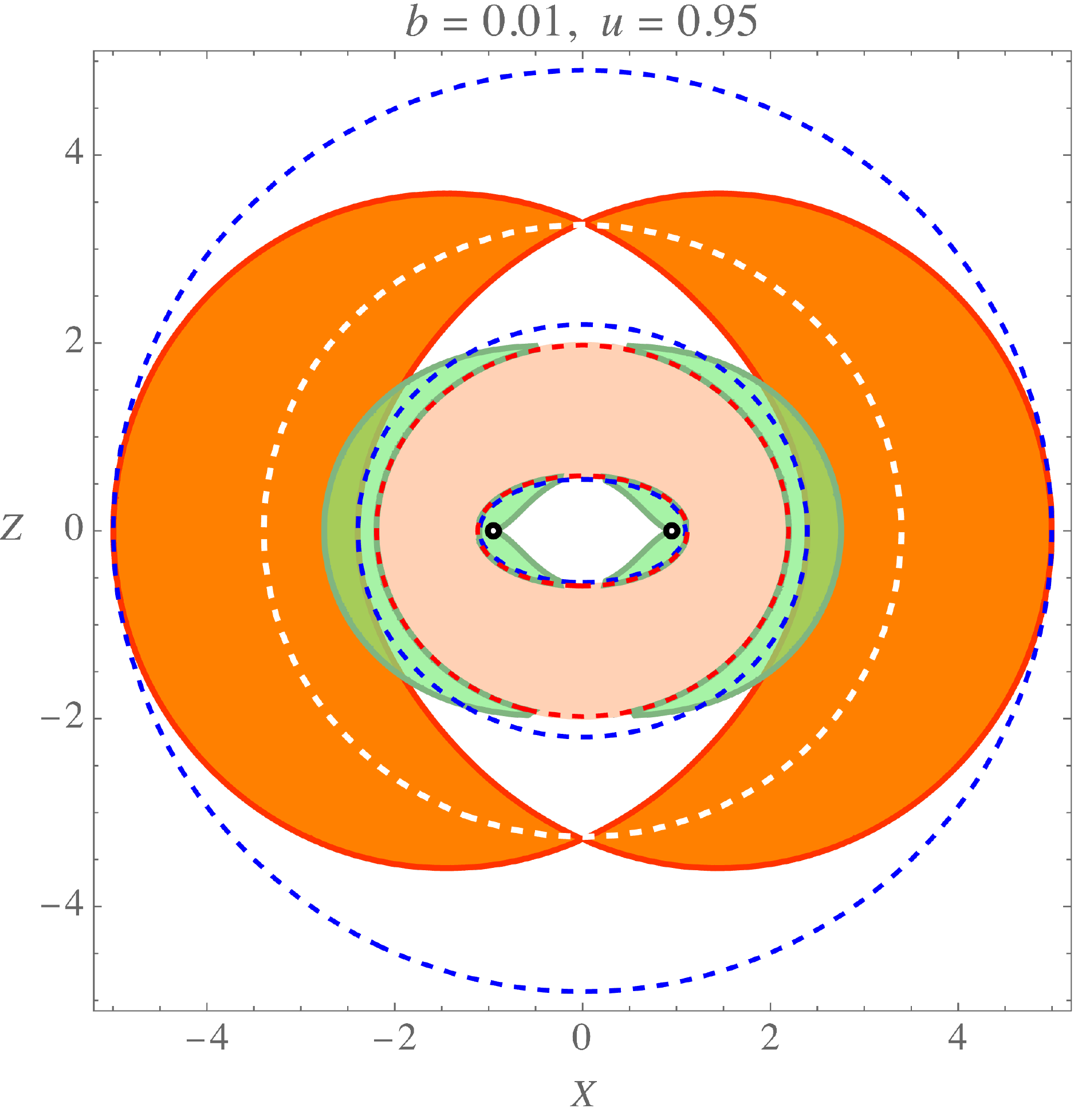}~~
    \includegraphics[width=4.5cm]{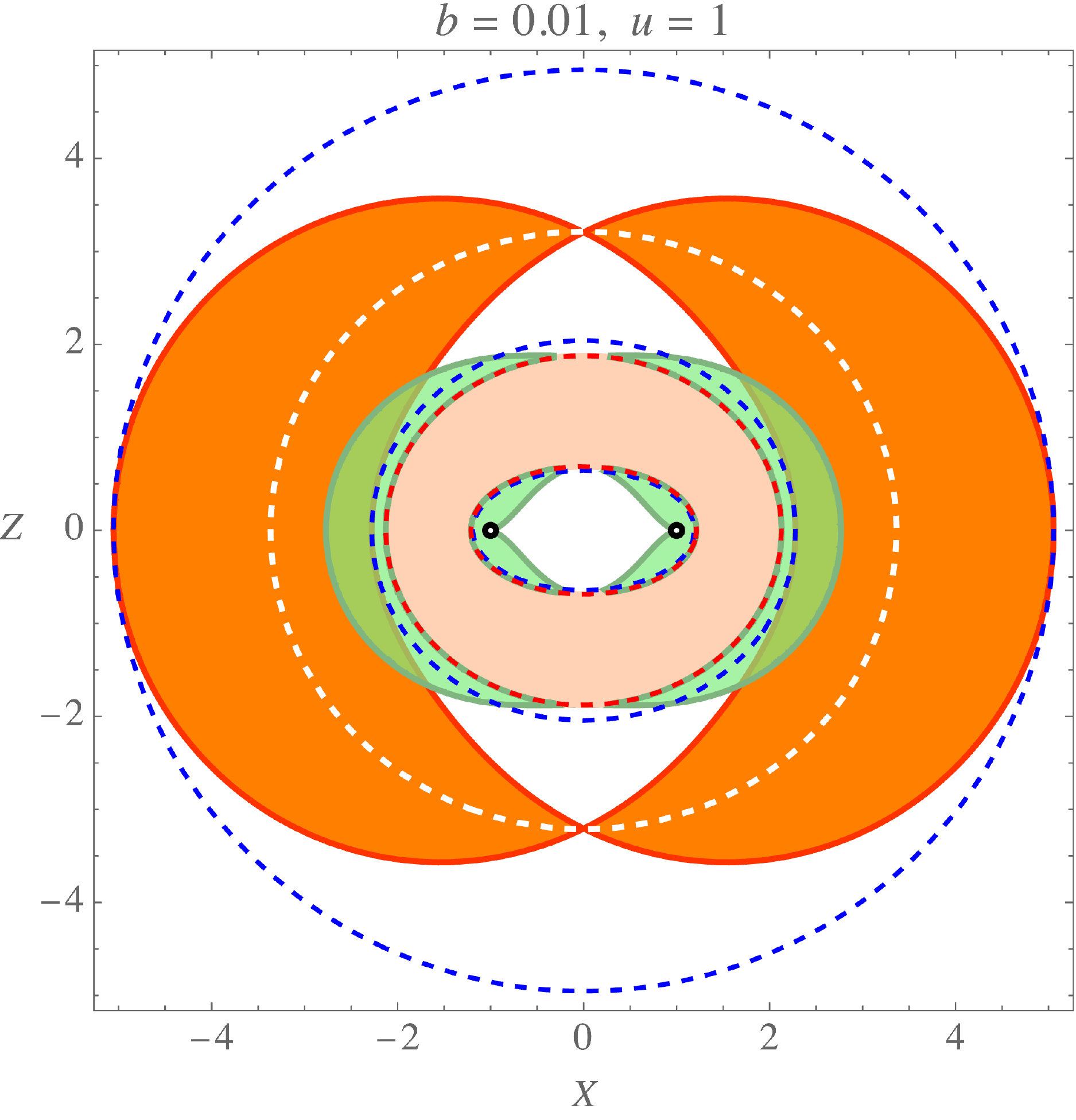}~~
    \includegraphics[width=4.5cm]{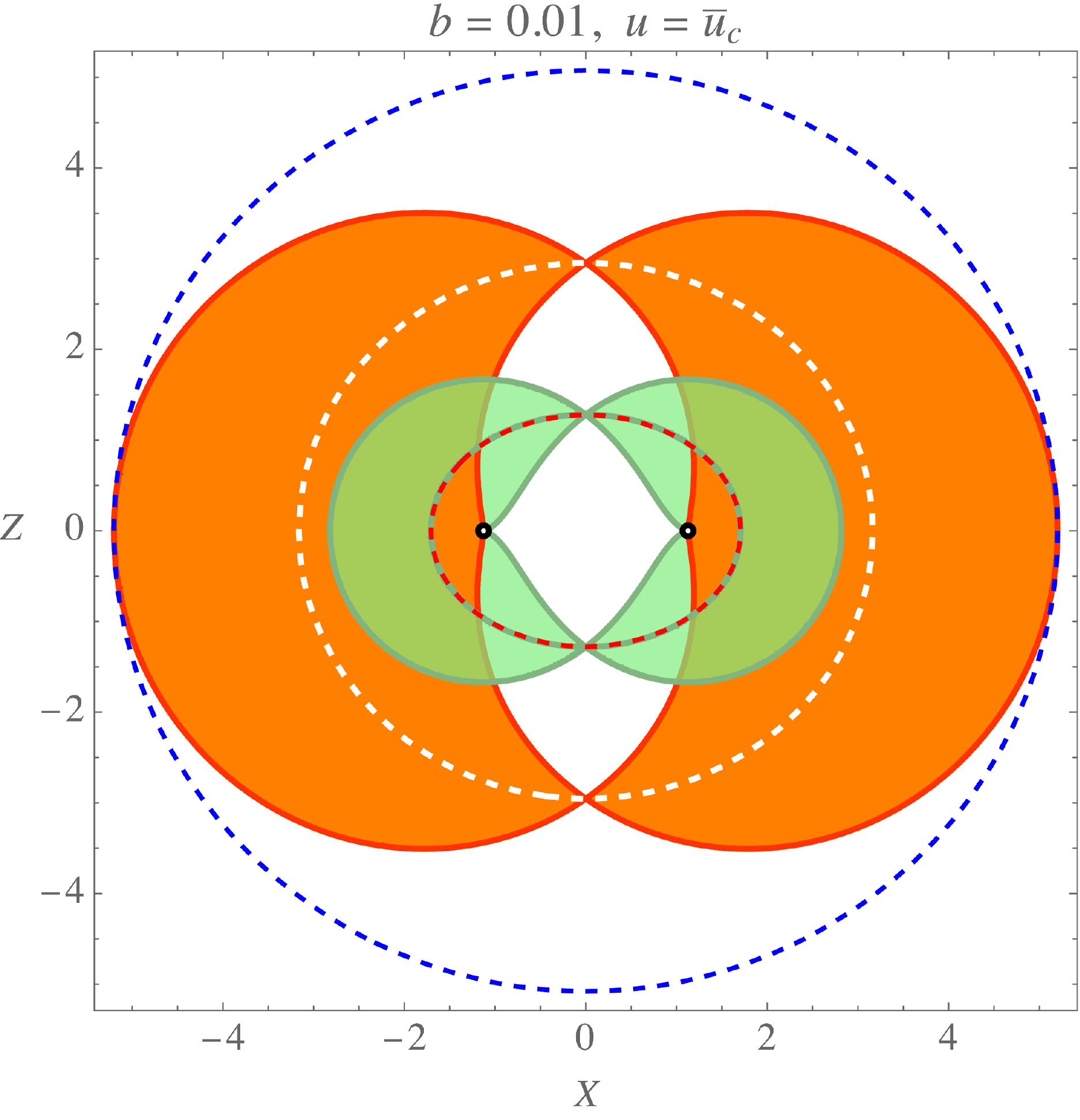}~~
    \includegraphics[width=4.5cm]{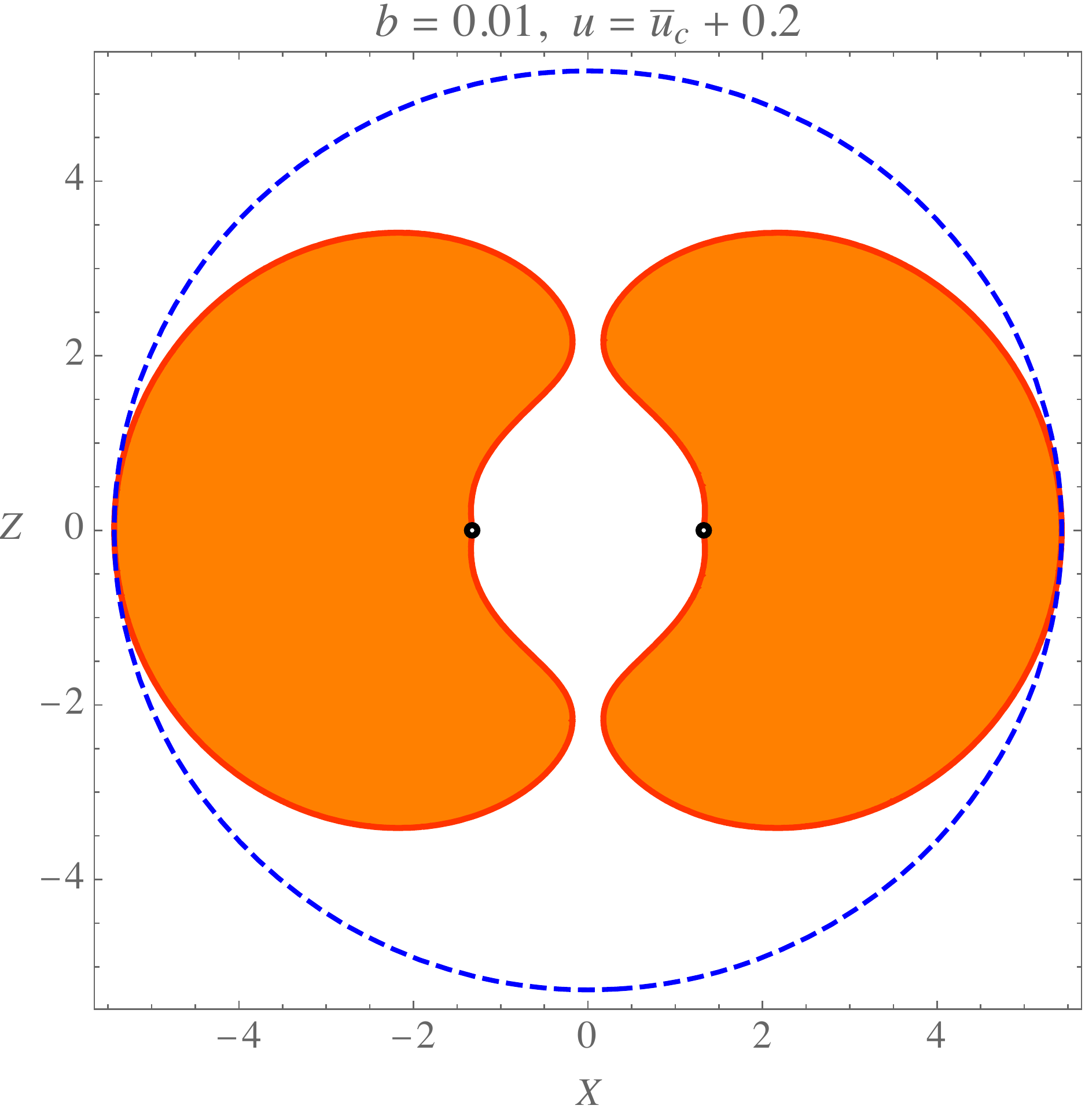}
     \includegraphics[width=4.5cm]{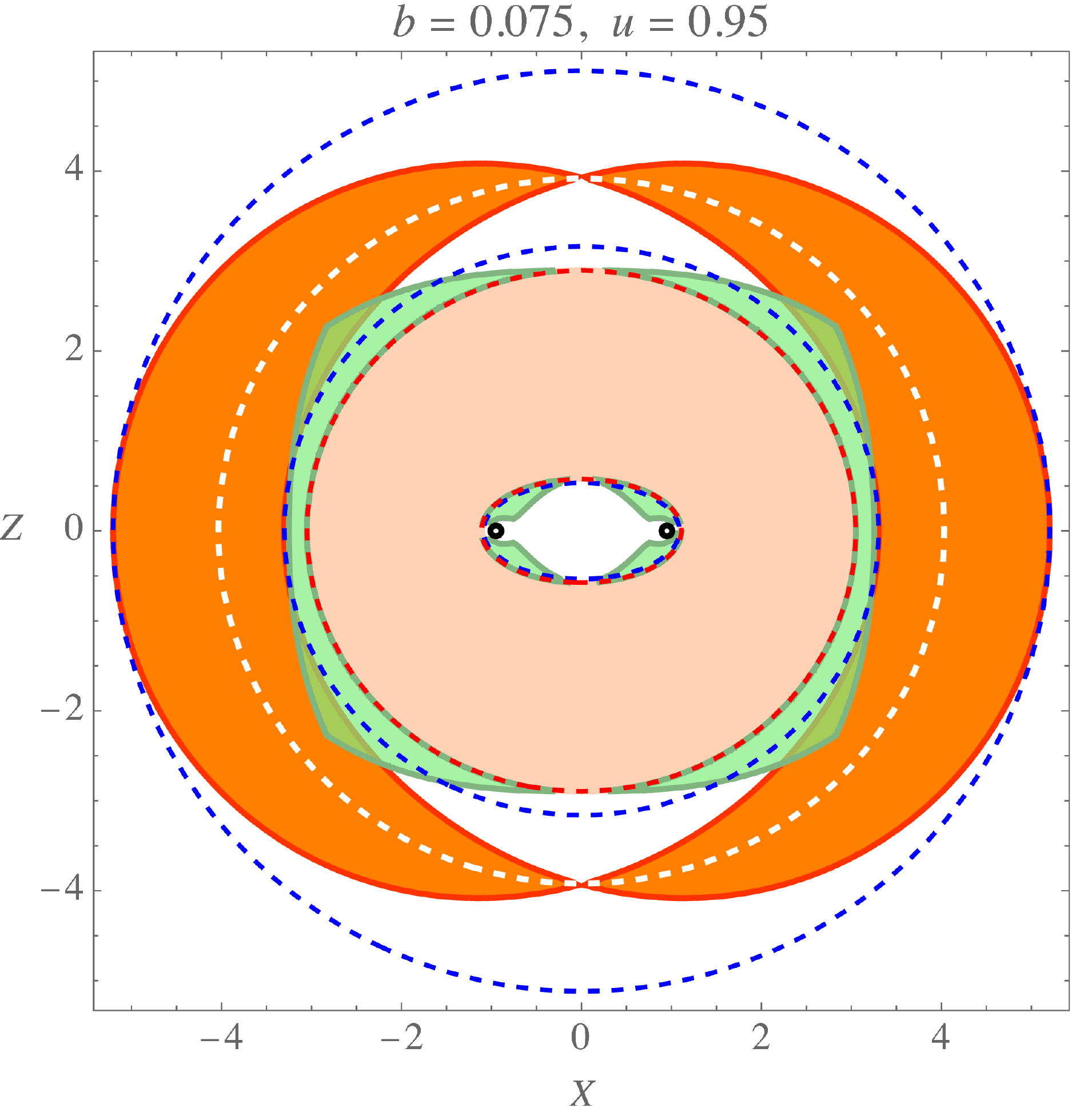}~~
    \includegraphics[width=4.5cm]{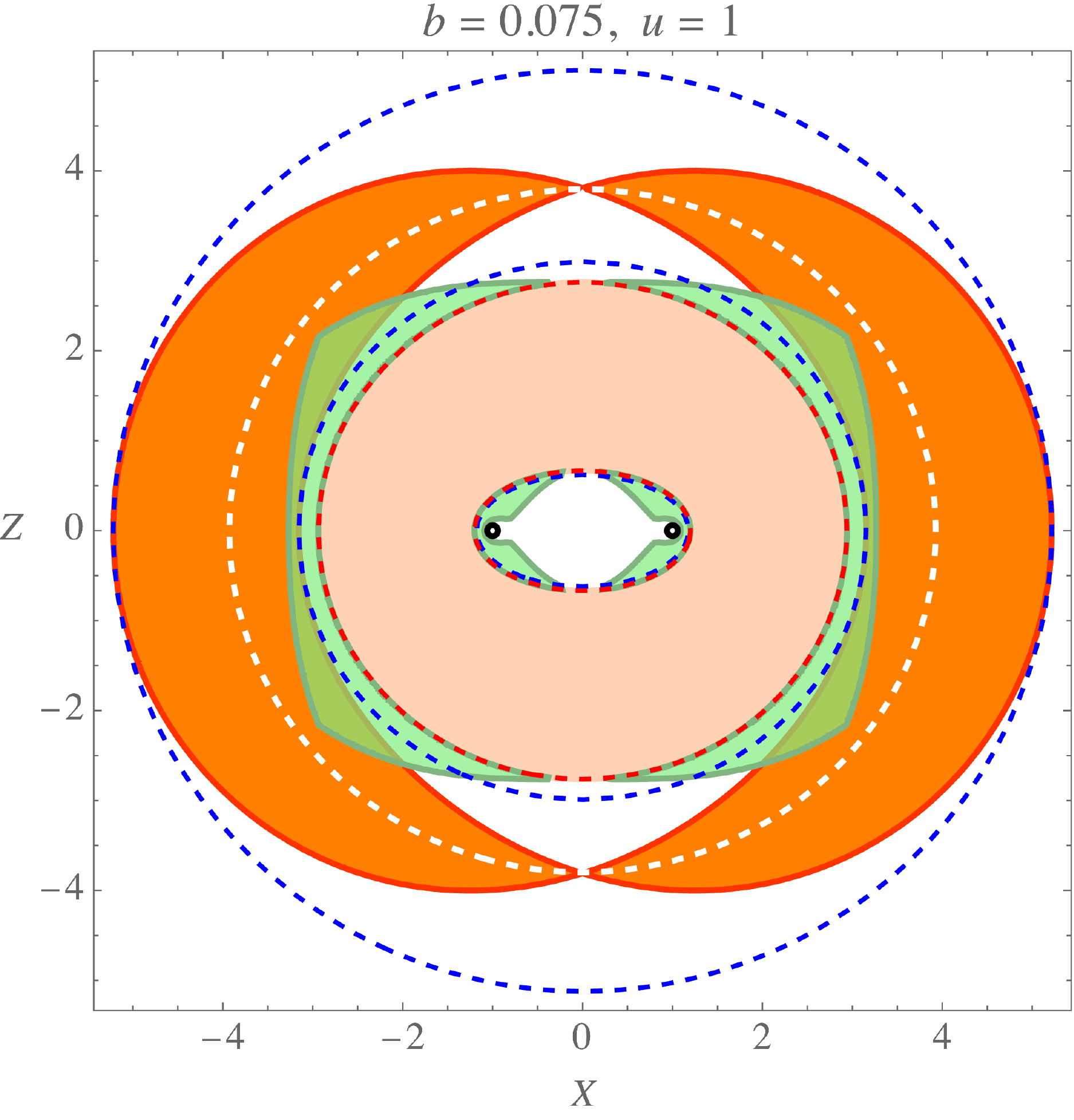}~~
    \includegraphics[width=4.5cm]{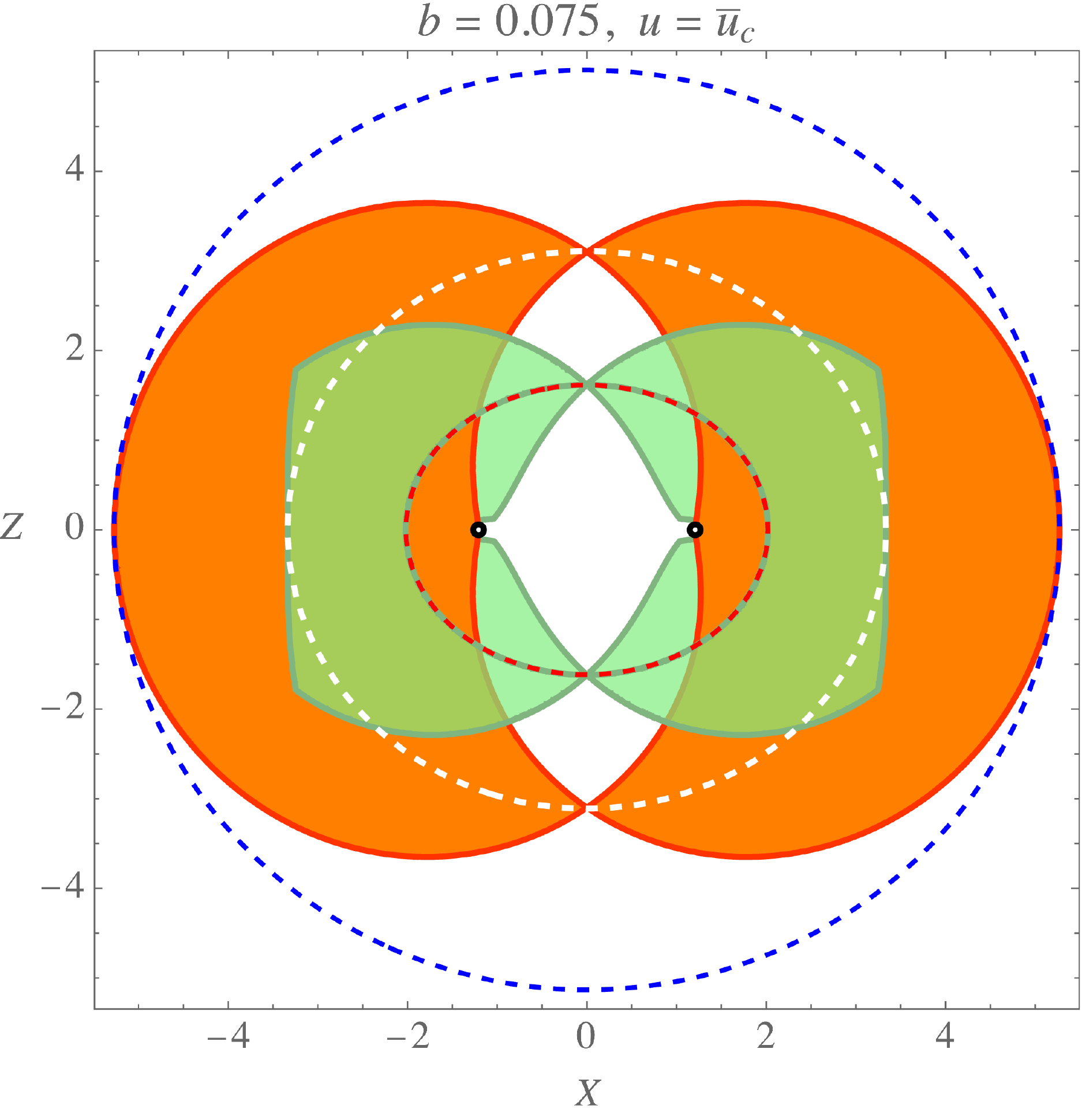}~~
    \includegraphics[width=4.5cm]{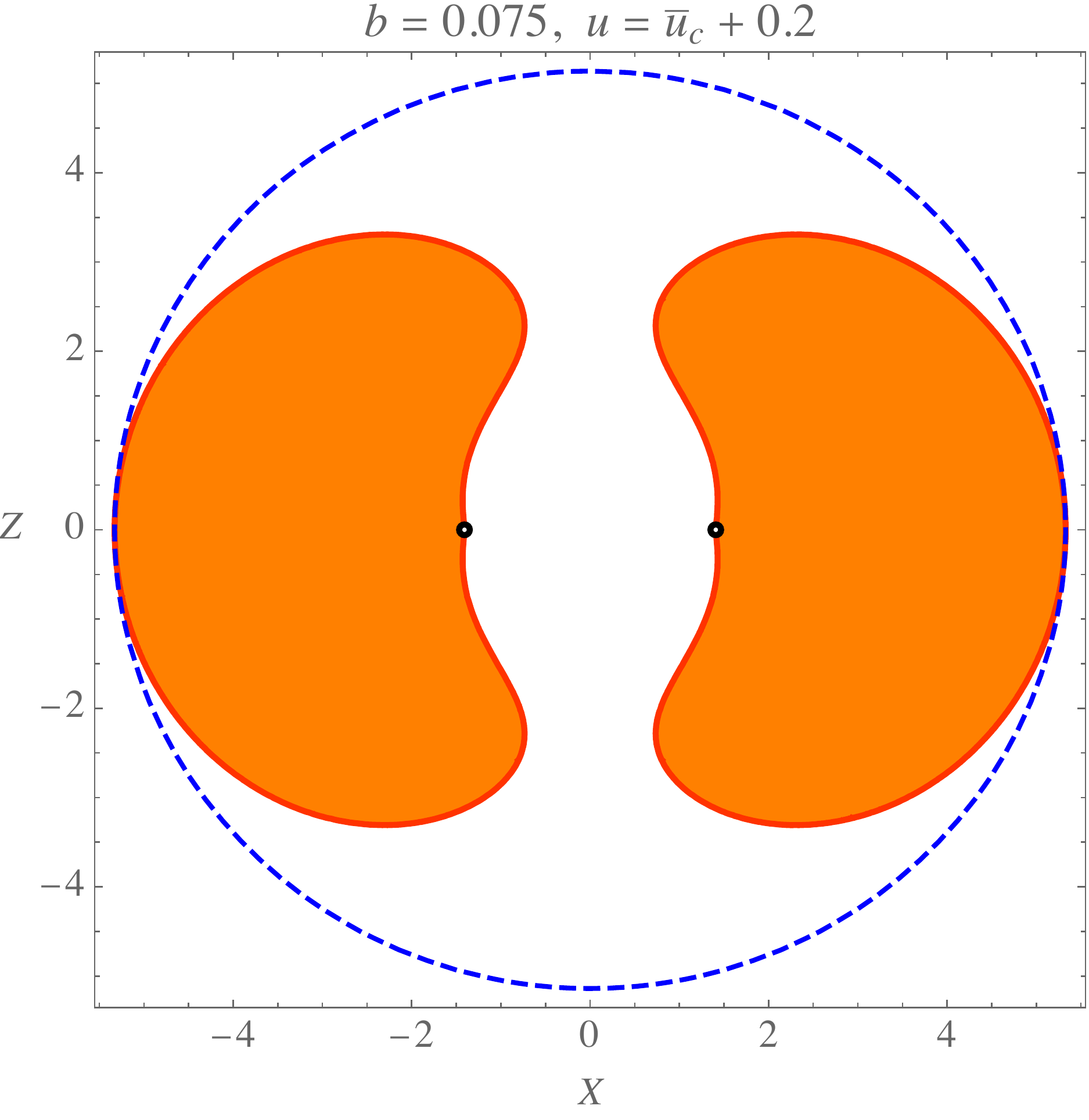}
    \caption{Examples of photon regions plotted in the Kerr-Schild coordinates in the polar plane $X$-$Z$, for $\alpha=0.2$. The diagrams correspond to two values $b = 0.01, 0.075$, and for each of the cases of $b$, they demonstrate two cases of sub-extremal black holes with $u=0.95, 1$, extremal black holes with $u=\bar{u}_c$, and super-extremal cases with $u=\bar{u}_c+0.2$. 
    %
    From smaller to larger, the dashed blue curves indicate respectively $x_{p_{\mathrm{int}}}$, $x_{p_-}$, and $x_{p_+}$. Same holds for the red dashed ones that indicate $x_-$ and $x_+$, with their separation filled with light color. Furthermore, the white dashed curve is the radius of polar orbits. The photon region which has entered the domain $0<x<x_{-}$, in the extremal cases, corresponds to the causality violation. The green regions correspond to the interior and exterior ergoregions. Finally, the black circles $\mathbf{\circ}$ indicate the cross-section of the ring singularity which is located on the equatorial plane, at $Z=0$ and $X=
    u$. For the super-extremal cases (naked singularities), the horizons disappear and the photon regions are connected directly to the ring singularity, although they do not fill the entire exterior regions. In this case, there is only one radius for the spherical orbits, which is $x_{p_+}$. However, since these regions do not connect on the axis of symmetry, $Z$, the shadow contour of the naked singularity suffers from disconnections. In this sense, the more the photon regions recede from the $Z$-axis, the more these disconnections can be observable (see Ref.~\cite{Gyulchev_observational_2020} for a rigorous discussion on the shadow of naked singularities).
    Furthermore, note that the unsmooth behavior of $r_{\mathrm{st}_+}$ for the cases of $b=0.075$, can be expected from the unsmooth behavior of characteristic hypersurfaces for this value of $b$, as seen in Fig.~\ref{fig:solutions}
    .}
    \label{fig:photonregions}
\end{figure}

As we mentioned, the planar circular orbits that correspond to the case of $\sQ=0$ and $\theta=\frac{\pi}{2}$, are on the equatorial plane. The generic existence of the spherical photon orbits, however, can be determined for $\sQ>0$ which results in $0<i<\frac{\pi}{2}$ (or $0<\nu<1$). This means that the $\theta$-coordinate also oscillates between these two angles. In fact, applying the condition $\Theta(\theta)\geq0$ to Eq.~\eqref{eq:Theta}, and exploiting the critical values for $\xi_p$ and $\eta_p$ given in Eqs.~\eqref{eq:xip} and \eqref{eq:etap}, result in the inequality
\begin{equation}
\left[4x\Delta(x)-
\Delta'(x)\Sigma(x,\theta)\right]^2\leq 16 u^2 x^2\Delta(x)\sin^2\theta.
    \label{eq:Theta>0}
\end{equation}
Now considering the Kerr-Schild coordinates
\begin{subequations}\label{eq:Kerr-Schild}
\begin{align}
    & X=
    \sqrt{x^2+u^2}\sin\theta\cos\phi,\label{eq:Kerr-Schilda}\\
    & Y=
    \sqrt{x^2+u^2}\sin\theta\sin\phi,\label{eq:Kerr-Schildb}\\
    & Z=
    x\cos\theta.\label{eq:Kerr-Schildc}
\end{align}
\end{subequations}
In the $X$-$Z$ plane, which is of our interest in this subsection, we let $\phi=0$, and therefore, $Y=0$. In Fig.~\ref{fig:photonregions}, we have applied the condition \eqref{eq:Theta>0} to demonstrate the photon regions (filled by photons on unstable spherical orbits), in the polar plane $X$-$Z$, for the same values of $b$ as used in Fig.~\ref{fig:solutions}. For each case, we have used the static limit radii given in Eqs.~\eqref{eq:rstp} and \eqref{eq:rstn}, to demonstrate the respected shape of the ergoregions. The plots correspond to the sub-extremal and extremal black holes, as well as the super-extremal cases (naked singularities), and only the exterior photon orbits have been taken into account. 



\section{Analytical solutions for the spherical photon orbits}\label{sec:analyticalSpherical}

It is often convenient to re-scale $E\lambda\rightarrow\lambda$. This is basically equivalent to letting $E=1$, which is considered to be the case for our further studies.

As it is inferred from Eq.~\eqref{eq:Theta}, photon orbits are, in general, allowed for  $\eta\geq0$. This of course includes spherical photon orbits. Below, we calculate the analytical solutions for the polar and azimuth angles that correspond to the spherical orbits.

\subsection{The latitudinal motion}\label{subsec:theta-motion}

In fact, one can recast Eq.~\eqref{eq:Theta} as \cite{fathi_analytical_2021}
\begin{equation}\label{eq:Theta_W)}
    \Theta(\theta) =  u^2\cos^2\theta\left[\left(1-\sqrt{\mathcal{W}(\theta)}\right)\left(1+\sqrt{\mathcal{W}(\theta)}\right)\right],
\end{equation}
where
\begin{equation}
\mathcal{W}(\theta) = \frac{1}{u^2}\left[\frac{\eta}{\cos^2\theta}-\frac{\xi^2}{\sin^2\theta}\right],
    \label{eq:W(theta)}
\end{equation}
is the angular gravitational potential. Accordingly and from Eq.~\eqref{eq:dtheta}, one can write the differential equation for the $\theta$-motion as
\begin{equation}
-\frac{\ed\mZ}{\ed\lambda} = \sqrt{\Theta_\mZ},
    \label{eq:dtheta_zeta}
\end{equation}
where we have defined $\mZ=\cos\theta$, and therefore
\begin{equation}
\Theta_\mZ = \eta - \chi_0\mZ^2-u^2\mZ^4.
    \label{eq:Thetazeta}
\end{equation}
where $\chi_0=\eta+\xi^2- u^2$. 
Naturally, for the planar orbits with $\eta_p=0$, the trajectories remain on the equatorial plane.

\subsubsection{Properties of the planar orbits}\label{subsubsec:Prop.Planar}

\begin{figure}[t]
\centering
    \includegraphics[width=9cm]{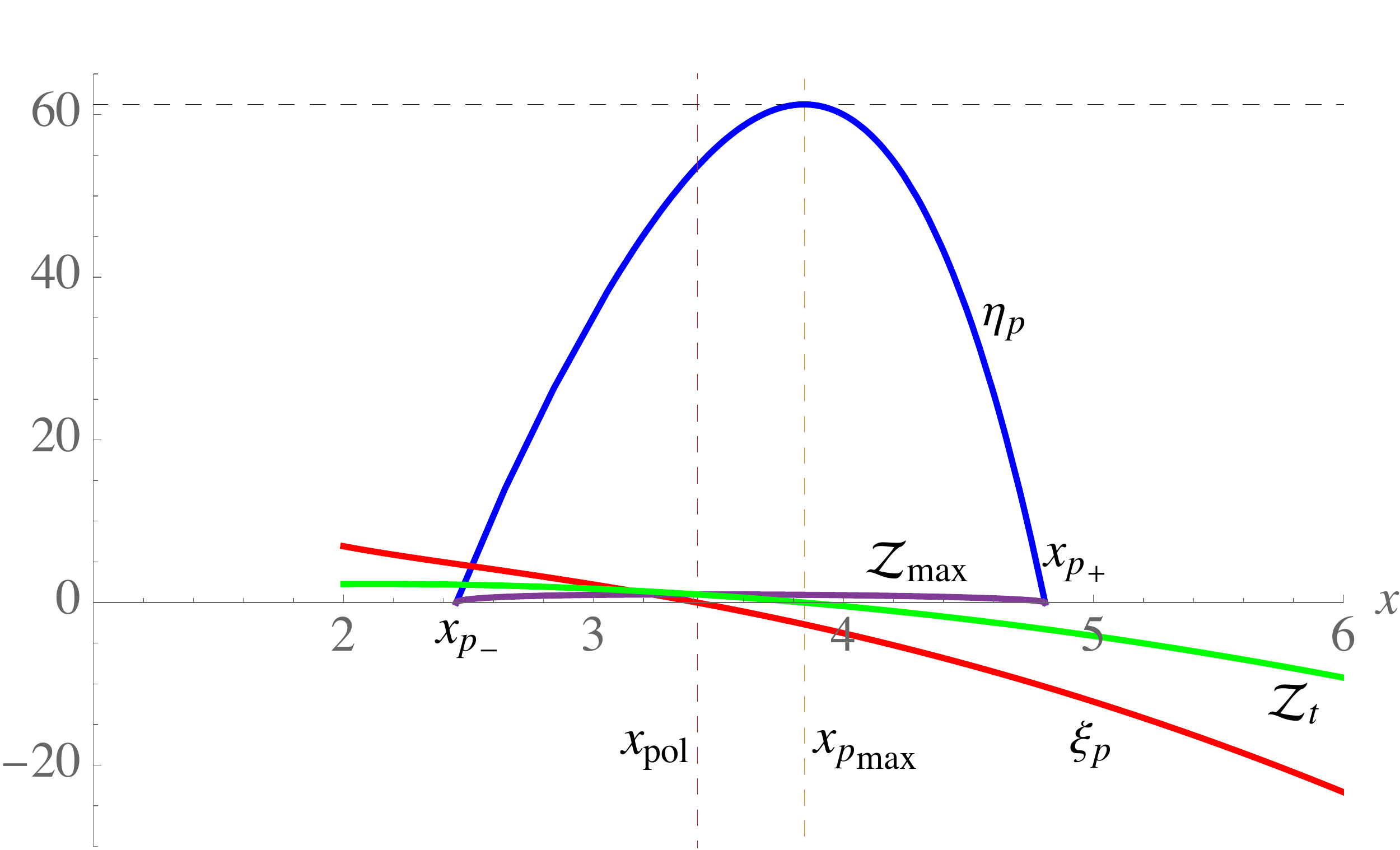}
    \caption{The behaviors of $\eta_p(x)$, $\xi_p(x)$, $\mathcal{Z}_{\max}$ and $\mathcal{Z}_t$, plotted for $u=0.85$, $\alpha=0.2$ and $b=0.01$. }
    \label{fig:propPlanar}
\end{figure}

As mentioned above, the planar orbits that correspond to $i=0$, are indeed circles on the equatorial plane. In general, the impact parameter $\eta_p$ is confined between its values at the two radii of planar orbits, $x_{p_-}$ and $x_{p_+}$ (where it vanishes). As it can be seen in Fig.~\ref{fig:propPlanar}, the $\eta_p$ parameter increases monotonically from the point $x_{p_-}$ and reaches its maximum at
\begin{equation}
x_{p_{\max}} = \frac{1-\alpha -\sqrt{\alpha ^2-2 \alpha -6 b+1}}{b}.
    \label{eq:xpmax}
\end{equation}
As before, the above value is well-defined only for $b\neq0$. For the case of a Kerr black hole (i.e. $\alpha=\beta=0$), the equation $\eta_p'(x)=0$ results in  $x_{p_{\max}}=3$ (which is indeed that of the Schwarzschild black hole). After this point, $\eta_p$ decreases monotonically until it reaches its second zero at $x_{p_+}$. At the point $x_{\mathrm{pol}}$, the impact parameter $\xi_p$ switches from positive values to negative values. At this point, the angular momentum of the photons is zero (i.e., orbits are along the axis of symmetry and $i=\frac{\pi}{2}$). Note that, there is a relation between $\xi_p$ and the maximum latitude $\mZ_{\max}$ reachable by the photons, which is the angular value where $\Theta_\mZ=0$. This equation gives the two values 
\begin{subequations}\label{eqzetamaxmin}
\begin{align}
 &   \mZ_{\max}^2=\frac{\chi_0}{2
 u^2 }\left(\sqrt{1+{4
 u^2{\eta}\over \chi_0^2}}-1\right),\label{zetamax}\\
&   \mZ_{\min}=-\mZ_{\max},\label{eq:zetamin}
\end{align}
\end{subequations}
that confine the $\mZ$-parameter. This way, the $\theta$-parameter oscillates in the domain $\theta\in\left[\theta_{\min},\theta_{\max}\right]$, where $\theta_{\min}=\arccos(\mZ_{\max})$ and $\theta_{\max}=\arccos(\mZ_{\min})$. 
So, letting $\eta=\eta_p(x)$, we can plot the radial profile of $\mZ_{\max}$, and compare it with the behavior of $\xi_p$ (see Fig.~\ref{fig:propPlanar}). Note that, the $\phi$-coordinate changes its sign during each orbit. This sign change can be determined by mean of the equation of motion \eqref{eq:dphi}. However, to determine the sign change of the $\phi$-coordinate in terms of the latitudinal evolution, one can solve the equation $\dot\phi=0$ for $\mZ$, which gives the latitudinal turning points of the azimuth angle. After doing the proper substitutions, this equation provides the value
\begin{equation}
\mathcal{Z}_t = \frac{x^2 \left[b x^2+2 (\alpha -1) x+6\right]}{u^2 \left[3 b x^2+2 (\alpha +1) x+2\right]},
    \label{eq:Zt}
\end{equation}
whose radial profile has been shown in Fig.~\ref{fig:propPlanar}. As it is observed from the figure, the physically reliable segments are where $|\mZ_t|< |\mZ_{\max}|$. According to the figure, this inequality holds when $x_{\mathrm{pol}}<x<x_{p_{\max}}$, corresponding to $\xi_{p_m}<\xi_p<0$, where
\begin{equation}
\xi_{p_m}\equiv\xi_p(x_{p_{\max}})=\frac{u \left[(2-\alpha) \left(\alpha +\sqrt{(1-\alpha)^2-6 b}-1\right)+4 b\right]}{(1-\alpha) \left(\alpha +\sqrt{(1-\alpha)^2-6 b}-1\right)+4 b}.
    \label{eq:xi_xpmax}
\end{equation}
Such orbits, therefore, do not move in a fixed azimuth direction. We continue by considering the more general cases.

\subsubsection{The case of $\eta>0$ (non-planar orbits)}\label{subsub:eta>0}

The equation of motion \eqref{eq:dtheta_zeta} can be integrated directly to provide the analytical solution for the evolution of the $\theta$-coordinate. This yields
\begin{equation}\label{eq:thetasol>0}
\theta(\lambda)=
\arccos\left(\mZ_{\max} -\frac{3}{12\wp\left(\kappa_0 \lambda\right)+\psi_0}\right),
\end{equation}
in which $\wp(\dots;g_2,g_3)\equiv\wp(\dots)$ is the Weierstra{\ss}ian $\wp$ function, with the invariants $g_2$ and $g_3$. In Eq.~\eqref{eq:thetasol>0}, we have defined
\begin{subequations}
\begin{align}
    & \kappa_0 = 
    u \sqrt{2 \mZ_{\max}\left(\mZ_0^2+\mZ_{\max}^2\right)},\label{eq:kappa0}\\
    & \psi_0 = \frac{\mZ_0^2+5 \mZ_{\max}^2}{2\mZ_{\max}\left(\mZ_0^2+\mZ_{\max}^2\right)},\label{eq:psi0}
\end{align}
\end{subequations}
with 
\begin{equation}\label{eq:zeta0}
    \mZ_{0}^2={\chi_0\over2
    u^2}\left(\sqrt{1+{4
    u^2{\eta}\over \chi_0^2}}+1\right),
\end{equation}
and 
\begin{subequations}\label{eq:g2,3}
\begin{align}
 &   g_2 = \frac{\mZ_0^4+\mZ_{\max}^4-14 \mZ_0^2 \mZ_{\max}^2}{48 \mZ_{\max}^2 \left(\mZ_0^2+\mZ_{\max}^2\right)^2},\label{eq:g2}\\
& g_3=\frac{33 \mZ_0^4 \mZ_{\max}^2-33 \mZ_0^2 \mZ_{\max}^4+\mZ_0^6-z_{\max}^6}{1728 \mZ_{\max}^3 \left(\mZ_0^2+\mZ_{\max}^2\right)^3}.\label{eq:g3} 
\end{align}
\end{subequations}
As an example, in Fig.~\ref{fig:W,theta}, the profile of $\mathcal{W}(\theta)$ together with the corresponding evolution of the $\theta$-coordinate, have been plotted for specific values of the black hole parameters and the inclination, which lead to the two real values $x_p=x_{p_{1,2}}$ from Eq.~\eqref{eq:p8=0}. 
\begin{figure}[t]
\centering
    \includegraphics[width=8cm]{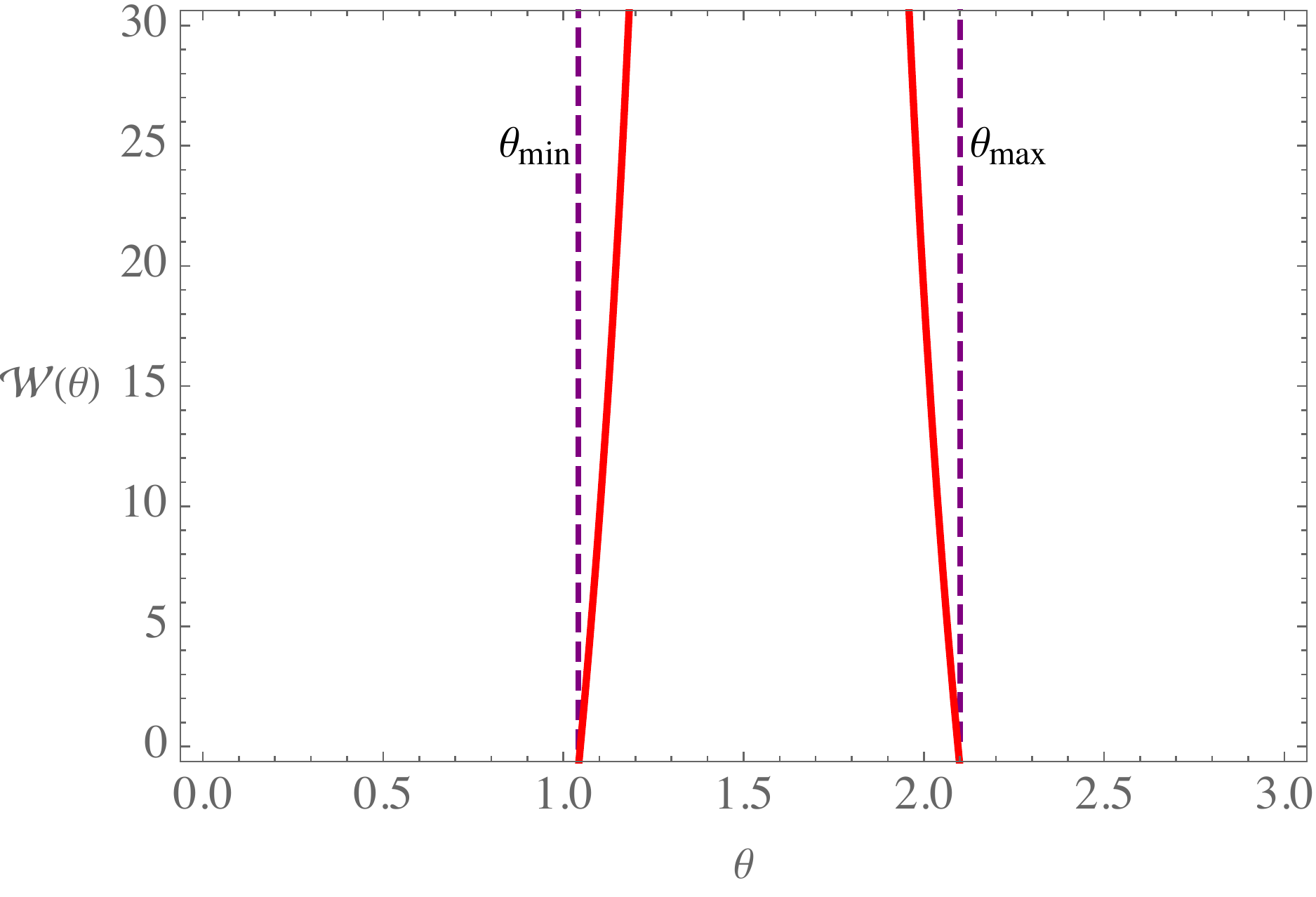}\qquad
    \includegraphics[width=8cm]{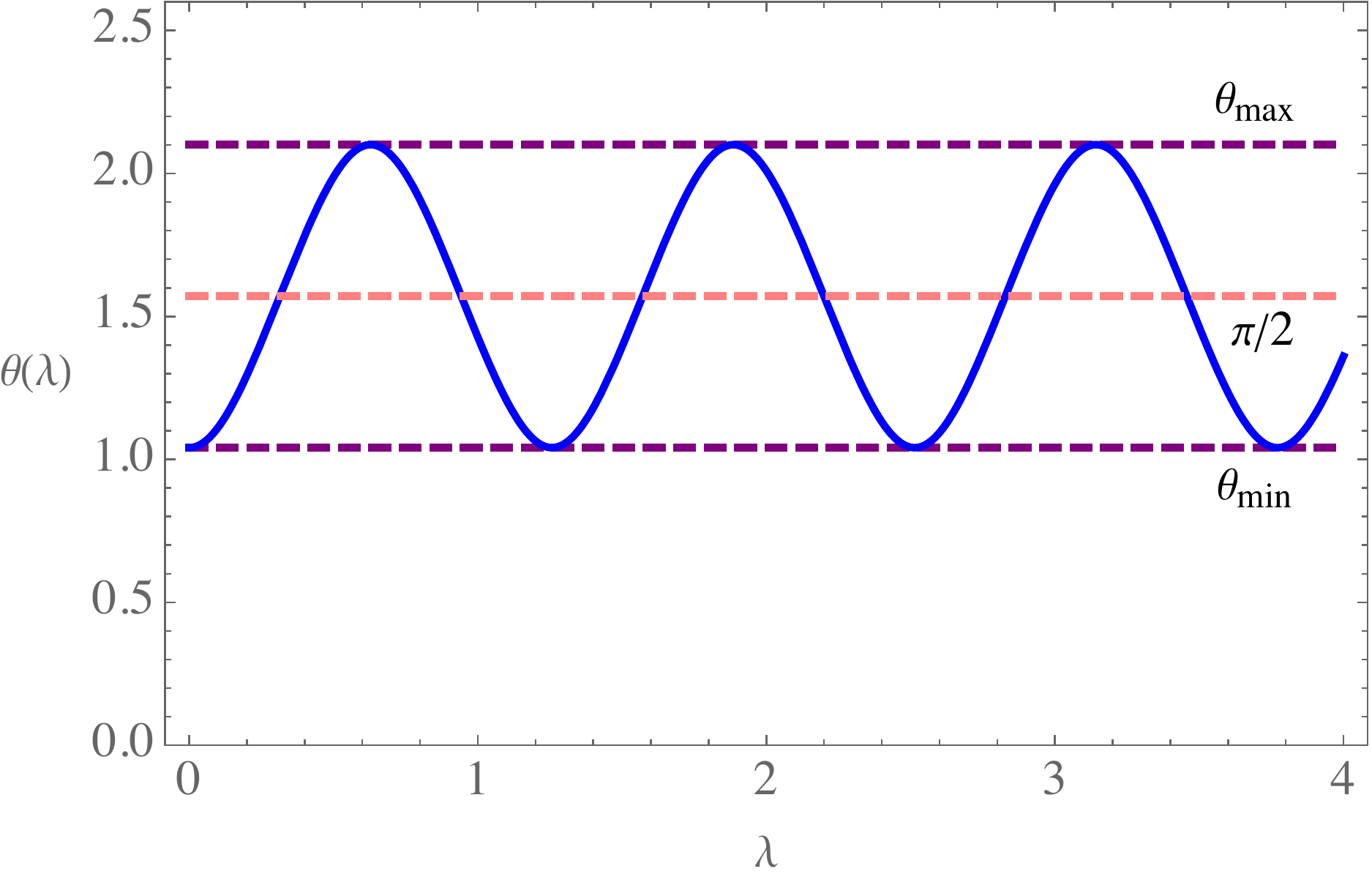}~(a)
    \includegraphics[width=8cm]{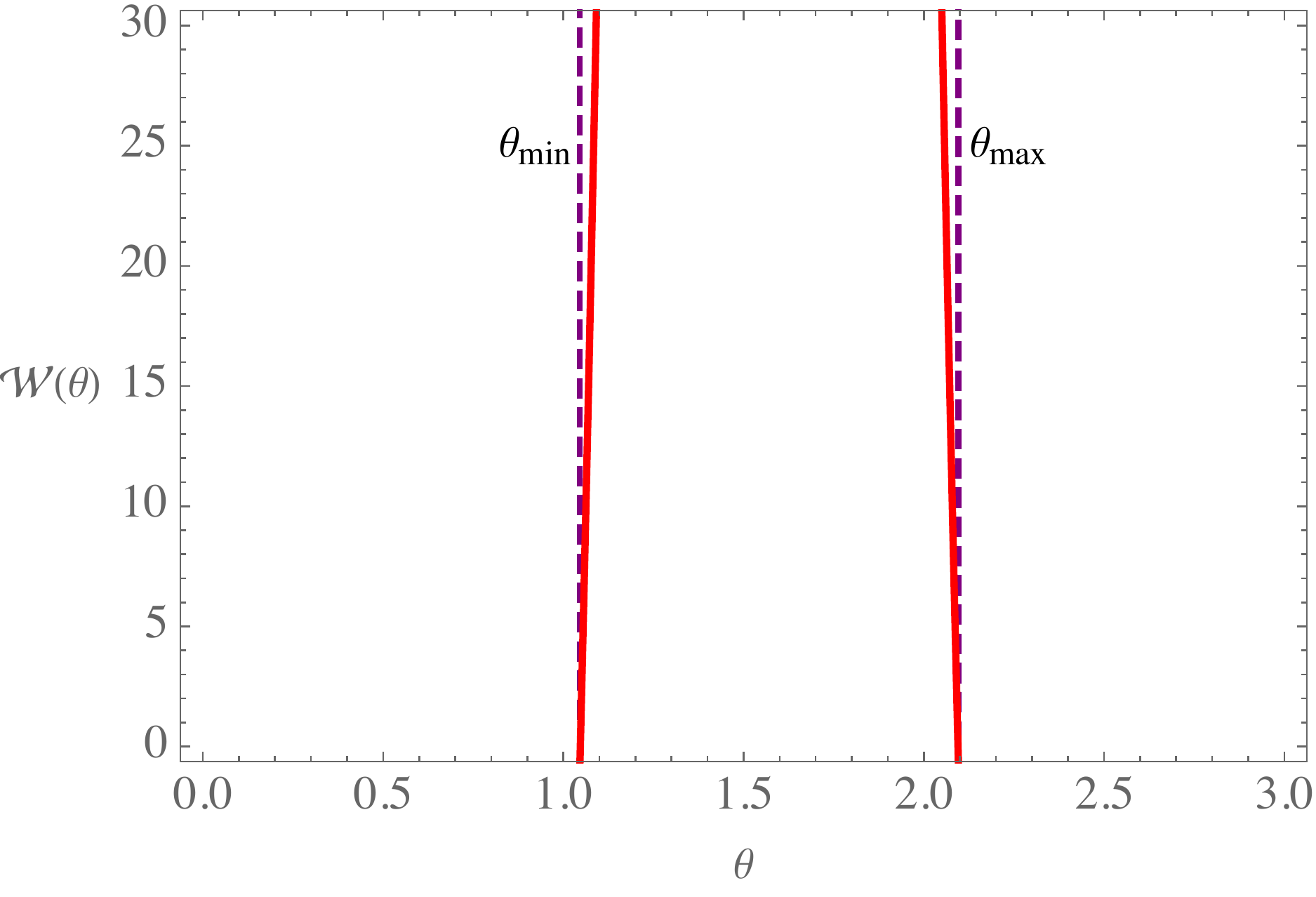}\qquad
    \includegraphics[width=8cm]{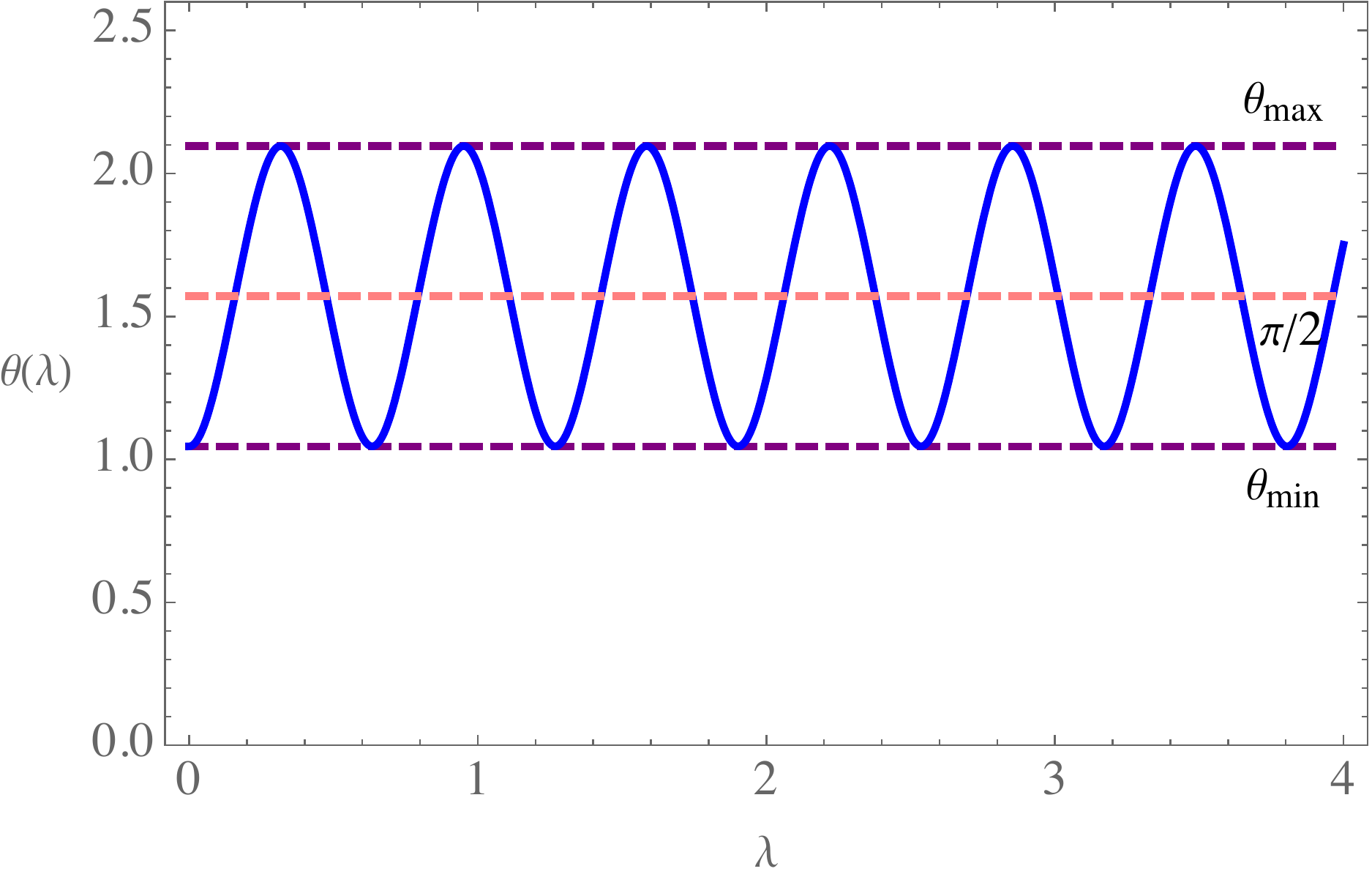}~(b)
    \caption{The profile of $\mathcal{W}(\theta)$ and the behavior of $\theta(\lambda)$ plotted for the spherical photon orbits with 
    the inclination angle $i=30^\circ$ (or $\nu=\frac{1}{4}$). The octic \eqref{eq:p8=0} has been solved numerically for $u=0.85$, $\alpha=0.2$ and $b=0.01$, leading to the two real radii (a) $x_{p_1} = 2.536$ (prograde) and (b) $x_{p_2}=4.610$ (retrograde), that correspond respectively to $\xi_{p_1}=4.367, \eta_{p_1}=6.358$ and $\xi_{p_2}=-8.601, \eta_{p_2}=24.659$. Where the orbits pass the $\theta=\frac{\pi}{2}$ plane, nodes will appear in the way of the spherical orbits. 
    It is also evident the that period of the latitudinal oscillations for the prograde orbits is larger compared to that for the retrograde ones.
    %
   }
    \label{fig:W,theta}
\end{figure}

\subsubsection{Period of the latitudinal motion}\label{subsub:nodes}

As it can be observed from the latitudinal motion in Fig.~\ref{fig:W,theta}, the temporal evolution of the $\theta$-coordinate includes points, at which, the periodic (wave-like) motion passes the $\theta=\frac{\pi}{2}$ line. These points are the so-called {\textit{nodes}}, and for each full oscillation of the function, there are two of them. In fact, by applying Eq.~\eqref{eq:thetasol>0}, the Mino time for the nodes can be obtained as
\begin{equation}
\lambda_{\mathrm{nod}} = \frac{1}{\kappa_0}\ss\left(\frac{1}{4\mZ_{\max}}-\frac{\psi_0}{12}\right),
    \label{eq:lambdaNode}
\end{equation}
where $\ss(\cdots)\equiv\wp^{-1}(\cdots;g_2,g_3)$. On the other hand, as it can also be inferred from the figures, the period of the oscillations of the polar angle $\theta$ differs between the prograde and the retrograde orbits. In fact, by applying Eq.~\eqref{eq:dtheta_zeta}, it is straightforward to calculate the period of the latitudinal oscillations for the cycle $\theta_{\min}\rightarrow\theta_{\max}\rightarrow\theta_{\min}$ (or $\mZ_{\max}\rightarrow\mZ_{\min}\rightarrow\mZ_{\max}$). Using the definition in Eq.~\eqref{eq:Thetazeta}, the general relation for this period is obtained as
\begin{equation}
\mathcal{T}_\lambda = \frac{2\pi\mZ_{\max}}{\sqrt{\eta_p(x)}},
    \label{eq:periodZ}
\end{equation}
for oscillations around $\mZ=0$. For the particular cases of Fig.~\ref{fig:W,theta}, it is found that $\mathcal{T}_{\lambda_1} = 1.259$ for prograde orbits, and $\mathcal{T}_{\lambda_2} = 0.634$ for retrograde ones.

\subsection{The azimuth motion}\label{subsec:phi-motion}

The evolution of the azimuth angle in Eq.~\eqref{eq:dphi}, can be recast as
\begin{equation}
\phi(\lambda) = \bar{\mathcal{C}}_p(x)\Phi_{\theta_1}(\lambda)+\Phi_{\theta_2}(\lambda),
    \label{eq:dphi_new}
\end{equation}
for the spherical photon orbits, in which 
\begin{subequations}\label{eq:cpPhitheta}
\begin{align}
    & \bar{\mathcal{C}}_p(x)=\frac{
    u}{\Delta(x)}\left[
    \left(x^2+u^2\right)-
    u\xi_p(x)-\Delta(x)\right],\label{eq:cp}\\
    & \Phi_{\theta_1}(\lambda) = \int^{\theta(\lambda)}\frac{\ed\theta}{\sqrt{\Theta(\theta)}},\label{eq:Phitheta1}\\
    & \Phi_{\theta_2}(\lambda) = \xi_p\int^{\theta(\lambda)}\frac{\ed\theta}{\sin^2\theta\sqrt{\Theta(\theta)}}.\label{eq:Phitheta2}
\end{align}
\end{subequations}
At any fixed radius of spherical orbits, we have $\bar{\mathcal{C}}_p(x)\rightarrow\bar{\mathcal{C}}_p(x_p)$. 
The integral \eqref{eq:Phitheta1} can be inferred directly from Eq.~\eqref{eq:thetasol>0}, yielding
\begin{equation}
\Phi_{\theta_1}(\lambda) = \frac{\ss\left(\mathcal{U}_\theta+\frac{1}{4}{\psi_0}\right)}{
\kappa_0},
\label{eq:Phitheta1-sol}
\end{equation}
with $\kappa_0$ and $\psi_0$ given in Eqs.~\eqref{eq:kappa0} and \eqref{eq:psi0}, and 
\begin{equation}
 \mathcal{U}_\theta = \frac{1}{4\left(\mathcal{Z}_{\max}-\cos\theta\right)}-\frac{\psi_0}{3},
    \label{eq:Utheta}
\end{equation}
where $\mZ_{\max}$ is given in Eq.~\eqref{zetamax}. Applying the same analytical methods, we get 
\begin{equation}
\Phi_\theta(\lambda) = \mathcal{K}_0\left[
\mathcal{K}_1\mathscr{F}_1(\mathcal{U}_\theta)-\mathcal{K}_2\mathscr{F}_2(\mathcal{U}_\theta)-\kappa_0\lambda
\right],
    \label{eq:Phitheta-sol}
\end{equation}
where
\begin{equation}
\mathscr{F}_j(\mathcal{U}_\theta) = \frac{1}{\wp'(\upsilon_j)}\left[
\ln\left(
\frac{\sigma\left(\ss(\mathcal{U}_\theta)-\upsilon_j\right)}{\sigma\left(\ss(\mathcal{U}_\theta)+\upsilon_j\right)}
\right)+2\ss(\mathcal{U}_\theta)\zeta(\upsilon_j)
\right],\qquad j=1,2,
    \label{eq:Fj}
\end{equation}
in which $\wp'(\upsilon)\equiv\frac{\ed}{\ed\upsilon}\wp(\upsilon;g_2,g_3)$,  and the Weierstra{\ss} invariants $g_{2,3}$ are the same as those in Eqs.~\eqref{eq:g2,3}. Here, $\sigma(\cdots)$ and $\zeta(\cdots)$ are, respectively, the Weierstra{\ss}ian Sigma and Zeta functions, with the same invariants \cite{handbookElliptic}. Furthermore, 
\begin{subequations}\label{eq:upsilonUtheta}
\begin{align}
    & \upsilon_1=\ss\left(-\frac{\psi_0}{12}-\frac{1}{4|1-\mZ_{\max}|}\right),\label{eq:upsilon1}\\
    & \upsilon_2=\ss\left(-\frac{\psi_0}{12}+\frac{1}{4|1+\mZ_{\max}|}\right).\label{eq:upsilon2}
\end{align}
\end{subequations}
In Eq.~\eqref{eq:Phitheta-sol}, we have defined
\begin{subequations}\label{eq:K0,1,2}
\begin{align}
    & \mathcal{K}_0 = \frac{\xi_p^2}{
    u(1-\mZ_{\max})(1+\mZ_{\max})\sqrt{2\mZ_{\max}\left(\mZ_{\max}^2+\mZ_0^2\right)}},\label{eq:K0,1,2a}\\
    & \mathcal{K}_1 = \frac{1+\mZ_{\max}}{8(1-\mZ_{\max})},\label{eq:K0,1,2b}\\
     & \mathcal{K}_2 = \frac{1-\mZ_{\max}}{8(1+\mZ_{\max})}.\label{eq:K0,1,2b}
\end{align}
\end{subequations}
For the sake of convenience in the simulation of the orbits, we let $\phi(\theta_{\min}) = 0$ as the initial condition. 
It is then necessary to interpolate $\theta\rightarrow\theta(\lambda)$ in the expression of $\mathcal{U}_\theta$ in Eq.~\eqref{eq:Utheta}.

\section{Explicit examples of orbits}\label{sec:examples}

\begin{figure}[t]
\centering
    \includegraphics[width=8cm]{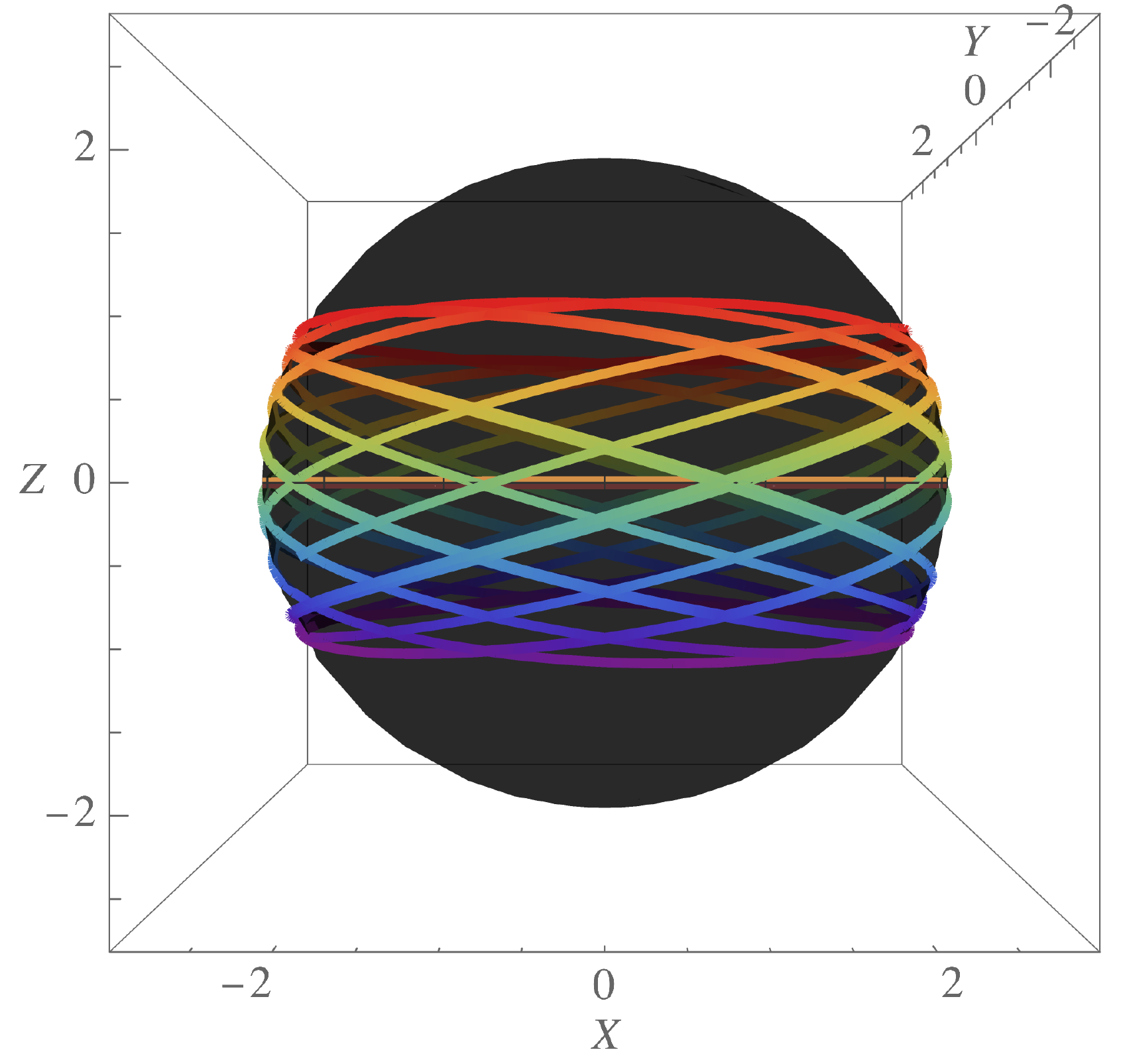}\qquad
    \includegraphics[width=8cm]{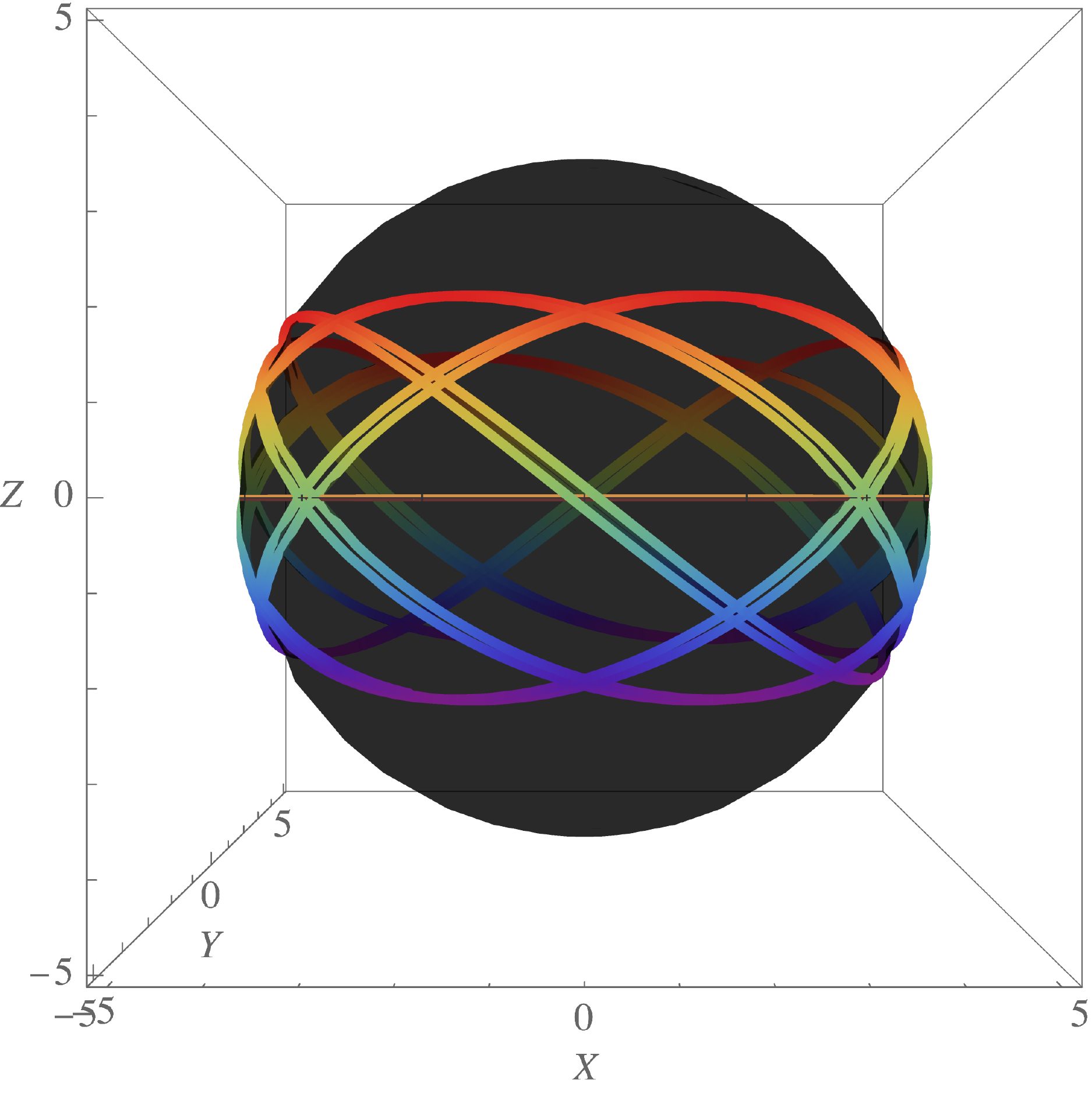}
        \caption{The spherical photon orbits on the radii $x_{p_1}$ and $x_{p_2}$ (from left to right), whose profiles of the polar angle have been depicted in Fig.~\ref{fig:W,theta}. The smooth interior black surface is the closure of the points which are swept by $x_p$, and is cut into half by a yellow circle which indicates the $\theta=\frac{\pi}{2}$ plane. By doing a comparison with the profiles of the $\theta$-coordinate in the cases (a) and (b) of Fig.~\ref{fig:W,theta}, one can confirm that the limits of the latitudinal oscillations in the orbits are the same as those given in the profiles. On the other hand, as expected from the $\theta$-profiles, the period of oscillations of the prograde orbits (left) are larger than those of the retrograde ones (right), as it can also be inferred by comparing the rapidity of the changes between the maximum and minimum latitudes in the above orbits.}
    \label{fig:orbitsx1x2}
\end{figure}

\begin{figure}[t]
\centering
   \includegraphics[width=4cm]{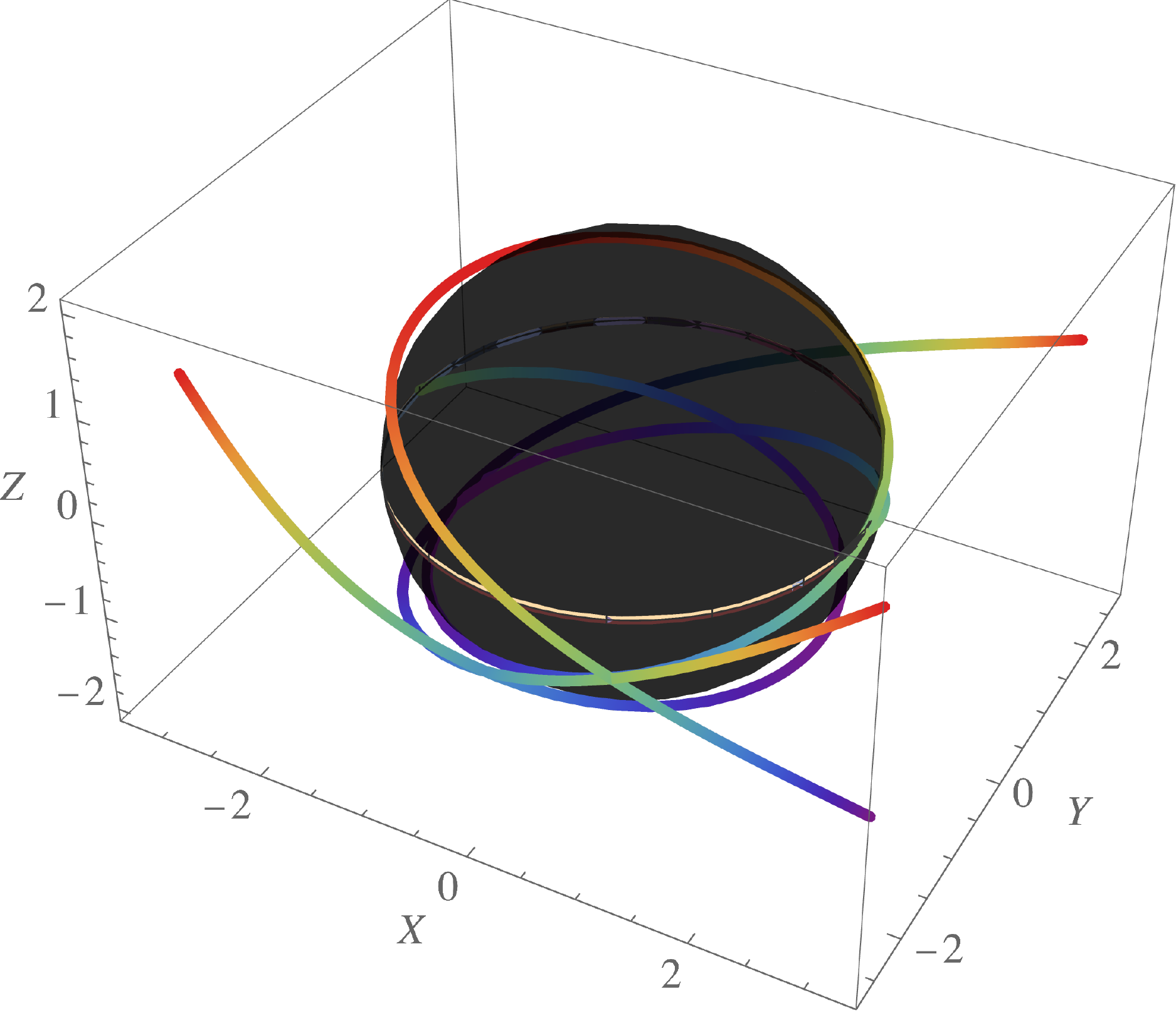} ~$(s_1)$
    \includegraphics[width=4cm]{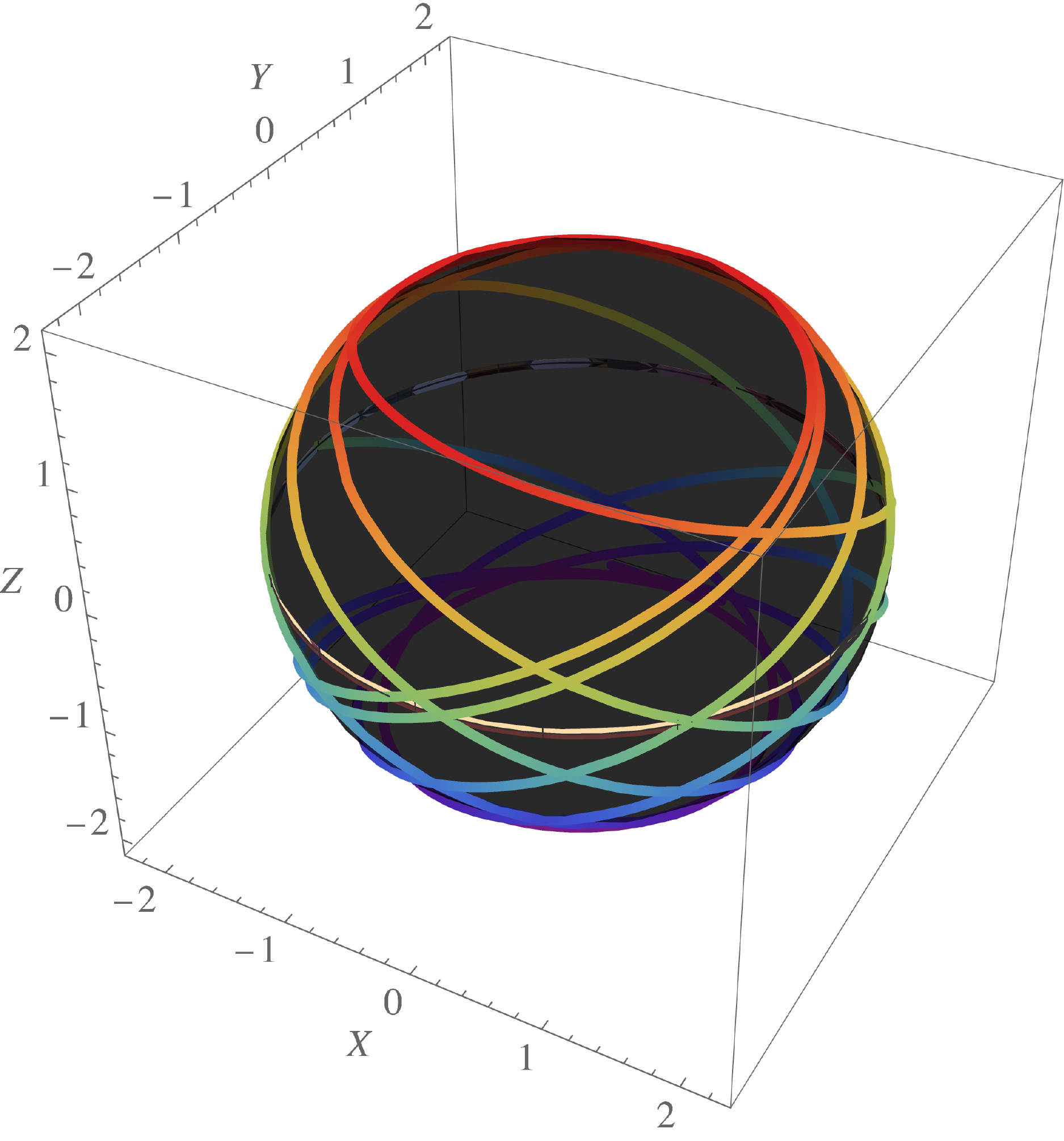}~$(s_2)$\qquad
    \includegraphics[width=4cm]{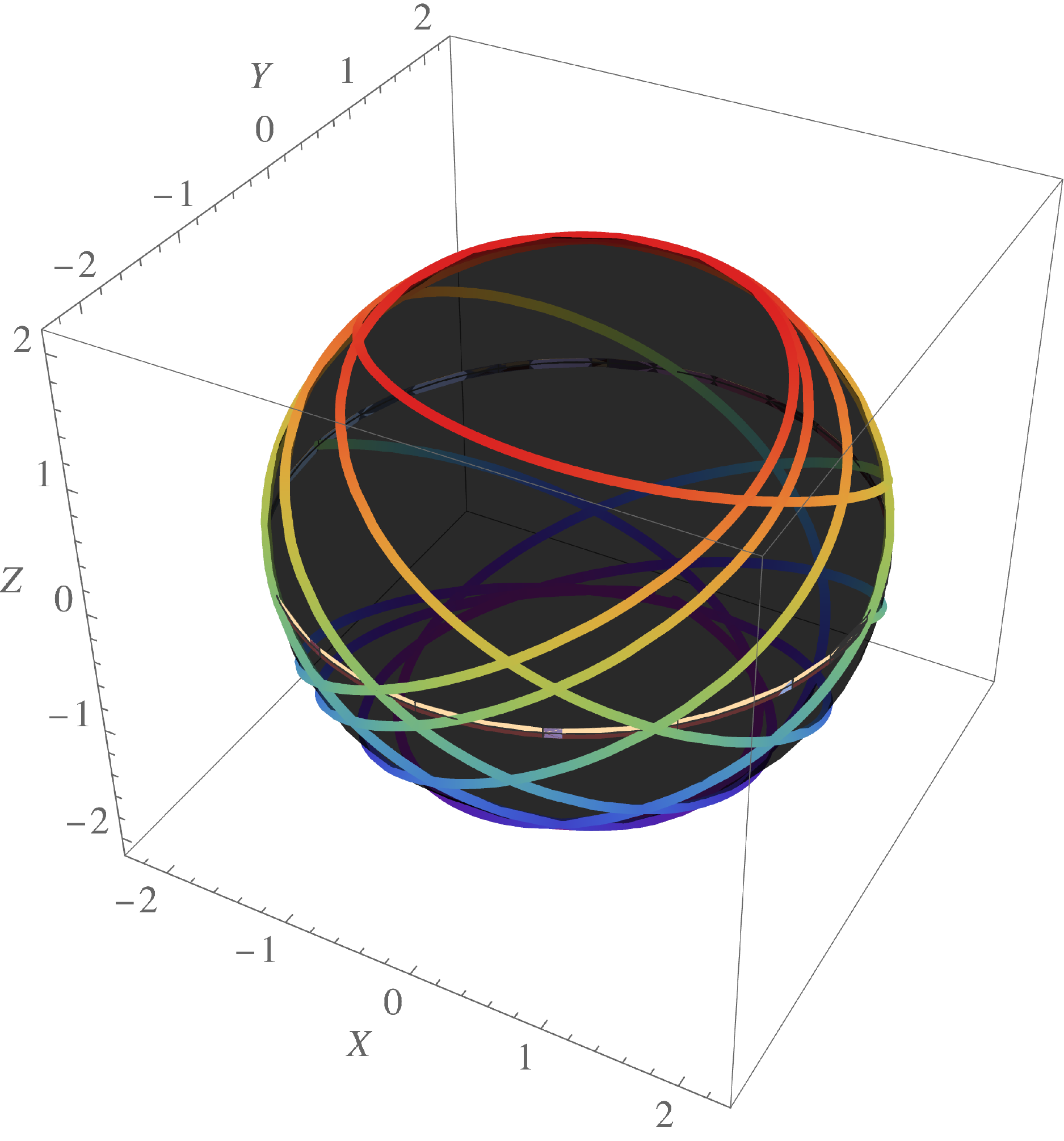}~$(s_3)$\qquad
    \includegraphics[width=4cm]{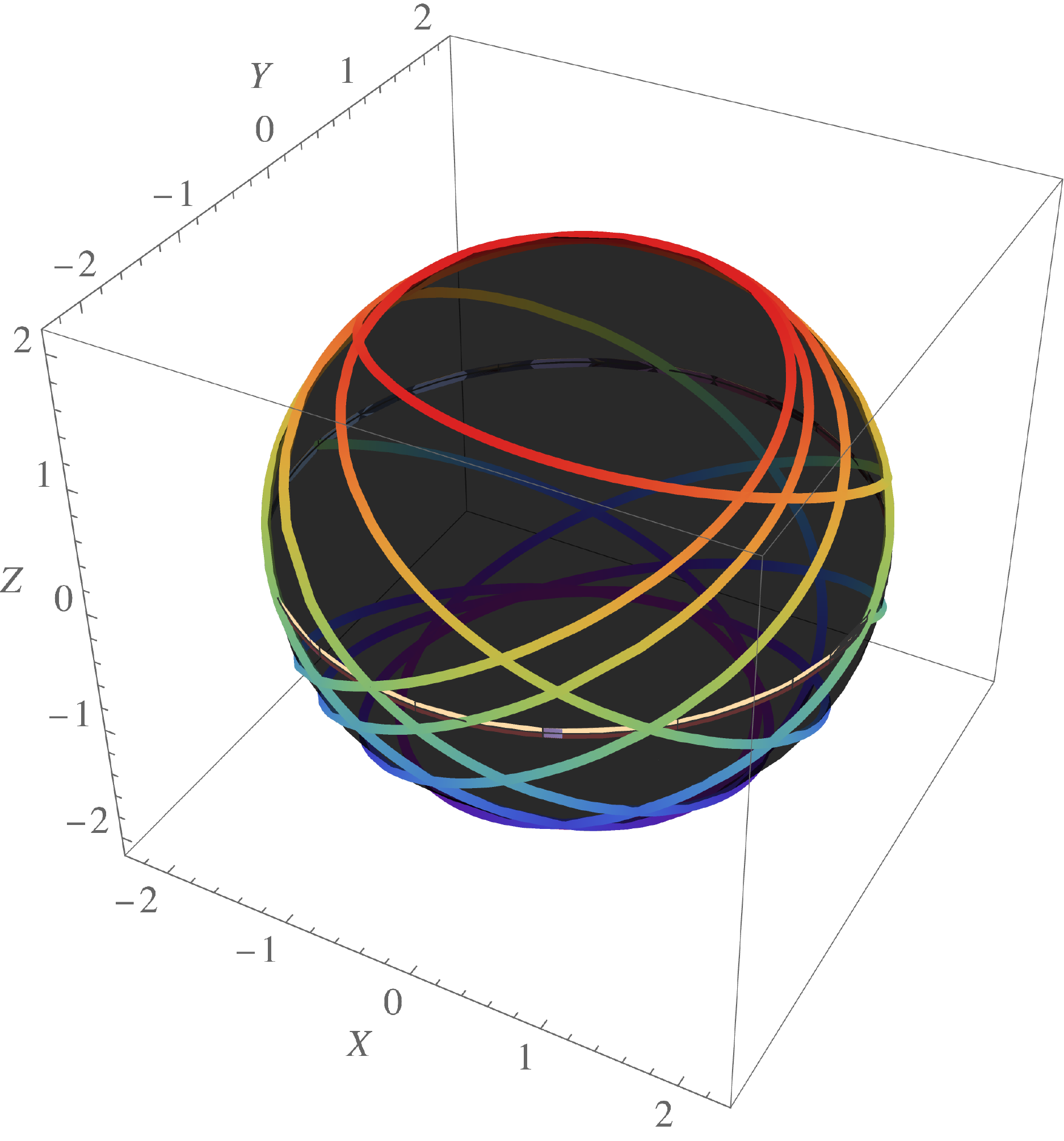}~$(s_4)$\qquad
    \includegraphics[width=4cm]{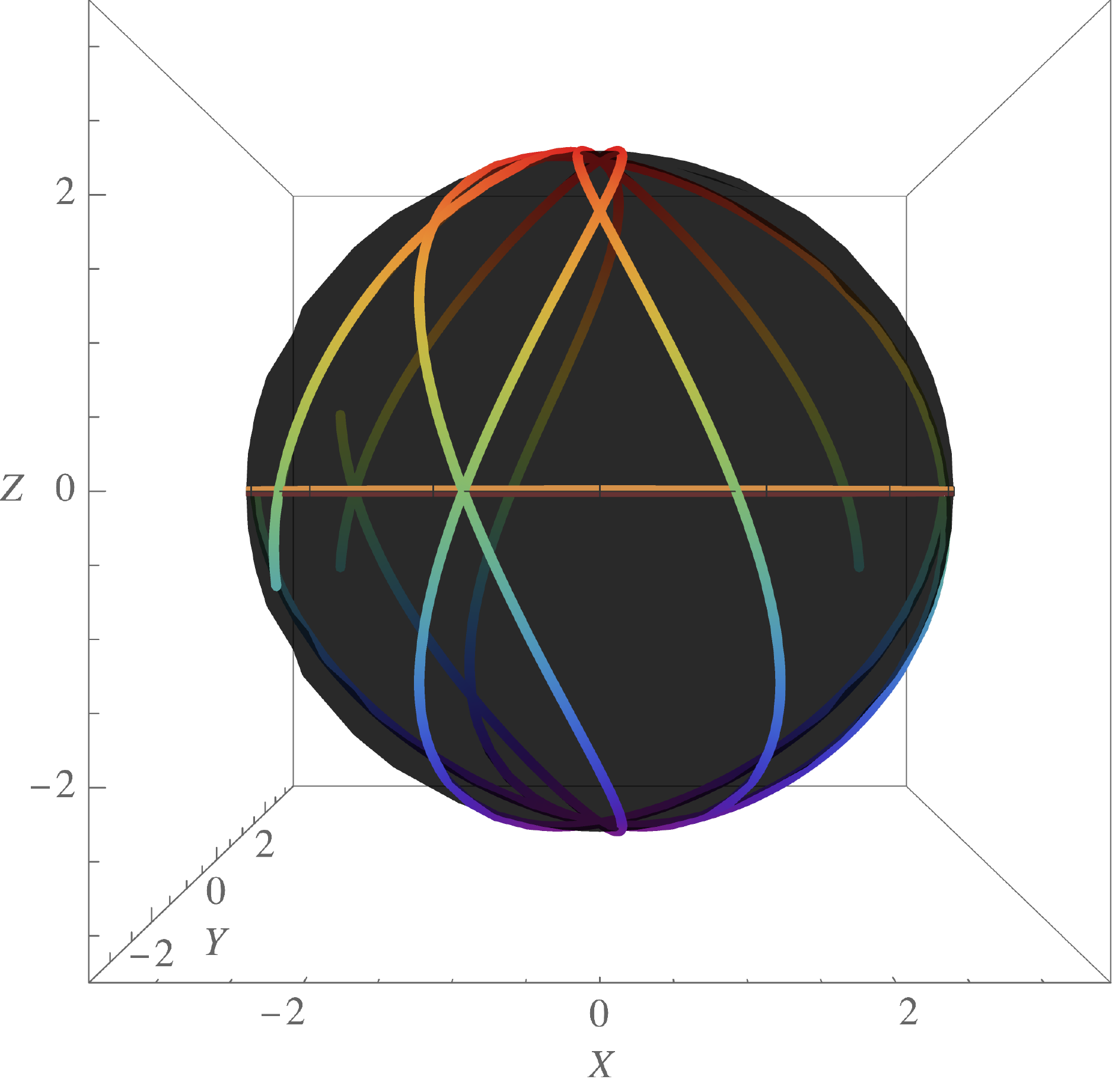}~$(s_5)$\qquad
    \includegraphics[width=4cm]{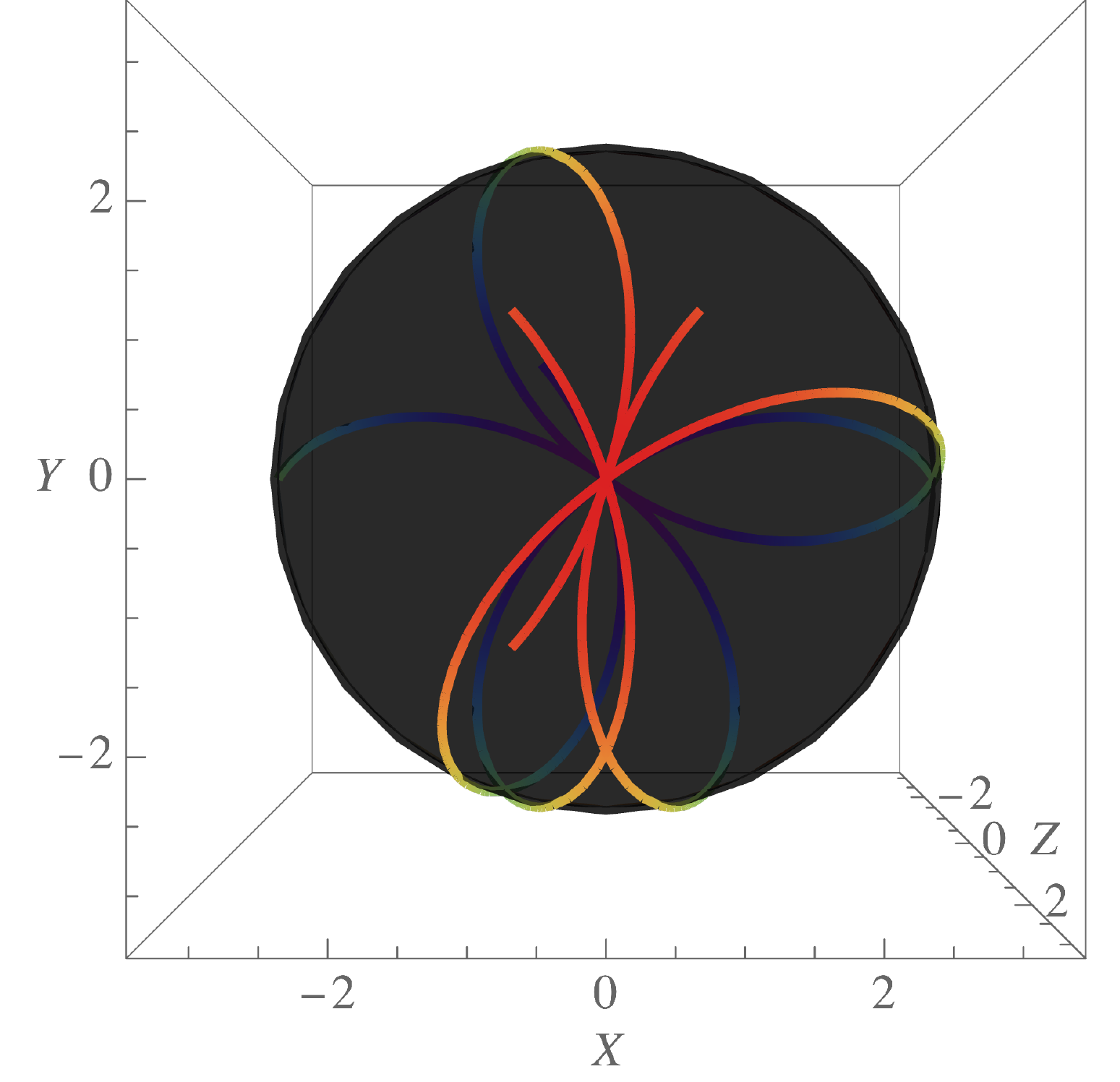}~$(s_5)$ top~view\qquad
    \includegraphics[width=4cm]{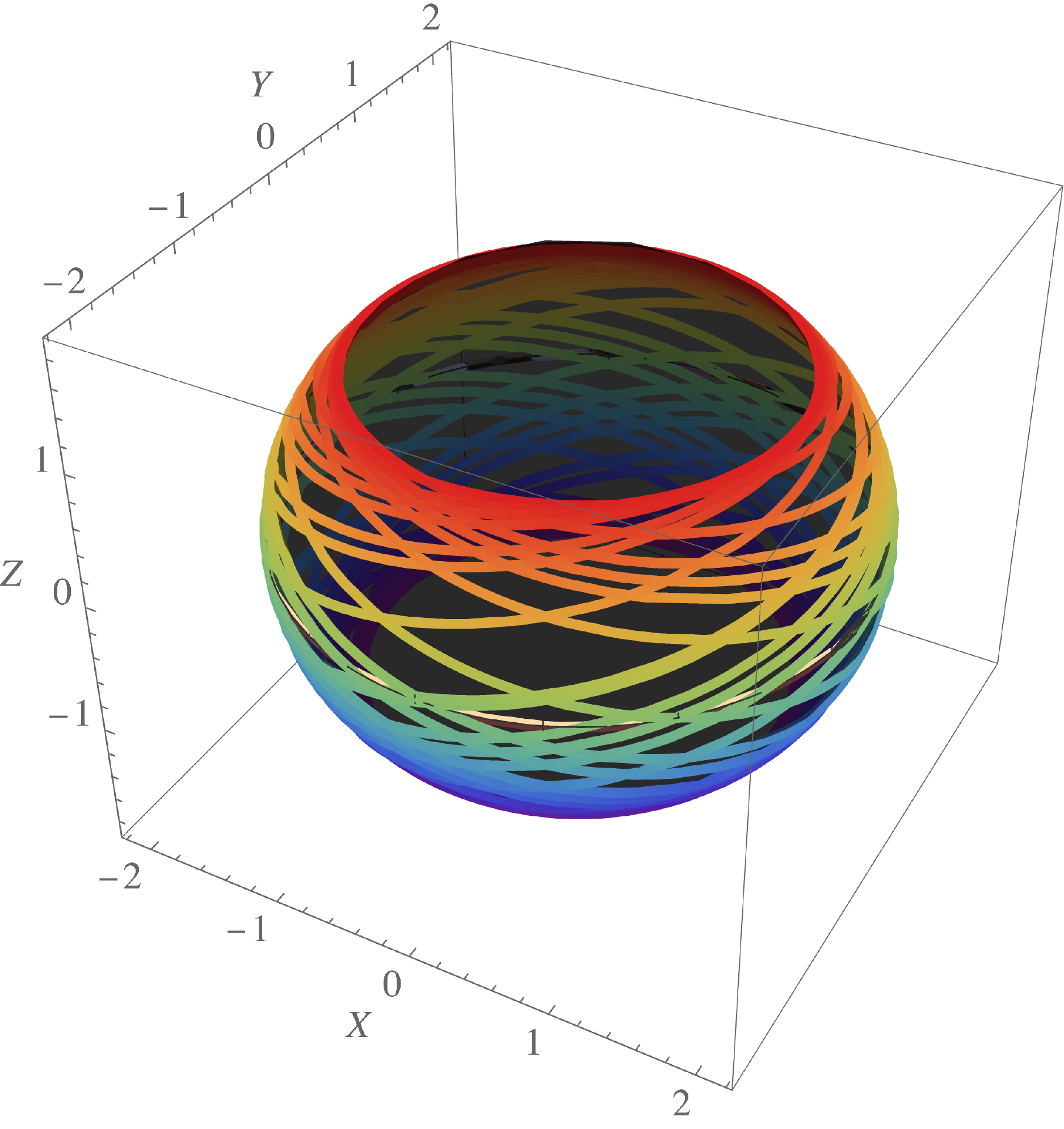}~$(f_1)$\qquad
    \includegraphics[width=4cm]{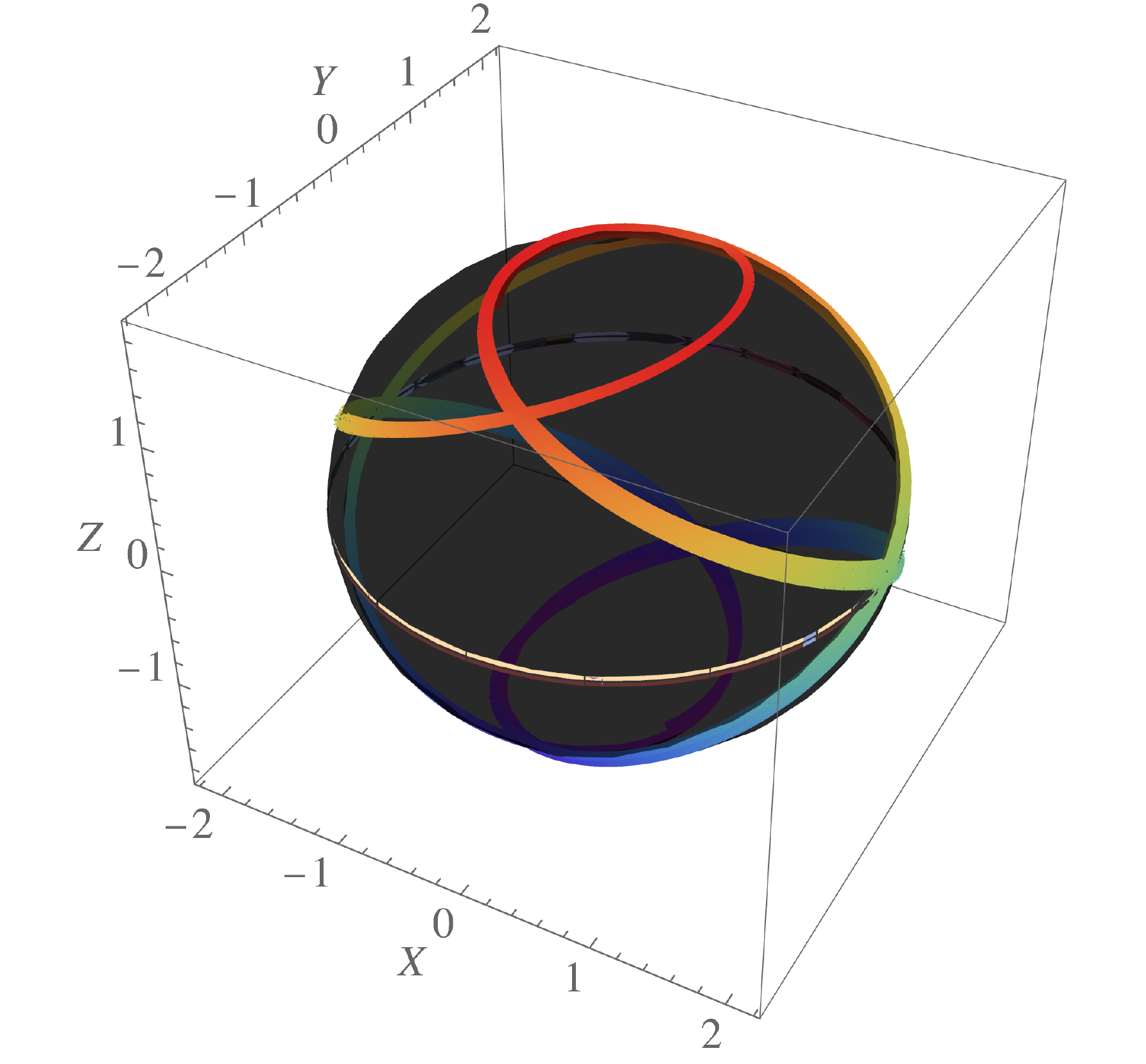}~$(f_2)$\qquad
    \includegraphics[width=4cm]{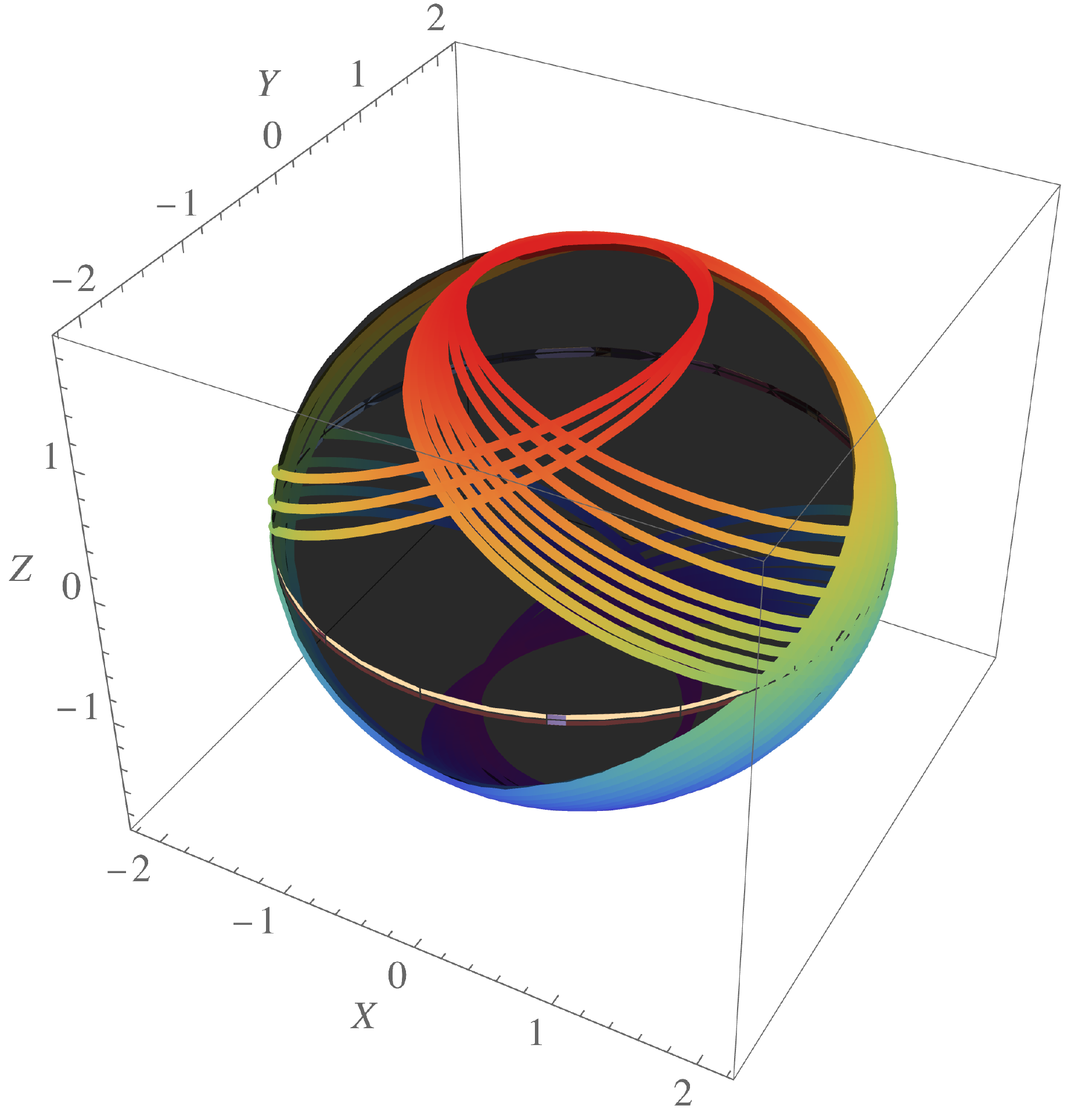}~$(f_3)$\qquad
    \includegraphics[width=4cm]{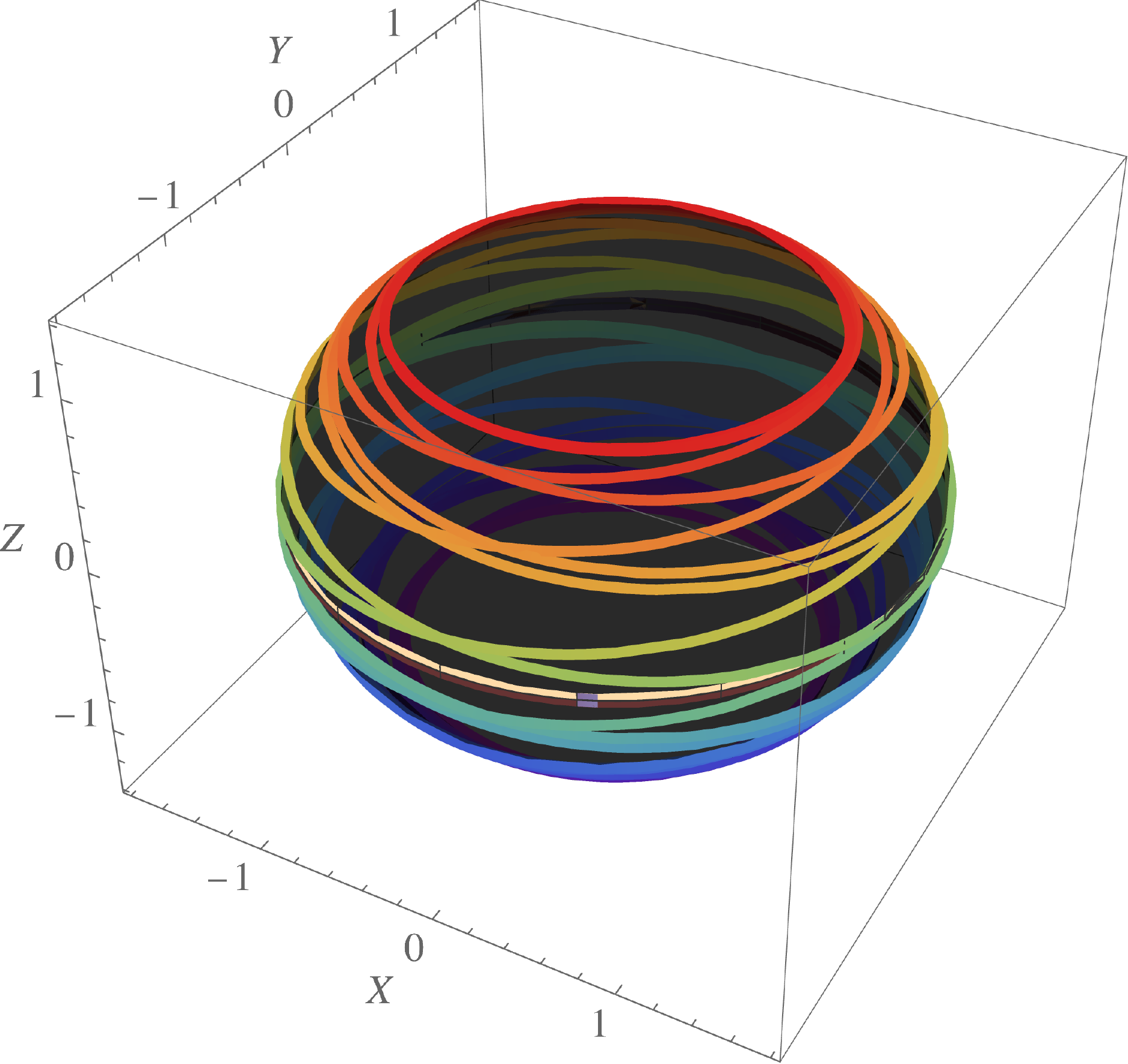}~$(e_1)$\qquad
    \includegraphics[width=4cm]{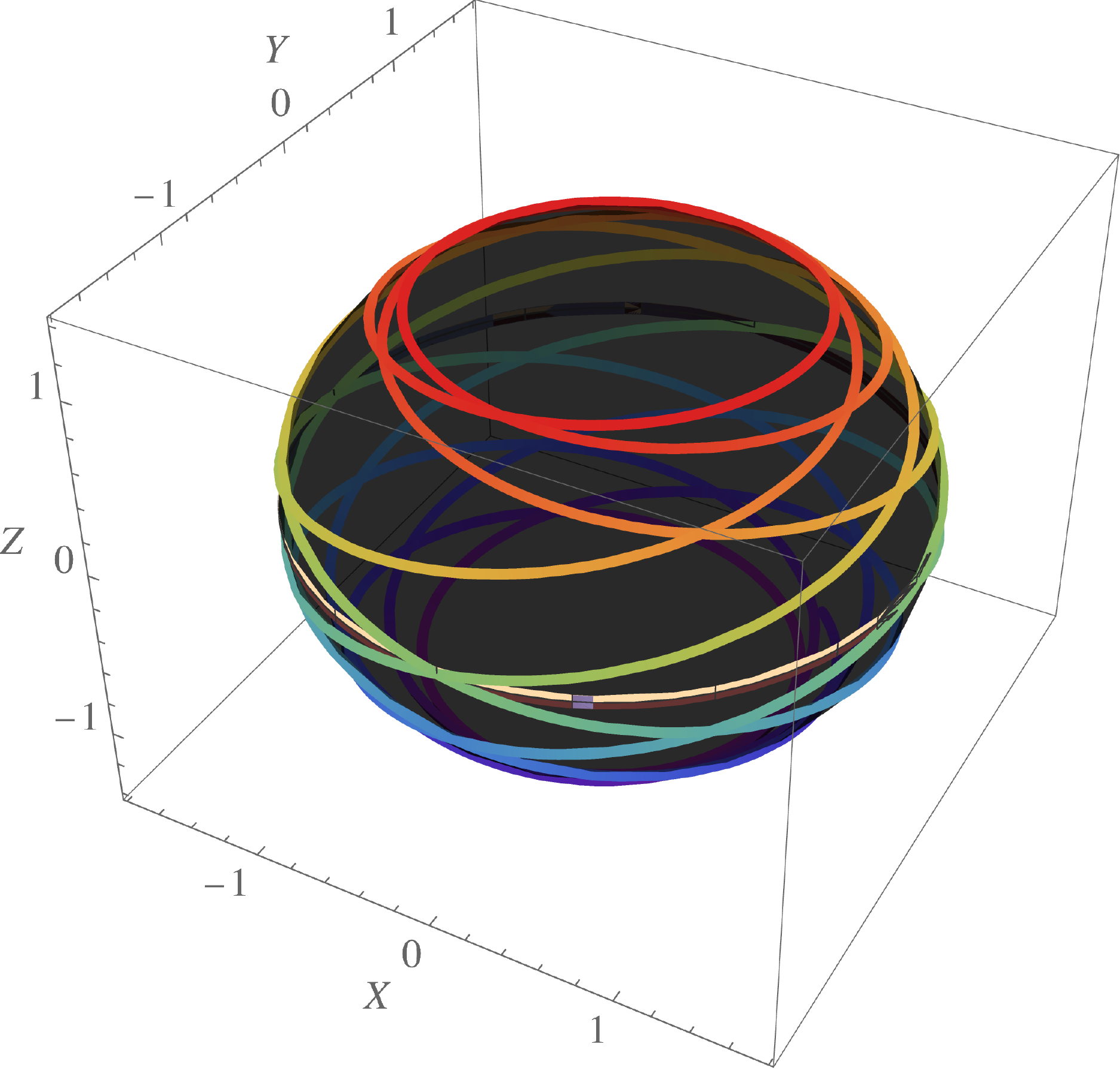}~$(e_2)$\qquad
    \includegraphics[width=4cm]{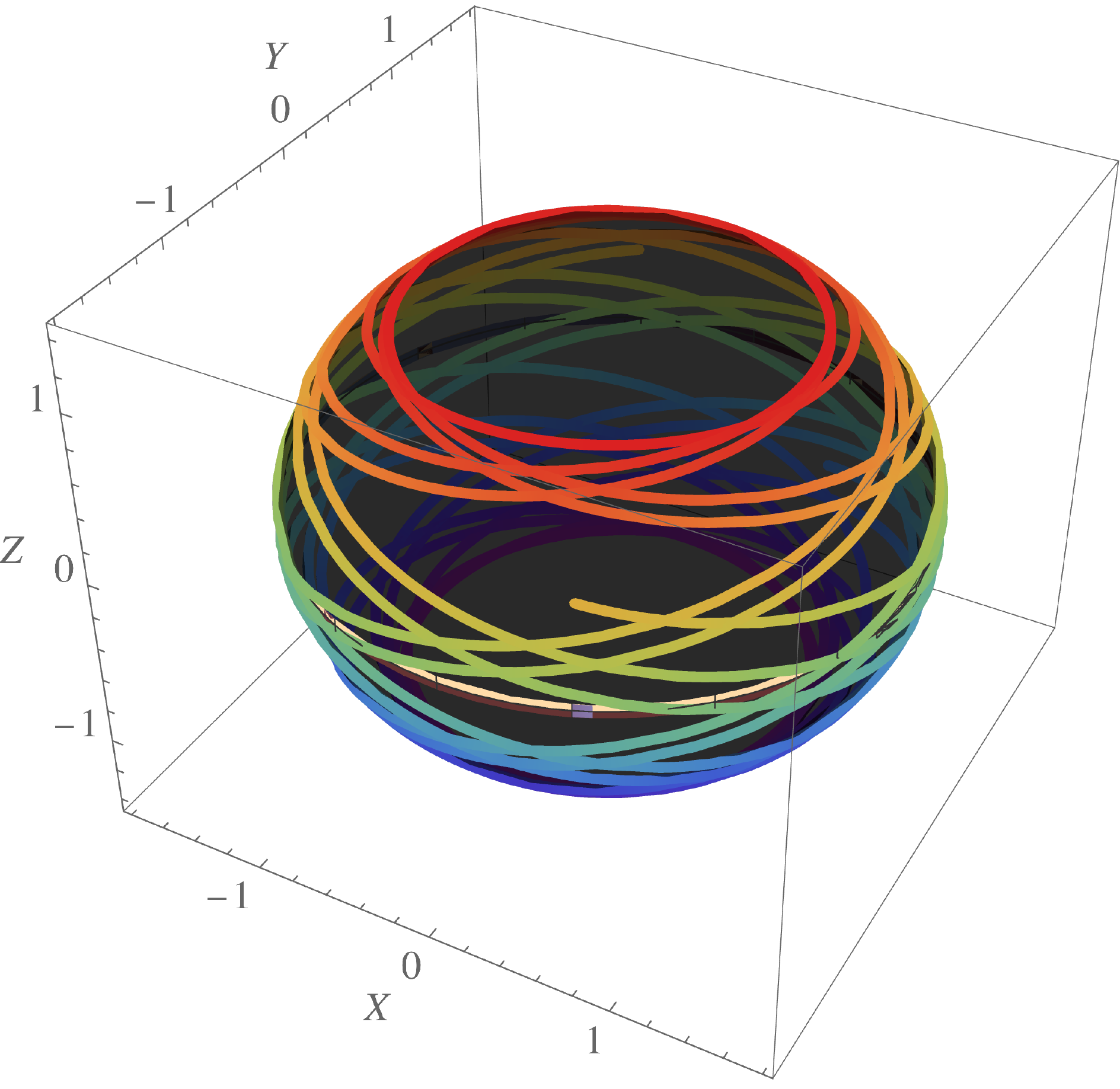}~$(e_3)$\qquad
    \includegraphics[width=6.2cm]{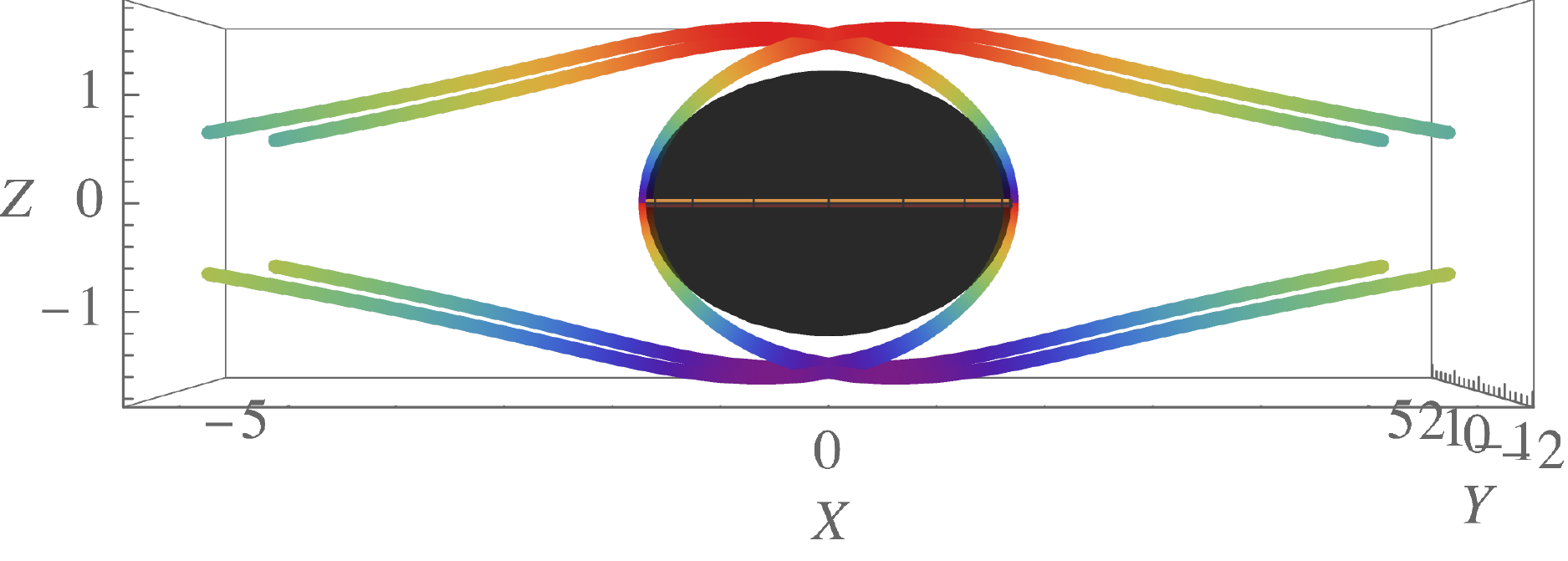}~$(se)$ front~view\qquad
     \includegraphics[width=6.2cm]{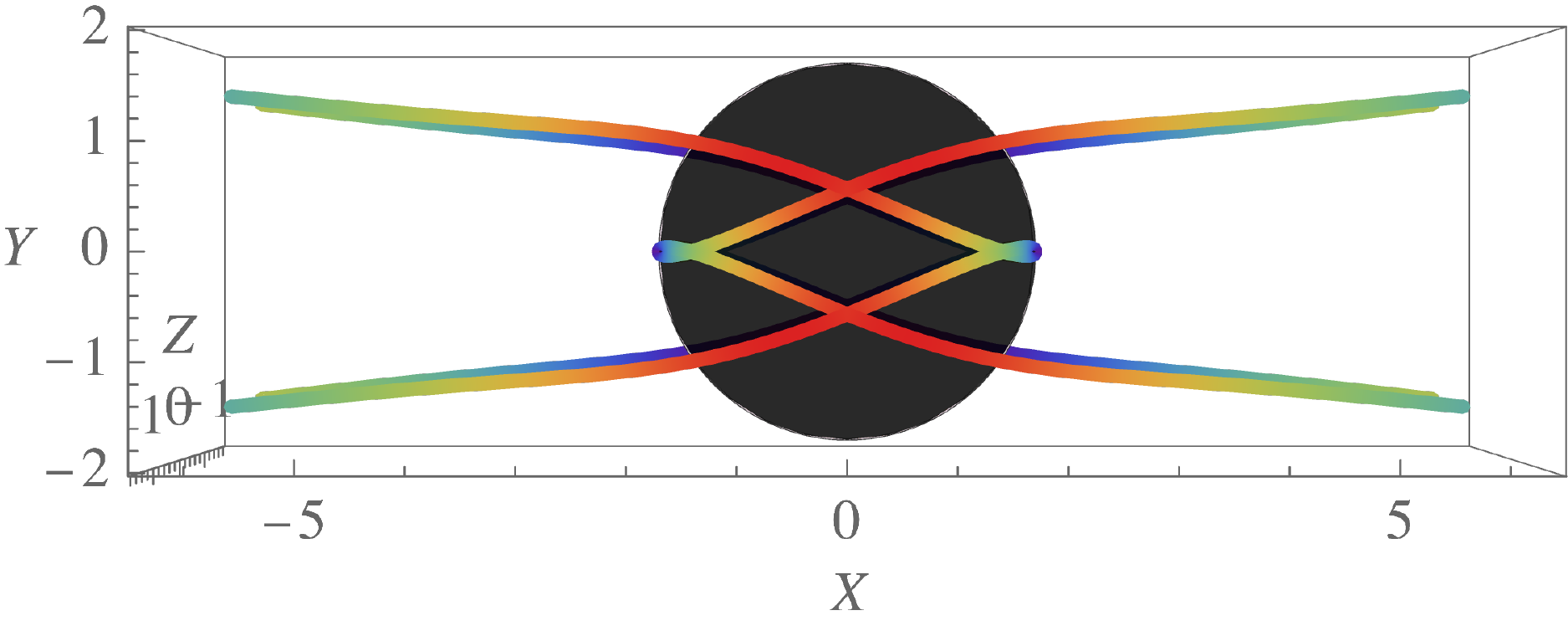}~$(se)$ top~view
        \caption{Some examples of the spherical photon orbits, for the particular categories of Table \ref{table:2}.}
    \label{fig:orbitsGeneral}
\end{figure}

\begin{table}[t]
\centering
\begin{tabular}{c | c | c | c | c | c | c} 
 \hline
 $u$ & $\alpha$ & $b$ & $i^\circ$ & $x_{p}$ 
 & $(\xi_p,\eta_p)$ & case\\ [0.3ex] 
 \hline\hline
  & $10^{-1}$ & $10^{-2}$ & 17 &  1.918 & (7.444, 58.885)& $(s_1)$\\
  & $10^{-2}$ & $10^{-4}$ & 30 &  1.939 & (7.307, 55.226)&$(s_2)$\\
   0.85 & $10^{-4}$ & $10^{-6}$ & 60 & 1.952 & (7.224, 53.066)&$(s_3)$\\
  & $10^{-6}$ & $10^{-8}$ & 89 & 1.956 & (7.198, 52.397)&$(s_4)$\\
  & $10^{-1}$ & $10^{-2}$ & 90 & 2.985 & (0, 36.308): polar~orbit &$(s_5)$\\
 \hline
 & $10^{-1}$ & $10^{-2}$ & 30 & 1.751 & (8.878,101.566)&$(f_1)$\\
1 & $10^{-2}$ & $10^{-4}$ & 45 & 1.754 & (8.840, 100.339)&$(f_2)$\\
 & $10^{-6}$ & $10^{-8}$ & 75 & 1.757 & (8.811, 99.391)&$(f_3)$\\
 \hline
 & $10^{-1}$ & $10^{-2}$ & 30 & 1.281 & ($9.092\times10^{1}$, $1.917\times10^{4}$)&$(e_1)$\\
 $\bar{u}_c$ & $10^{-4}$ & $10^{-6}$ & 45 & 1.302 & ($10.835\times10^{1}$, $2.625\times10^{4}$)&$(e_2)$\\
 & $10^{-6}$ & $10^{-8}$ & 85 & 1.339 & (41.842, $3.684\times10^{3}$)&$(e_3)$\\
 \hline
 $\bar{u}_c+0.2$& $10^{-1}$ & $10^{-2}$ & 60 & 1.323 &(55.661, $6.684\times10^{3}$)&$(se)$\\
 [1ex] 
 \hline
\end{tabular}
\caption{Some examples for the values of $x_p$ together with their corresponding pair $(\xi_p,\eta_p)$, given for a variety of characteristic parameters for the black hole, as well as different inclinations. Each case has been indicated by a letter with a numerical subscript, which will be referred to in the simulations of the orbits. The examples contain sub-extremal, extremal and super-extremal cases.}
\label{table:2}
\end{table}
In this section, we apply the exact analytical solutions for the $\theta(\lambda)$ and $\phi(\lambda)$, to simulate some specific examples of the spherical orbits in the spacetime. The trajectories that are presented in this section are each based on specific initial conditions, and the solutions are then evolved in order to generate the desired trajectories.
First of all, in Fig.~\ref{fig:orbitsx1x2}, the spherical orbits corresponding to the radii dealt with in Fig.~\ref{fig:W,theta}, have been plotted. The plots have been done in the  Kerr-Schild coordinates \eqref{eq:Kerr-Schild}. 
Furthermore, in Fig.~\ref{fig:orbitsGeneral}, the data given in Table \ref{table:2} have been used to simulate several spherical orbits on fast and extremal black holes, as well as on the naked singularity. The radii included in this tables have been obtained by solving, numerically, the octic \eqref{eq:p8=0}, for different initial data for the black hole.
It is evident that some of the orbits exhibit profound instability. It is therefore worth studying the stability of the orbits.

\subsection{Stability of the orbits}

\begin{figure}[t]
\centering
    \includegraphics[width=11cm]{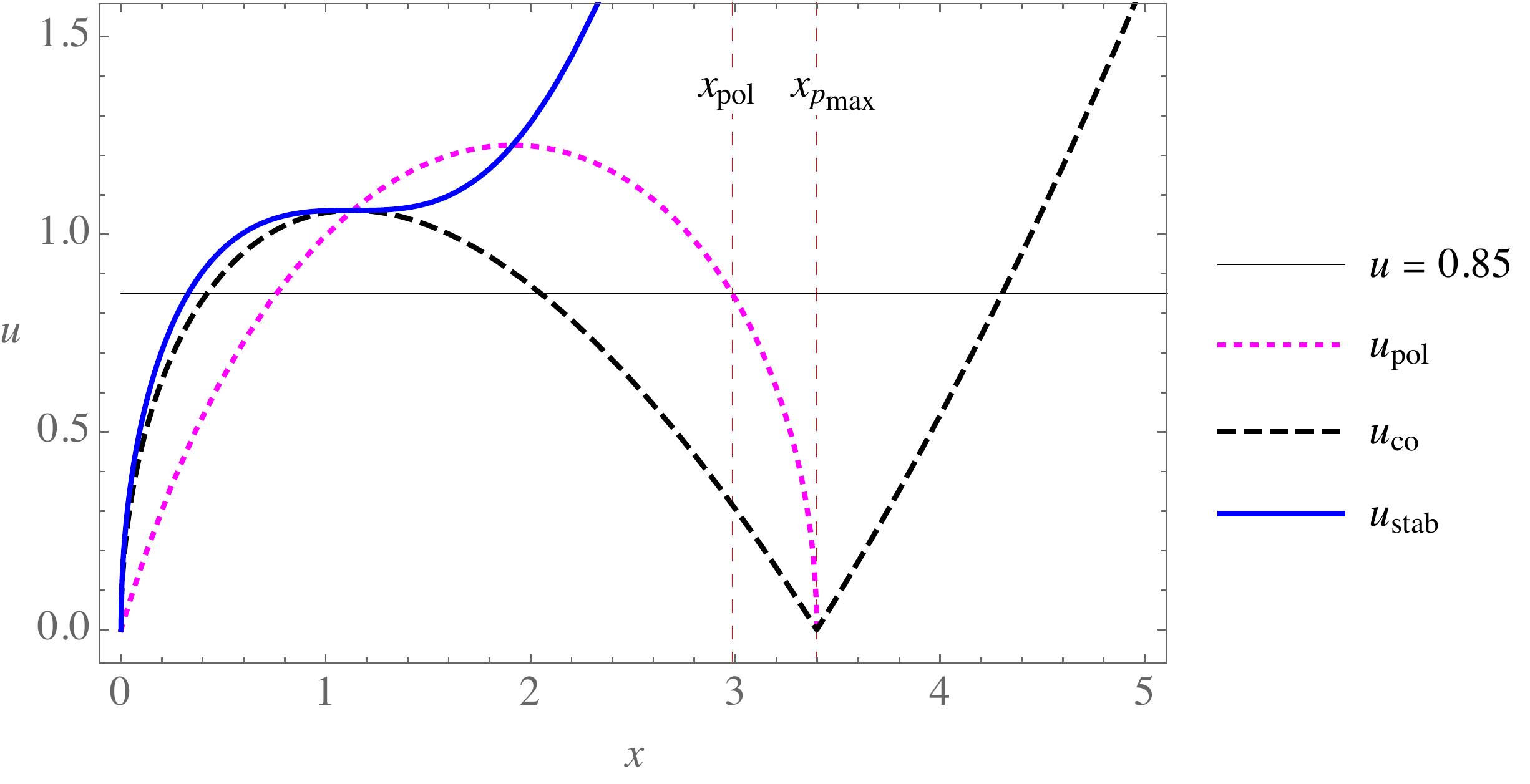}
    \caption{The radial profiles of the spin parameter for the circular, polar and marginally stable orbits, in accordance with the case $(s_1)$ in Table \ref{table:2}.}
    \label{fig:u_stab}
\end{figure}

In fact, condition $R(x)=R'(x)=0$, with
\begin{equation}
\mathcal{R}(x) = \left[(x^2+u^2)-u\xi\right]^2-\Delta(x)\left[(u-\xi)^2-\eta\right],
    \label{eq:R_x}
\end{equation}
can be regarded as the instability condition against radial perturbations for the photon orbits, once it is accompanied by the extra condition $R''(x)<0$, that indicates the existence of a maximum in the radial effective potential. In this sense, the orbits become marginally stable when $R(x)=R'(x)=R''(x)=0$. On the other hand, this situation can be recovered at $x_{p_{\max}}$ in Eq.~\eqref{eq:xpmax}, where the extremum of $\eta_p(x)$ occurs. However, regarding the stability of the orbits in terms of the spin parameter, one can solve the equation $\eta_p'(x)=0$ for $u$, that yields
\begin{equation}
u_{\mathrm{stab}} = \sqrt{\frac{x \left[3 b^2 x^4-6 (1-\alpha) b x^3+4 x^2 \Big(b+(1-\alpha)^2\Big)-12 (1-\alpha) x+12\right]}{4-6 b x^2}}.
    \label{eq:u_stab}
\end{equation}
Furthermore, as discussed in subsection \ref{subsubsec:Prop.Planar}, the circular orbits occur when $\eta_p= 0$. This condition results in the spin parameter
\begin{equation}
u_{\mathrm{co}} =\pm \frac{\sqrt{x} \left[b x^2+2 (\alpha -1) x+6\right]}{\sqrt{16-8 b x^2}}.
    \label{eq:u_co}
\end{equation}
Finally, the polar orbits (where $\xi_p=0$) correspond to
\begin{equation}
u_{\mathrm{pol}} =\frac{x \sqrt{b x^2-2 (1-\alpha) x+6}}{\sqrt{3 b x^2+2 (1+\alpha) x+2}}.
    \label{eq:u_pol}
\end{equation}
For the case of Kerr black holes, the above values reduce correctly to $u_{\mathrm{stab}} = \sqrt{(x-1)^3+1}$, $u_{\mathrm{co}} = \pm\frac{\sqrt{x}}{2}(3-x)$, and $x_{\mathrm{pol}} = x \sqrt{\frac{3-x}{x+1}}$. The behaviors of the spin parameters in Eqs.~\eqref{eq:u_stab}--\eqref{eq:u_pol} have been plotted in Fig.~\ref{fig:u_stab}, for the special case of $(s_1)$ in Table \ref{table:2}. Accordingly, and by inspecting the conditions discussed above, it turns out that only the cases $(f_2)$ and $(f_3)$ among all of the explicit categories in Fig.~\ref{fig:orbitsGeneral}, are marginally stable, regarding their radii of the orbits in the context of the considered spin parameter. The others are indeed unstable, and hence, they either fall onto the event horizon, or escape from the black hole and contribute in the formation of the photon ring and the shadow.

The discussion that has been made so far, covers some important concepts of the spherical photon orbits on a stationary black hole, and for our specif case of study, we have provided various analytical and numerical results. We therefore leave our study at this point, and summarize our results in the next section.

\section{Summary and conclusion}\label{sec:conclusion}

There is no doubt that the astrophysical objects become observable due to the electromagnetic radiations (photons) that we receive from them. Black holes in particular, are not observable by their own and their strong gravitational lensig, which make them observable, becomes noticeable whenever they pass a luminous background. In this sense, they can even trap photons in their exterior geometry and force them to make orbits of constant radius. In this study, we aimed at the scrutinization of such orbits on a rotating black hole which is associated with a particular type of quintessence and cloud of strings. Under certain circumstances, the orbits become unstable, tending to either fall onto the event horizon or escaping from the black hole. Imposing these conditions, we obtained an octic equation that governed the general radii of spherical photon orbits. Solving the octic requires rather peculiar treatments and is left to a future study. Instead, we focused on the determination of the radii of planar orbits, by reducing the aforementioned octic to a quintic. Applying a series of reductions (as explained in the appendices), we reduced the quintic to its Bring-Jerrard form which has known analytical solutions in terms of the generalized hypergeometric functions. We then presented some numerical examples for the included characteristic constant $K$ in this reduced quintic. The determination of the polar orbits was then done by means of solving a quartic equation. Furthermore, considering definite initial conditions for the black hole, we demonstrated the profile of these radii versus changes in the spin parameter for two distinct values for the quintessential parameter $b$. We observed that by the raise in $b$, the radius of retrograde photon orbits come closer to that of the polar orbits. The existence of polar orbits is only an artifact of the frame-dragging caused by the rotation of the black hole, as it also generates a photon region occupied with photon orbits with non-zero inclinations. Hence, the impact of the quintessence becomes apparent in hampering the external growth of this region, and as a result, the photon region becomes narrower by the raise in $b$. This is also conceivable in the demonstration of the photon regions that we did in Fig.~\ref{fig:photonregions}. In this part, we also indicated the peculiar behavior of the ergoregions for the larger quintessence. We then switched to the derivation of analytical solutions for the evolution of the polar and azimuth angles. The corresponding integrals are of elliptic nature, so that we could express their solutions in terms of the three main Weierstra{\ss}ian elliptic functions. For the case of latitudinal motion, we observed that the prograde and retrograde orbits show different oscillation periods. This was also inferred from the analytic expression for the periods of latitudinal motion and the simulation of the spherical orbits performed for a particular example. Considering a variety of initial values for the black hole parameters, and by the help of the analytical solutions we had at hand, we exemplified numerous spherical orbits for sub-extremal, extremal and super-extremal cases. Moreover, we highlighted that, although we had applied the necessary conditions for the presence of instability in the orbits, an extra condition must also be satisfied as the sufficient condition. Taking this into account, we performed an analytical study on the stability of the orbits. It turned out that among the presented examples, only two cases were marginally stable and all the others were unstable. As it is well-known, such unstable orbits are of crucial importance in astrophysical observations of black holes, because they carry information from the near-horizon regions to the distant observers. For a future study, we have in mind the analytical study of photon trajectories in all of their possible forms. This, potentially, is an interesting subject of investigation, since it makes it possible to perform a more rigorous assessment of the strong gravitational lensing for the black hole, when the cosmological parameters are present.

\section*{Acknowledgements}
M. Fathi has been supported by the Agencia Nacional de Investigaci\'{o}n y Desarrollo (ANID) through DOCTORADO Grant No. 2019-21190382, and No. 2021-242210002. J.R. Villanueva was partially supported by the Centro de Astrof\'isica de Valpara\'iso (CAV).


\appendix

\section{Reduction of the quintic to the Bring-Jerrard form}\label{app:A}

Let us first recast the quintic \eqref{eq:p5=0} as
\begin{equation}\label{eq:A1}
    x^5+a_1 x^4 + a_2 x^3+ a_3 x^2+a_4 x +a_5 = 0,
\end{equation}
by defining $a_j=\bar{m}_5^{-1}{\bar{m}_{5-j}}$. We now proceed with transforming Eq.~\eqref{eq:A1} to the principal quintic form that is missing the $x^4$ and $x^3$ terms, by means of the quadratic Tschirnhausen transformation
\begin{equation}
y = x^2+b_1 x + b_2.
    \label{eq:A2}
\end{equation}
Applying a simple code in the software {\it{Mathematica}}, we can eliminate $x$ between Eqs.~\eqref{eq:A1} and \eqref{eq:A2}, which results in 
\begin{equation}
y^5+c_1y^4+c_2y^3+c_3y^2+c_4y+c_5=0,
    \label{eq:A3}
\end{equation}
where
\begin{subequations}\label{eq:A4}
\begin{align}
    & c_1 = a_1 b_1-a_1^2+2 a_2-5 b_2,\label{eq:A41}\\
    & c_2 = 4 a_1^2 b_2-a_2 a_1 b_1-4 a_1 b_1 b_2+a_2 b_1^2+3 a_3 b_1-8 a_2 b_2-2 a_3 a_1+a_2^2+2 a_4+10 b_2^2,\label{eq:A42}\\
    & c_3 = a_3 b_1^3-a_1 a_3 b_1^2+4 a_4 b_1^2-3 a_2 b_2 b_1^2+6 a_1 b_2^2 b_1+a_2 a_3 b_1-3 a_1 a_4 b_1+5 a_5 b_1+3 a_1 a_2 b_2 b_1-9 a_3 b_2 b_1\nonumber\\
    &-6 a_1^2 b_2^2+12 a_2 b_2^2-3 a_2^2 b_2+6 a_1 a_3 b_2-6 a_4 b_2-a_3^2+2 a_2 a_4-2 a_1 a_5-10 b_2^3,\label{eq:A43}\\
    & c_4 = a_4 b_1^4-a_1 a_4 b_1^3+5 a_5 b_1^3-2 a_3 b_2 b_1^3+3 a_2 b_2^2 b_1^2+a_2 a_4 b_1^2-4 a_1 a_5 b_1^2+2 a_1 a_3 b_2 b_1^2-8 a_4 b_2 b_1^2-4 a_1 b_2^3 b_1\nonumber\\
    &-3 a_1 a_2 b_2^2 b_1+9 a_3 b_2^2 b_1-a_3 a_4 b_1+3 a_2 a_5 b_1-2 a_2 a_3 b_2 b_1+6 a_1 a_4 b_2 b_1-10 a_5 b_2 b_1\nonumber\\
    &+4 a_1^2 b_2^3-8 a_2 b_2^3+3 a_2^2 b_2^2-6 a_1 a_3 b_2^2+6 a_4 b_2^2+2 a_3^2 b_2-4 a_2 a_4 b_2+4 a_1 a_5 b_2+a_4^2-2 a_3 a_5+5 b_2^4,\label{eq:A44}\\
    & c_5 = a_5 b_1^5-a_1 a_5 b_1^4+a_2 a_5 b_1^3-a_3 a_5 b_1^2+a_4 a_5 b_1-a_5^2.\label{eq:A45}
\end{align}
\end{subequations}
The two unknowns $b_{1,2}$ allow for the elimination of $c_{1,2}$. In fact, one can see that the equations $c_1=c_2=0$ result in two quadratics, solving which, provide the values
\begin{eqnarray}
&& b_1=\frac{4 a_1^3-13 a_2 a_1\pm\sqrt{5} \sqrt{8 a_3 a_1^3+\left(16 a_4-3 a_2^2\right) a_1^2-38 a_2 a_3 a_1+12 a_2^3+45 a_3^2-40 a_2 a_4}+15 a_3}{4 a_1^2-10 a_2},\label{eq:A5}\\
&& b_2 = \frac{5 a_2 a_1^2+\left(15a_3\pm\sqrt{5} \sqrt{8 a_3 a_1^3+\left(16 a_4-3 a_2^2\right) a_1^2-38 a_2 a_3 a_1+12 a_2^3+45 a_3^2-40 a_2 a_4}\right) a_1-20 a_2^2}{20 a_1^2-50 a_2}.\label{eq:A6}
\end{eqnarray}
Applying these values in the coefficients in Eq.~\eqref{eq:A4}, the quintic \eqref{eq:A3} reduces to the principal form
\begin{equation}
y^5+\fu y^2+\fv y+\fw =0,
    \label{eq:A7}
\end{equation}
in which
\setlength{\abovedisplayskip}{3pt}
\begin{eqnarray}
 \fu &=&
     \frac{1}{40 \left(2 a_1^2-5 a_2\right)^3}\left[
    -90 \sqrt{5} a_1 \sqrt{8 a_3 a_1^3+\left(16 a_4-3 a_2^2\right) a_1^2-38 a_2 a_3 a_1+12 a_2^3+45 a_3^2-40 a_2 a_4} a_2^4\right.\nonumber\\
&&    \left.+48 \sqrt{5} \sqrt{8 a_3 a_1^3+\left(16 a_4-3 a_2^2\right) a_1^2-38 a_2 a_3 a_1+12 a_2^3+45 a_3^2-40 a_2 a_4} a_1 \left(2 a_1^2-5 a_2\right) a_2^3\right.\nonumber\\
 &&   \left.+1350 \sqrt{5} \sqrt{8 a_3 a_1^3+\left(16 a_4-3 a_2^2\right) a_1^2-38 a_2 a_3 a_1+12 a_2^3+45 a_3^2-40 a_2 a_4} a_3 a_2^3\right.\nonumber\\
&&   \left.-6 \sqrt{5} a_1 \left(2 a_1^2-5 a_2\right)^2 \sqrt{8 a_3 a_1^3+\left(16 a_4-3 a_2^2\right) a_1^2-38 a_2 a_3 a_1+12 a_2^3+45 a_3^2-40 a_2 a_4} a_2^2\right.\nonumber\\
&&   \left.+320 \sqrt{5} \sqrt{8 a_3 a_1^3+\left(16 a_4-3 a_2^2\right) a_1^2-38 a_2 a_3 a_1+12 a_2^3+45 a_3^2-40 a_2 a_4} \left(5 a_2-2 a_1^2\right) a_3 a_2^2\right.\nonumber\\
&&    \left.-2700 \sqrt{5} a_1 a_3^2 \sqrt{8 a_3 a_1^3+\left(16 a_4-3 a_2^2\right) a_1^2-38 a_2 a_3 a_1+12 a_2^3+45 a_3^2-40 a_2 a_4} a_2\right.\nonumber\\
&&    \left.-46 \sqrt{5} \left(2 a_1^2-5 a_2\right)^2 a_3 \sqrt{8 a_3 a_1^3+\left(16 a_4-3 a_2^2\right) a_1^2-38 a_2 a_3 a_1+12 a_2^3+45 a_3^2-40 a_2 a_4} a_2\right.\nonumber\\
&&    \left.+280 \sqrt{5} \sqrt{8 a_3 a_1^3+\left(16 a_4-3 a_2^2\right) a_1^2-38 a_2 a_3 a_1+12 a_2^3+45 a_3^2-40 a_2 a_4} a_1 \left(5 a_2-2 a_1^2\right) a_4 a_2\right.\nonumber\\
 &&   \left.+4500 \sqrt{5} \sqrt{8 a_3 a_1^3+\left(16 a_4-3 a_2^2\right) a_1^2-38 a_2 a_3 a_1+12 a_2^3+45 a_3^2-40 a_2 a_4} a_3^3\right.\nonumber\\
 &&   \left.+520 \sqrt{5} \sqrt{8 a_3 a_1^3+\left(16 a_4-3 a_2^2\right) a_1^2-38 a_2 a_3 a_1+12 a_2^3+45 a_3^2-40 a_2 a_4} a_1 \left(2 a_1^2-5 a_2\right) a_3^2\right.\nonumber\\
&&    \left.+8 \sqrt{5} \sqrt{8 a_3 a_1^3+\left(16 a_4-3 a_2^2\right) a_1^2-38 a_2 a_3 a_1+12 a_2^3+45 a_3^2-40 a_2 a_4} \left(2 a_1^2-5 a_2\right)^3 a_3\right.\nonumber\\
&&    \left.+675 \left(a_2^6-8 a_1 a_3 a_2^4+60 a_3^2 a_2^3-80 a_1 a_3^3 a_2+100 a_3^4\right)+40 a_1^2 \left(2 a_1^2-5 a_2\right)^3 a_4\right.\nonumber\\
 &&   \left.+92 \sqrt{5} \sqrt{8 a_3 a_1^3+\left(16 a_4-3 a_2^2\right) a_1^2-38 a_2 a_3 a_1+12 a_2^3+45 a_3^2-40 a_2 a_4} a_1 \left(2 a_1^2-5 a_2\right)^2 a_4\right.\nonumber\\
 &&   \left.+1400 \sqrt{5} \sqrt{8 a_3 a_1^3+\left(16 a_4-3 a_2^2\right) a_1^2-38 a_2 a_3 a_1+12 a_2^3+45 a_3^2-40 a_2 a_4} \left(2 a_1^2-5 a_2\right) a_3 a_4\right.\nonumber\\
&&    \left.+5 \left(2 a_1^2-5 a_2\right)^3 \left(a_2^3-26 a_4 a_2+44 a_3^2\right)+135 \left(2 a_1^2-5 a_2\right) \left(-3 a_2^5+20 \left(a_1 a_3+a_4\right) a_2^3-70 a_3^2 a_2^2\right.\right.\nonumber\\
&&    \left.\left.-80 a_1 a_3 a_4 a_2+40 a_3^2 \left(2 a_1 a_3+5 a_4\right)\right)\right.\nonumber\\
&&    \left.+100 \sqrt{5} \sqrt{8 a_3 a_1^3+\left(16 a_4-3 a_2^2\right) a_1^2-38 a_2 a_3 a_1+12 a_2^3+45 a_3^2-40 a_2 a_4} \left(2 a_1^2-5 a_2\right)^2 a_5\right.\nonumber\\
&&    \left.-20 a_1 \left(2 a_1^2-5 a_2\right)^3 \left(a_2 a_3-6 a_5\right)+5 \left(2 a_1^2-5 a_2\right)^2 \left(9 a_2^4-6 \left(8 a_1 a_3+33 a_4\right) a_2^2\right.\right.\nonumber\\
&&     \left.\left.-2 \left(137 a_3^2+30 a_1 a_5\right) a_2
   +4 \left(80 a_4^2+137 a_1 a_3 a_4+75 a_3 a_5\right)\right)
    \right.\Big],
    \label{eq:A8}
     \end{eqnarray}
\begin{eqnarray*}
    \fv &=& \frac{a_4 \left(4 a_1^3-13 a_2 a_1+\sqrt{5} \sqrt{8 a_3 a_1^3+\left(16 a_4-3 a_2^2\right) a_1^2-38 a_2 a_3 a_1+12 a_2^3+45 a_3^2-40 a_2 a_4}+15 a_3\right)^4}{\left(4 a_1^2-10 a_2\right)^4}\nonumber\\
    &&+\frac{5 a_5 \left(4 a_1^3-13 a_2 a_1+\sqrt{5} \sqrt{8 a_3 a_1^3+\left(16 a_4-3 a_2^2\right) a_1^2-38 a_2 a_3 a_1+12 a_2^3+45 a_3^2-40 a_2 a_4}+15 a_3\right)^3}{\left(4 a_1^2-10 a_2\right)^3}\nonumber\\
    &&-\frac{a_1 a_4 \left(4 a_1^3-13 a_2 a_1+\sqrt{5} \sqrt{8 a_3 a_1^3+\left(16 a_4-3 a_2^2\right) a_1^2-38 a_2 a_3 a_1+12 a_2^3+45 a_3^2-40 a_2 a_4}+15 a_3\right)^3}{\left(4 a_1^2-10 a_2\right)^3}\nonumber\\
    &&+\frac{a_2 a_4 \left(4 a_1^3-13 a_2 a_1+\sqrt{5} \sqrt{8 a_3 a_1^3+\left(16 a_4-3 a_2^2\right) a_1^2-38 a_2 a_3 a_1+12 a_2^3+45 a_3^2-40 a_2 a_4}+15 a_3\right)^2}{\left(4 a_1^2-10 a_2\right)^2}\nonumber\\
    &&-\frac{a_1 a_5 \left(4 a_1^3-13 a_2 a_1+\sqrt{5} \sqrt{8 a_3 a_1^3+\left(16 a_4-3 a_2^2\right) a_1^2-38 a_2 a_3 a_1+12 a_2^3+45 a_3^2-40 a_2 a_4}+15 a_3\right)^2}{\left(2 a_1^2-5 a_2\right)^2}\nonumber\\
    &&+\frac{3 a_2 a_5 \left(4 a_1^3-13 a_2 a_1+\sqrt{5} \sqrt{8 a_3 a_1^3+\left(16 a_4-3 a_2^2\right) a_1^2-38 a_2 a_3 a_1+12 a_2^3+45 a_3^2-40 a_2 a_4}+15 a_3\right)}{4 a_1^2-10 a_2}
\end{eqnarray*}    
\begin{eqnarray*}
    &&-\frac{a_3 a_4 \left(4 a_1^3-13 a_2 a_1+\sqrt{5} \sqrt{8 a_3 a_1^3+\left(16 a_4-3 a_2^2\right) a_1^2-38 a_2 a_3 a_1+12 a_2^3+45 a_3^2-40 a_2 a_4}+15 a_3\right)}{4 a_1^2-10 a_2}\nonumber\\
    &&+\frac{2 \left(5 a_2 a_1^2+\left(\sqrt{5} \sqrt{8 a_3 a_1^3+\left(16 a_4-3 a_2^2\right) a_1^2-38 a_2 a_3 a_1+12 a_2^3+45 a_3^2-40 a_2 a_4}+15 a_3\right) a_1-20 a_2^2\right) a_3^2}{20 a_1^2-50 a_2}\nonumber\\
    &&-\frac{4 a_2 a_4 \left(5 a_2 a_1^2+\left(\sqrt{5} \sqrt{8 a_3 a_1^3+\left(16 a_4-3 a_2^2\right) a_1^2-38 a_2 a_3 a_1+12 a_2^3+45 a_3^2-40 a_2 a_4}+15 a_3\right) a_1-20 a_2^2\right)}{20 a_1^2-50 a_2}\nonumber\\
    &&-\frac{a_2 a_3}{10 \left(2 a_1^2-5 a_2\right)^2}\left[\left(4 a_1^3-13 a_2 a_1+\sqrt{5} \sqrt{8 a_3 a_1^3+\left(16 a_4-3 a_2^2\right) a_1^2-38 a_2 a_3 a_1+12 a_2^3+45 a_3^2-40 a_2 a_4}+15 a_3\right)\right.\nonumber\\
    &&\left.\times \left(5 a_2 a_1^2+\left(\sqrt{5} \sqrt{8 a_3 a_1^3+\left(16 a_4-3 a_2^2\right) a_1^2-38 a_2 a_3 a_1+12 a_2^3+45 a_3^2-40 a_2 a_4}+15 a_3\right) a_1-20 a_2^2\right)\right]\nonumber\\
    &&+\frac{3 a_1 a_4}{10 \left(2 a_1^2-5 a_2\right)^2}\left[\left(4 a_1^3-13 a_2 a_1+\sqrt{5} \sqrt{8 a_3 a_1^3+\left(16 a_4-3 a_2^2\right) a_1^2-38 a_2 a_3 a_1+12 a_2^3+45 a_3^2-40 a_2 a_4}+15 a_3\right)\right.\nonumber\\
    &&\left.\times\left(5 a_2 a_1^2+\left(\sqrt{5} \sqrt{8 a_3 a_1^3+\left(16 a_4-3 a_2^2\right) a_1^2-38 a_2 a_3 a_1+12 a_2^3+45 a_3^2-40 a_2 a_4}+15 a_3\right) a_1-20 a_2^2\right)\right]\nonumber\\
    &&+\frac{a_1 a_3}{20 \left(2 a_1^2-5 a_2\right)^3} \left[\left(4 a_1^3-13 a_2 a_1+\sqrt{5} \sqrt{8 a_3 a_1^3+\left(16 a_4-3 a_2^2\right) a_1^2-38 a_2 a_3 a_1+12 a_2^3+45 a_3^2-40 a_2 a_4}+15 a_3\right)^2\right.\nonumber\\
    &&\left.\times\left(5 a_2 a_1^2+\left(\sqrt{5} \sqrt{8 a_3 a_1^3+\left(16 a_4-3 a_2^2\right) a_1^2-38 a_2 a_3 a_1+12 a_2^3+45 a_3^2-40 a_2 a_4}+15 a_3\right) a_1-20 a_2^2\right)\right]\nonumber\\
    &&-\frac{a_4}{5 \left(2 a_1^2-5 a_2\right)^3}\left[\left(4 a_1^3-13 a_2 a_1+\sqrt{5} \sqrt{8 a_3 a_1^3+\left(16 a_4-3 a_2^2\right) a_1^2-38 a_2 a_3 a_1+12 a_2^3+45 a_3^2-40 a_2 a_4}+15 a_3\right)^2\right.\nonumber\\
    &&\left.\times\left(5 a_2 a_1^2+\left(\sqrt{5} \sqrt{8 a_3 a_1^3+\left(16 a_4-3 a_2^2\right) a_1^2-38 a_2 a_3 a_1+12 a_2^3+45 a_3^2-40 a_2 a_4}+15 a_3\right) a_1-20 a_2^2\right)\right]\nonumber\\
    &&-\frac{a_3}{40 \left(2 a_1^2-5 a_2\right)^4}\left[\left(4 a_1^3-13 a_2 a_1+\sqrt{5} \sqrt{8 a_3 a_1^3+\left(16 a_4-3 a_2^2\right) a_1^2-38 a_2 a_3 a_1+12 a_2^3+45 a_3^2-40 a_2 a_4}+15 a_3\right)^3\right.\nonumber\\
    &&\left.\times\left(5 a_2 a_1^2+\left(\sqrt{5} \sqrt{8 a_3 a_1^3+\left(16 a_4-3 a_2^2\right) a_1^2-38 a_2 a_3 a_1+12 a_2^3+45 a_3^2-40 a_2 a_4}+15 a_3\right) a_1-20 a_2^2\right)\right]\nonumber\\
    && +\frac{3 a_2^2 \left(5 a_2 a_1^2+\left(\sqrt{5} \sqrt{8 a_3 a_1^3+\left(16 a_4-3 a_2^2\right) a_1^2-38 a_2 a_3 a_1+12 a_2^3+45 a_3^2-40 a_2 a_4}+15 a_3\right) a_1-20 a_2^2\right)^2}{\left(20 a_1^2-50 a_2\right)^2}\nonumber\\
    && -\frac{6 a_1 a_3 \left(5 a_2 a_1^2+\left(\sqrt{5} \sqrt{8 a_3 a_1^3+\left(16 a_4-3 a_2^2\right) a_1^2-38 a_2 a_3 a_1+12 a_2^3+45 a_3^2-40 a_2 a_4}+15 a_3\right) a_1-20 a_2^2\right)^2}{\left(20 a_1^2-50 a_2\right)^2}\nonumber\\
    &&+\frac{6 a_4 \left(5 a_2 a_1^2+\left(\sqrt{5} \sqrt{8 a_3 a_1^3+\left(16 a_4-3 a_2^2\right) a_1^2-38 a_2 a_3 a_1+12 a_2^3+45 a_3^2-40 a_2 a_4}+15 a_3\right) a_1-20 a_2^2\right)^2}{\left(20 a_1^2-50 a_2\right)^2}\nonumber\\
       &&\qquad-\frac{3 a_1 a_2}{200 \left(2 a_1^2-5 a_2\right)^3}\left[\left(4 a_1^3-13 a_2 a_1+\sqrt{5} \sqrt{8 a_3 a_1^3+\left(16 a_4-3 a_2^2\right) a_1^2-38 a_2 a_3 a_1+12 a_2^3+45 a_3^2-40 a_2 a_4}+15 a_3\right)\right.\nonumber\\
    &&\qquad\left.\times\left(5 a_2 a_1^2+\left(\sqrt{5} \sqrt{8 a_3 a_1^3+\left(16 a_4-3 a_2^2\right) a_1^2-38 a_2 a_3 a_1+12 a_2^3+45 a_3^2-40 a_2 a_4}+15 a_3\right) a_1-20 a_2^2\right)^2\right]\nonumber\\
    &&\qquad +\frac{9 a_3}{200 \left(2 a_1^2-5 a_2\right)^3} \left[\left(4 a_1^3-13 a_2 a_1+\sqrt{5} \sqrt{8 a_3 a_1^3+\left(16 a_4-3 a_2^2\right) a_1^2-38 a_2 a_3 a_1+12 a_2^3+45 a_3^2-40 a_2 a_4}+15 a_3\right)\right.\nonumber\\
    &&\qquad \left.\times\left(5 a_2 a_1^2+\left(\sqrt{5} \sqrt{8 a_3 a_1^3+\left(16 a_4-3 a_2^2\right) a_1^2-38 a_2 a_3 a_1+12 a_2^3+45 a_3^2-40 a_2 a_4}+15 a_3\right) a_1-20 a_2^2\right)^2\right]\nonumber\\
    &&\qquad+\frac{3 a_2}{400 \left(2 a_1^2-5 a_2\right)^4}\left[\left(4 a_1^3-13 a_2 a_1+\sqrt{5} \sqrt{8 a_3 a_1^3+\left(16 a_4-3 a_2^2\right) a_1^2-38 a_2 a_3 a_1+12 a_2^3+45 a_3^2-40 a_2 a_4}+15 a_3\right)^2\right.\nonumber\\
    &&\qquad\left.\times\left(5 a_2 a_1^2+\left(\sqrt{5} \sqrt{8 a_3 a_1^3+\left(16 a_4-3 a_2^2\right) a_1^2-38 a_2 a_3 a_1+12 a_2^3+45 a_3^2-40 a_2 a_4}+15 a_3\right) a_1-20 a_2^2\right)^2\right]\nonumber\\
    &&\qquad+\frac{4 a_1^2 \left(5 a_2 a_1^2+\left(\sqrt{5} \sqrt{8 a_3 a_1^3+\left(16 a_4-3 a_2^2\right) a_1^2-38 a_2 a_3 a_1+12 a_2^3+45 a_3^2-40 a_2 a_4}+15 a_3\right) a_1-20 a_2^2\right)^3}{\left(20 a_1^2-50 a_2\right)^3}
\end{eqnarray*}
\begin{eqnarray*}
    &&\qquad-\frac{8 a_2 \left(5 a_2 a_1^2+\left(\sqrt{5} \sqrt{8 a_3 a_1^3+\left(16 a_4-3 a_2^2\right) a_1^2-38 a_2 a_3 a_1+12 a_2^3+45 a_3^2-40 a_2 a_4}+15 a_3\right) a_1-20 a_2^2\right)^3}{\left(20 a_1^2-50 a_2\right)^3}\nonumber\\
    &&\qquad-\frac{a_1}{500 \left(2 a_1^2-5 a_2\right)^4}\left[\left(4 a_1^3-13 a_2 a_1+\sqrt{5} \sqrt{8 a_3 a_1^3+\left(16 a_4-3 a_2^2\right) a_1^2-38 a_2 a_3 a_1+12 a_2^3+45 a_3^2-40 a_2 a_4}+15 a_3\right)\right.
    \end{eqnarray*}
    \begin{eqnarray}
    &&\qquad\left.\times\left(5 a_2 a_1^2+\left(\sqrt{5} \sqrt{8 a_3 a_1^3+\left(16 a_4-3 a_2^2\right) a_1^2-38 a_2 a_3 a_1+12 a_2^3+45 a_3^2-40 a_2 a_4}+15 a_3\right) a_1-20 a_2^2\right)^3\right]\nonumber\\
    &&\qquad+\frac{5 \left(5 a_2 a_1^2+\left(\sqrt{5} \sqrt{8 a_3 a_1^3+\left(16 a_4-3 a_2^2\right) a_1^2-38 a_2 a_3 a_1+12 a_2^3+45 a_3^2-40 a_2 a_4}+15 a_3\right) a_1-20 a_2^2\right)^4}{\left(20 a_1^2-50 a_2\right)^4}\nonumber\\
    &&\qquad+\frac{3 a_2 \left(4 a_1^3-13 a_2 a_1+\sqrt{5} \sqrt{8 a_3 a_1^3+\left(16 a_4-3 a_2^2\right) a_1^2-38 a_2 a_3 a_1+12 a_2^3+45 a_3^2-40 a_2 a_4}+15 a_3\right) a_5}{4 a_1^2-10 a_2}-2 a_3 a_5\nonumber\\
    &&\qquad-\frac{a_1 \left(4 a_1^3-13 a_2 a_1+\sqrt{5} \sqrt{8 a_3 a_1^3+\left(16 a_4-3 a_2^2\right) a_1^2-38 a_2 a_3 a_1+12 a_2^3+45 a_3^2-40 a_2 a_4}+15 a_3\right)^2 a_5}{\left(2 a_1^2-5 a_2\right)^2}\nonumber\\
    &&\qquad+\frac{5 \left(4 a_1^3-13 a_2 a_1+\sqrt{5} \sqrt{8 a_3 a_1^3+\left(16 a_4-3 a_2^2\right) a_1^2-38 a_2 a_3 a_1+12 a_2^3+45 a_3^2-40 a_2 a_4}+15 a_3\right)^3 a_5}{\left(4 a_1^2-10 a_2\right)^3}\nonumber\\
    &&\qquad+\frac{4 a_1 \left(5 a_2 a_1^2+\left(\sqrt{5} \sqrt{8 a_3 a_1^3+\left(16 a_4-3 a_2^2\right) a_1^2-38 a_2 a_3 a_1+12 a_2^3+45 a_3^2-40 a_2 a_4}+15 a_3\right) a_1-20 a_2^2\right) a_5}{20 a_1^2-50 a_2}\nonumber\\
    &&\qquad-\frac{a_5}{2 \left(2 a_1^2-5 a_2\right)^2}\left[\left(4 a_1^3-13 a_2 a_1+\sqrt{5} \sqrt{8 a_3 a_1^3+\left(16 a_4-3 a_2^2\right) a_1^2-38 a_2 a_3 a_1+12 a_2^3+45 a_3^2-40 a_2 a_4}+15 a_3\right)\right.\nonumber\\
    &&\qquad\left.\times\left(5 a_2 a_1^2+\left(\sqrt{5} \sqrt{8 a_3 a_1^3+\left(16 a_4-3 a_2^2\right) a_1^2-38 a_2 a_3 a_1+12 a_2^3+45 a_3^2-40 a_2 a_4}+15 a_3\right) a_1-20 a_2^2\right)\right],
    \label{eq:A9}
\end{eqnarray}
\begin{eqnarray}\label{eq:A10}
  \fw &=& \frac{a_4 \left(4 a_1^3-13 a_2 a_1+\sqrt{5} \sqrt{8 a_3 a_1^3+\left(16 a_4-3 a_2^2\right) a_1^2-38 a_2 a_3 a_1+12 a_2^3+45 a_3^2-40 a_2 a_4}+15 a_3\right) a_5}{4 a_1^2-10 a_2}-a_5^2\nonumber\\
  && -\frac{a_3 \left(4 a_1^3-13 a_2 a_1+\sqrt{5} \sqrt{8 a_3 a_1^3+\left(16 a_4-3 a_2^2\right) a_1^2-38 a_2 a_3 a_1+12 a_2^3+45 a_3^2-40 a_2 a_4}+15 a_3\right)^2 a_5}{\left(4 a_1^2-10 a_2\right)^2}\nonumber\\
  && +\frac{a_2 \left(4 a_1^3-13 a_2 a_1+\sqrt{5} \sqrt{8 a_3 a_1^3+\left(16 a_4-3 a_2^2\right) a_1^2-38 a_2 a_3 a_1+12 a_2^3+45 a_3^2-40 a_2 a_4}+15 a_3\right)^3 a_5}{\left(4 a_1^2-10 a_2\right)^3}\nonumber\\
  &&-\frac{a_1 \left(4 a_1^3-13 a_2 a_1+\sqrt{5} \sqrt{8 a_3 a_1^3+\left(16 a_4-3 a_2^2\right) a_1^2-38 a_2 a_3 a_1+12 a_2^3+45 a_3^2-40 a_2 a_4}+15 a_3\right)^4 a_5}{\left(4 a_1^2-10 a_2\right)^4}\nonumber\\
  && +\frac{\left(4 a_1^3-13 a_2 a_1+\sqrt{5} \sqrt{8 a_3 a_1^3+\left(16 a_4-3 a_2^2\right) a_1^2-38 a_2 a_3 a_1+12 a_2^3+45 a_3^2-40 a_2 a_4}+15 a_3\right)^5 a_5}{\left(4 a_1^2-10 a_2\right)^5}.
\end{eqnarray}
Now, to transform the principal quintic \eqref{eq:A7} to its Bring-Jerrard form, we use the quartic Tschirnhausen transformation
\begin{equation}\label{eq:A11}
z = y^4+\fp y^3+\fq y^2+\fr y+ \fs.
\end{equation}
Eliminating $y$ between Eqs.~\eqref{eq:A7} and \eqref{eq:A11}, we get to the quintic
\begin{equation}\label{eq:A12}
    z^5+d_1z^4+d_2z^3+d_3z^2+d_4z+d_5=0,
\end{equation}
in which
\begin{subequations}\label{eq:A13}
\begin{align}
    & d_1 = 3 \fp \fu-5 \fs+4 \fv,\label{eq:A13a}\\
    & d_2 = 10 \fs^2 - 12 \fp \fs \fu + 3 \fp^2 \fu^2 - 3 \fq \fu^2 + 2 \fq^2 \fv - 16 \fs \fv + 5 \fp \fu \fv + 6 \fv^2 + 5 \fp \fq \fw - 4 \fu \fw + \fr (3 \fq \fu + 4 \fp \fv + 5 \fw),\label{eq:A13b}\\
    & d_3 = 7 \fp^2 \fq \fu \fw-4 \fp^2 \fq \fv^2+5 \fp^2 \fr \fu \fv-9 \fp^2 \fs \fu^2+\fp^2 \fu^2 \fv+\fp^3 \fu^3-3 \fp^3 \fv \fw-5 \fp^2 \fw^2-\fp \fq^2 \fu \fv+3 \fp \fq \fr \fu^2-15 \fp \fq \fs \fw\nonumber\\
    &\qquad-3 \fp \fq \fu^3+2 \fp \fq \fv \fw+5 \fp \fr^2 \fw-12 \fp \fr \fs \fv-\fp \fr \fu \fw+8 \fp \fr \fv^2+18 \fp \fs^2 \fu-15 \fp \fs \fu \fv-\fp \fu^2 \fw+\fp \fu \fv^2+5 \fq^2 \fr \fw\nonumber\\
    &\qquad-6 \fq^2 \fs \fv-\fq^3 \fu^2-8 \fq^2 \fu \fw+4 \fq^2 \fv^2+4 \fq \fr^2 \fv-9 \fq \fr \fs \fu-2 \fq \fr \fu \fv+9 \fq \fs \fu^2-2 \fq \fu^2 \fv-5 \fq \fw^2-3 \fr^2 \fu^2+\fr^3 \fu\nonumber\\
    &\qquad-15 \fr \fs \fw+3 \fr \fu^3+11 \fr \fv \fw+24 \fs^2 \fv-10 \fs^3+12 \fs \fu \fw-18 \fs \fv^2-\fu^4-8 \fu \fv \fw+4 \fv^3,\label{eq:A13c}\\
    & d_4 = 5 \fs^4 - 2 \fr^3 \fs \fu + 9 \fq \fr \fs^2 \fu - 12 \fp \fs^3 \fu + 2 \fq^3 \fs \fu^2 - 
 6 \fp \fq \fr \fs \fu^2 + 6 \fr^2 \fs \fu^2 + 9 \fp^2 \fs^2 \fu^2 - 9 \fq \fs^2 \fu^2 - 
 2 \fp^3 \fs \fu^3 + 6 \fp \fq \fs \fu^3 \nonumber\\
 &\qquad- 6 \fr \fs \fu^3 + 2 \fs \fu^4 + \fr^4 \fv - 
 8 \fq \fr^2 \fs \fv + 6 \fq^2 \fs^2 \fv + 12 \fp \fr \fs^2 \fv - 16 \fs^3 \fv - \fq^3 \fr \fu \fv + 
 3 \fp \fq \fr^2 \fu \fv - 3 \fr^3 \fu \fv + 2 \fp \fq^2 \fs \fu \fv \nonumber\\
 &\qquad- 10 \fp^2 \fr \fs \fu \fv + 
 4 \fq \fr \fs \fu \fv + 15 \fp \fs^2 \fu \fv + \fp^3 \fr \fu^2 \fv - 3 \fp \fq \fr \fu^2 \fv + 
 3 \fr^2 \fu^2 \fv - 2 \fp^2 \fs \fu^2 \fv + 4 \fq \fs \fu^2 \fv - \fr \fu^3 \fv + \fq^4 \fv^2 \nonumber\\
 &\qquad- 4 \fp \fq^2 \fr \fv^2 + 2 \fp^2 \fr^2 \fv^2 + 4 \fq \fr^2 \fv^2 + 8 \fp^2 \fq \fs \fv^2 - 
 8 \fq^2 \fs \fv^2 - 16 \fp \fr \fs \fv^2 + 18 \fs^2 \fv^2 - \fp^3 \fq \fu \fv^2 + 
 3 \fp \fq^2 \fu \fv^2 + \fp^2 \fr \fu \fv^2 \nonumber\\
 &\qquad- 5 \fq \fr \fu \fv^2 - 2 \fp \fs \fu \fv^2 + \fq \fu^2 \fv^2 +
  \fp^4 \fv^3 - 4 \fp^2 \fq \fv^3 + 2 \fq^2 \fv^3 + 4 \fp \fr \fv^3 - 8 \fs \fv^3 -  \fp \fu \fv^3 + \fv^4 + 5 \fq \fr^3 \fw - 10 \fq^2 \fr \fs \fw \nonumber\\
 &\qquad- 10 \fp \fr^2 \fs \fw + 
 15 \fp \fq \fs^2 \fw + 15 \fr \fs^2 \fw - 2 \fq^4 \fu \fw + 6 \fp \fq^2 \fr \fu \fw + 
 3 \fp^2 \fr^2 \fu \fw - 9 \fq \fr^2 \fu \fw - 14 \fp^2 \fq \fs \fu \fw + 16 \fq^2 \fs \fu \fw  \nonumber\\
  &\qquad+2 \fp \fr \fs \fu \fw - 12 \fs^2 \fu \fw + 2 \fp^3 \fq \fu^2 \fw - 6 \fp \fq^2 \fu^2 \fw +
 6 \fq \fr \fu^2 \fw + 2 \fp \fs \fu^2 \fw - 2 \fq \fu^3 \fw + \fp \fq^3 \fv \fw - 7 \fp^2 \fq \fr \fv \fw + 
 3 \fq^2 \fr \fv \fw \nonumber\\
 &\qquad+ 13 \fp \fr^2 \fv \fw + 6 \fp^3 \fs \fv \fw - 4 \fp \fq \fs \fv \fw - 
 22 \fr \fs \fv \fw - 3 \fp^4 \fu \fv \fw + 11 \fp^2 \fq \fu \fv \fw - 4 \fq^2 \fu \fv \fw - 
 10 \fp \fr \fu \fv \fw + 16 \fs \fu \fv \fw \nonumber\\
 &\qquad+ 3 \fp \fu^2 \fv \fw + \fp^3 \fv^2 \fw - 3 \fp \fq \fv^2 \fw + 
 7 \fr \fv^2 \fw - 4 \fu \fv^2 \fw + 5 \fp^2 \fq^2 \fw^2 - 5 \fq^3 \fw^2 - 5 \fp^3 \fr \fw^2 - 
 5 \fp \fq \fr \fw^2 + 5 \fr^2 \fw^2 \nonumber\\
 &\qquad+ 10 \fp^2 \fs \fw^2 + 10 \fq \fs \fw^2 - 2 \fp^3 \fu \fw^2 + 
 4 \fp \fq \fu \fw^2 - 7 \fr \fu \fw^2 + 2 \fu^2 \fw^2 + \fp^2 \fv \fw^2 - 6 \fq \fv \fw^2 + 5 \fp \fw^3,\label{eq:A13d}\\
& d_5 = \fw^3 \fp^5-\fs \fv^3 \fp^4-2 \fr \fu \fw^2 \fp^4-\fq \fv \fw^2 \fp^4+\fr \fv^2 \fw \fp^4+3 \fs \fu \fv \fw \fp^4+\fs^2 \fu^3 \fp^3-5 \fq \fw^3 \fp^3+\fq \fs \fu \fv^2 \fp^3+5 \fr \fs \fw^2 \fp^3\nonumber\\
&\qquad+\fq^2 \fu \fw^2 \fp^3+2 \fs \fu \fw^2 \fp^3+\fr \fv \fw^2 \fp^3-\fr \fs \fu^2 \fv \fp^3+\fr^2 \fu^2 \fw \fp^3-2 \fq \fs \fu^2 \fw \fp^3-\fs \fv^2 \fw \fp^3-3 \fs^2 \fv \fw \fp^3-\fq \fr \fu \fv \fw \fp^3\nonumber\\
&\qquad+4 \fq \fs \fv^3 \fp^2+5 \fr \fw^3 \fp^2-\fu \fw^3 \fp^2-3 \fs^3 \fu^2 \fp^2-4 \fq \fs^2 \fv^2 \fp^2-2 \fr^2 \fs \fv^2 \fp^2-\fr \fs \fu \fv^2 \fp^2-5 \fq \fr^2 \fw^2 \fp^2-5 \fs^2 \fw^2 \fp^2\nonumber\\
&\qquad-5 \fq^2 \fs \fw^2 \fp^2+6 \fq \fr \fu \fw^2 \fp^2+4 \fq^2 \fv \fw^2 \fp^2-\fs \fv \fw^2 \fp^2+\fs^2 \fu^2 \fv \fp^2+5 \fr \fs^2 \fu \fv \fp^2-4 \fq \fr \fv^2 \fw \fp^2+7 \fq \fs^2 \fu \fw \fp^2-3 \fr^2 \fs \fu \fw \fp^2\nonumber\\
&\qquad+2 \fr^3 \fv \fw \fp^2+7 \fq \fr \fs \fv \fw \fp^2+\fr^2 \fu \fv \fw \fp^2-11 \fq \fs \fu \fv \fw \fp^2-3 \fq \fs^2 \fu^3 \fp-4 \fr \fs \fv^3 \fp+\fs \fu \fv^3 \fp+5 \fq^2 \fw^3 \fp-5 \fs \fw^3 \fp+\fv \fw^3 \fp\nonumber\\
&\qquad+3 \fq \fr \fs^2 \fu^2 \fp+8 \fr \fs^2 \fv^2 \fp+4 \fq^2 \fr \fs \fv^2 \fp+\fs^2 \fu \fv^2 \fp-3 \fq^2 \fs \fu \fv^2 \fp+5 \fr^3 \fw^2 \fp+2 \fr \fu^2 \fw^2 \fp+5 \fq^3 \fr \fw^2 \fp+5 \fq \fr \fs \fw^2 \fp\nonumber\\
&\qquad-3 \fq^3 \fu \fw^2 \fp-7 \fr^2 \fu \fw^2 \fp-4 \fq \fs \fu \fw^2 \fp-7 \fq \fr \fv \fw^2 \fp+\fq \fu \fv \fw^2 \fp+3 \fs^4 \fu \fp-4 \fr \fs^3 \fv \fp+3 \fq \fr \fs \fu^2 \fv \fp-5 \fs^3 \fu \fv \fp-\fq^2 \fs^2 \fu \fv \fp\nonumber\\
&\qquad-3 \fq \fr^2 \fs \fu \fv \fp-5 \fq \fs^3 \fw \fp+5 \fr^2 \fs^2 \fw \fp-3 \fq \fr^2 \fu^2 \fw \fp-\fs^2 \fu^2 \fw \fp+6 \fq^2 \fs \fu^2 \fw \fp+4 \fr^2 \fv^2 \fw \fp+3 \fq \fs \fv^2 \fw \fp-\fr \fu \fv^2 \fw \fp\nonumber\\
&\qquad+3 \fq \fr^3 \fu \fw \fp-\fr \fs^2 \fu \fw \fp-6 \fq^2 \fr \fs \fu \fw \fp-4 \fq^2 \fr^2 \fv \fw \fp+2 \fq \fs^2 \fv \fw \fp-3 \fs \fu^2 \fv \fw \fp-\fq^3 \fs \fv \fw \fp-13 \fr^2 \fs \fv \fw \fp+3 \fq^2 \fr \fu \fv \fw \fp\nonumber\\
&\qquad+10 \fr \fs \fu \fv \fw \fp-\fs^5-\fs^2 \fu^4-\fs \fv^4-\fw^4+3 \fr \fs^2 \fu^3+4 \fs^2 \fv^3-2 \fq^2 \fs \fv^3-5 \fq \fr \fw^3+2 \fq \fu \fw^3+3 \fq \fs^3 \fu^2-\fq^3 \fs^2 \fu^2\nonumber\\
&\qquad-3 \fr^2 \fs^2 \fu^2-6 \fs^3 \fv^2+4 \fq^2 \fs^2 \fv^2-\fq \fs \fu^2 \fv^2-\fq^4 \fs \fv^2-4 \fq \fr^2 \fs \fv^2+5 \fq \fr \fs \fu \fv^2-\fq^5 \fw^2-5 \fq^2 \fr^2 \fw^2-5 \fq \fs^2 \fw^2-\fq^2 \fu^2 \fw^2\nonumber\\
&\qquad-2 \fs \fu^2 \fw^2-\fq \fv^2 \fw^2+5 \fq^3 \fs \fw^2-5 \fr^2 \fs \fw^2+3 \fq^2 \fr \fu \fw^2+7 \fr \fs \fu \fw^2-2 \fq^3 \fv \fw^2+3 \fr^2 \fv \fw^2+6 \fq \fs \fv \fw^2-3 \fr \fu \fv \fw^2\nonumber\\
&\qquad-3 \fq \fr \fs^3 \fu+\fr^3 \fs^2 \fu+4 \fs^4 \fv-2 \fq^2 \fs^3 \fv+\fr \fs \fu^3 \fv+4 \fq \fr^2 \fs^2 \fv-2 \fq \fs^2 \fu^2 \fv-3 \fr^2 \fs \fu^2 \fv-\fr^4 \fs \fv-2 \fq \fr \fs^2 \fu \fv+3 \fr^3 \fs \fu \fv\nonumber\\
&\qquad+\fq^3 \fr \fs \fu \fv+\fr^5 \fw-5 \fr \fs^3 \fw-\fr^2 \fu^3 \fw+2 \fq \fs \fu^3 \fw+\fr \fv^3 \fw+5 \fq^2 \fr \fs^2 \fw+3 \fr^3 \fu^2 \fw-6 \fq \fr \fs \fu^2 \fw+2 \fq^2 \fr \fv^2 \fw-7 \fr \fs \fv^2 \fw\nonumber\\
&\qquad+4 \fs \fu \fv^2 \fw-5 \fq \fr^3 \fs \fw-3 \fr^4 \fu \fw+4 \fs^3 \fu \fw-\fq^3 \fr^2 \fu \fw-8 \fq^2 \fs^2 \fu \fw+2 \fq^4 \fs \fu \fw+9 \fq \fr^2 \fs \fu \fw+4 \fq \fr^3 \fv \fw+11 \fr \fs^2 \fv \fw\nonumber\\
&\qquad+\fq \fr \fu^2 \fv \fw+\fq^4 \fr \fv \fw-3 \fq^2 \fr \fs \fv \fw-5 \fq \fr^2 \fu \fv \fw-8 \fs^2 \fu \fv \fw+4 \fq^2 \fs \fu \fv \fw.\label{eq:A13e}
\end{align}
\end{subequations}
Similar to the previous step, it is now necessary to solve the equations $d_1=d_2=0$ and the extra equation $3 \fq \fu + 4 \fp \fv + 5 \fw=0$ extracted from Eq.~\eqref{eq:A13b}, for the parameters $\fp, \fq$ and $\fs$. These equations result in the values
\begin{eqnarray}
  \fp &=& \frac{1}{54 \fu^4+600 \fu \fv \fw-320 \fv^3}\left[
  \mp\left(27 \fu^3 \fv+375 \fu \fw^2-400 \fv^2 \fw\right)+\mathfrak{Q}\right],\label{eq:A14}\\
  \fq &=& \frac{1}{27 \fu^5 - 160 \fu \fv^3 + 300 \fu^2 \fv \fw}\left[
  18 \fu^3 \fv^2 - 45 \fu^4 \fw - 250 \fu \fv \fw\pm\frac{2}{3}\fv\mathfrak{Q}\right],\label{eq:A15}\\
  \fs &=& \frac{1}{270 \fu^4+3000 \fu \fv \fw-1600 \fv^3}\left[
  135 \fu^4 \fv-1125 \fu^2 \fw^2+3600 \fu \fv^2 \fw-1280 \fv^4\mp3\fu\mathfrak{Q}\right],\label{eq:A16}
\end{eqnarray}
where $\mathfrak{Q} = 3|\fu|\sqrt{5 \left(-27 \fu^4 \fv^2+2250 \fu^2 \fv \fw^2+108 \fu^5 \fw-1600 \fu \fv^3 \fw+256 \fv^5+3125 \fw^4\right)}$. Note that, this process leaves $\fr$ as a free parameter. This parameter can be however determined appropriately, by means of the equation $d_3 = 0$, which results in the cubic
\begin{equation}\label{eq:A17}
    e_3 \fr^3+e_2 \fr^2+e_1 \fr+e_0 = 0,
\end{equation}
where
\begin{subequations}\label{eq:A18}
\begin{align}
    & e_0 = 7 \fp^2 \fq \fu \fw-4 \fp^2 \fq \fv^2-9 \fp^2 \fs \fu^2+\fp^2 \fu^2 \fv+\fp^3 \fu^3-3 \fp^3 \fv \fw-5 \fp^2 \fw^2-\fp \fq^2 \fu \fv-15 \fp \fq \fs \fw-3 \fp \fq \fu^3+2 \fp \fq \fv \fw\nonumber\\
    &\qquad+6 \fs^2 (3 \fp \fu+4 \fv)-15 \fp \fs \fu \fv-\fp \fu^2 \fw+\fp \fu \fv^2-6 \fq^2 \fs \fv-\fq^3 \fu^2-8 \fq^2 \fu \fw+4 \fq^2 \fv^2+9 \fq \fs \fu^2-2 \fq \fu^2 \fv-5 \fq \fw^2\nonumber\\
    &\qquad+12 \fs \fu \fw-18 \fs \fv^2-\fu^4-8 \fu \fv \fw+4 \fv^3-10 \fs^3,\label{eq:A18a}\\
    & e_1 = -9 \fq \fs \fu + 3 \fp \fq \fu^2 + 3 \fu^3 - 12 \fp \fs \fv + 5 \fp^2 \fu \fv - 2 \fq \fu \fv +  8 \fp \fv^2 + 5 \fq^2 \fw - 15 \fs \fw - \fp \fu \fw + 11 \fv \fw,\label{eq:A18b}\\
    & e_2 = -3 \fu^2 + 4 \fq \fv + 5 \fp \fw,\label{eq:A18c}\\
    & e_3 = \fu,\label{eq:A18d}
\end{align}
\end{subequations}
whose solution, as it is well-known, can be expressed in terms of radicals. Now, applying these solutions for $\fr$, together with those expressed in Eqs.~\eqref{eq:A14}--\eqref{eq:A16} for $\fp$, $\fq$ and $\fs$, the values of $d_{4,5}$ in Eqs.~\eqref{eq:A13d} and \eqref{eq:A13e} are obtained. The expressions are, however, that huge that cannot be put in the paper. But we can be confident that the quintic \eqref{eq:A12} has been reduced to the Bring-Jerrard form
\begin{equation}
z^5 + d_4 z+ d_5 =0.
    \label{eq:A19}
\end{equation}
It is still possible to make more simplifications by defining
\begin{equation}
z\doteq\frac{\ft}{\ff}.
    \label{eq:A20}
\end{equation}
This way, the quintic \eqref{eq:A19} can be recast as
\begin{equation}
\ft^5+d_4 \ff^4 \ft+d_5\ff^5=0.
    \label{eq:A21}
\end{equation}
Now letting 
\begin{equation}
\ff=\left(\pm\frac{1}{d_4}\right)^{\frac{1}{4}},
\label{eq:A22}
\end{equation}
we get to the more simplified Bring-Jerrard form of the quintic
\begin{equation}
\ft^5\pm\ft+K=0,
    \label{eq:A23}
\end{equation}
where we have defined $K=d_5 \ff^5$.


\section{Derivation of the solutions to the $x$-parameter}\label{app:B}

Let us denote the solutions in the Eqs.~\eqref{eq:t1sol}--\eqref{eq:t5sol} by $\ft_j$ with $j=\overline{1,5}$. Based on the definition in Eq.~\eqref{eq:A20}, we have $z_j = \ff^{-1} \ft_j$. Then from Eq.~\eqref{eq:A11}, one needs to solve a quartic of the general form, in order to obtain an expression for $y_j$ in terms of $z_j$. This way, for each of the solutions for $z_j$, we have four solutions for $y_j$. To proceed with solving the quartic \eqref{eq:A11}, let us first apply the change of variable
\begin{equation}
y_j=W_j-\frac{\fp}{4},
    \label{eq:B1}
\end{equation}
which depresses the equation to
\begin{equation}
W_j^4+\mathcal{A}W_j^2+\mathcal{B}W_j+\mathcal{C} = 0,
    \label{eq:B2}
\end{equation}
where
\begin{subequations}\label{eq:B3}
\begin{align}
    & \mathcal{A} = \fq-\frac{3\fp^2}{8},\label{eq:B3a}\\
    & \mathcal{B} = \fr+\frac{\fp^3}{8}-\frac{\fp \fq}{2},\label{eq:B3b}\\
    & \mathcal{C} = (\fs-z_j)+\frac{\fp^2\fq}{16}-\frac{3\fp^4}{256}-\frac{\fp\fr}{4}.\label{eq:B3c}
\end{align}
\end{subequations}
The method of solving the suppressed quartic \eqref{eq:B2} has been given in the appendix C of Ref.~\cite{Fathi:2020sfw}. Pursuing this method, we obtain the four solutions
\begin{eqnarray}
W_{j1} &=& \tilde{\mathcal{A}} + \sqrt{\tilde{\mathcal{A}^2}-\tilde{\mathcal{B}}},\label{eq:B4}\\
W_{j2} &=& \tilde{\mathcal{A}} - \sqrt{\tilde{\mathcal{A}^2}-\tilde{\mathcal{B}}},\label{eq:B5}\\
W_{j3} &=& - \tilde{\mathcal{A}} + \sqrt{\tilde{\mathcal{A}^2}-\tilde{\mathcal{C}}},\label{eq:B6}\\
W_{j4} &=& - \tilde{\mathcal{A}} - \sqrt{\tilde{\mathcal{A}^2}-\tilde{\mathcal{C}}},\label{eq:B7}
\end{eqnarray}
in which
\begin{subequations}\label{eq:B8}
\begin{align}
    & \tilde{\mathcal{A}} = \sqrt{\tilde{\mathcal{U}}-\frac{\mathcal{A}}{6}},\label{eq:B8a}\\
    & \tilde{\mathcal{B}} = 2\tilde{\mathcal{A}}^2+\frac{\mathcal{A}}{2}+\frac{\mathcal{B}}{4\tilde{\mathcal{A}}},\label{eq:B8b}\\
    & \tilde{\mathcal{C}} = 2\tilde{\mathcal{A}}^2+\frac{\mathcal{A}}{2}-\frac{\mathcal{B}}{4\tilde{\mathcal{A}}},\label{eq:B8c}
\end{align}
\end{subequations}
where
\begin{equation}
\tilde{\mathcal{U}} = \sqrt{\frac{\tilde{\epsilon}_2}{3}}\cosh\left(
\frac{1}{3}\arccosh\left(3\tilde{\epsilon}_3\sqrt{\frac{3}{\tilde{\epsilon}_2^3}}\right)
\right),
    \label{eq:tildeU}
\end{equation}
with
\begin{subequations}
\begin{align}
  & \tilde{\epsilon}_2=\frac{\mathcal{A}^2}{12}+\mathcal{C},\label{eq:tildeepsilon2} \\
  & \tilde{\epsilon}_3 = \frac{\mathcal{A}^3}{216}-\frac{\mathcal{A}\mathcal{C}}{6}+\frac{\mathcal{B}^2}{16}. 
\end{align}
\end{subequations}
Finally, the solutions to the $y$-parameter are given as
\begin{equation}
\left(y_j\right)_i\equiv y_{ji} = W_{ji}-\frac{\fp}{4},
    \label{eq:B9}
\end{equation}
where $i=\overline{1,4}$. In this manner, the $y_{ji}$ solutions form a $5\times4$ matrix (or in other words, four sets of solutions to the quintic \eqref{eq:A7}, in accordance with the solutions to the quintic \eqref{eq:A12}). Now, in order to obtain the solutions for the $x$-parameter, as it is the original purpose of this discussion, we have to solve the quadratic equation \eqref{eq:A2}, for the known solutions $y_{ji}$. This results in the two solutions
\begin{eqnarray}
 \left(x_{ji}\right)_1 &=& \frac{-b_1+\sqrt{b_1^2-4\left(b_2-y_{ji}\right)}}{2},\label{eq:B10}\\
 \left(x_{ji}\right)_2 &=& \frac{-b_1-\sqrt{b_1^2-4\left(b_2-y_{ji}\right)}}{2},\label{eq:B11}
\end{eqnarray}
for $b_{1,2}$ given in Eqs.~\eqref{eq:A5} and \eqref{eq:A6}. In this sense, for each of the $y_{ji}$ solutions, there are two solutions for the $x$-parameter, which can be abbreviated as $x_{jil}$ with $l=1,2$. These solutions form a $5\times 4$ matrix of $2\times1$ matrices, in the form
\begin{equation}\label{eq:B12}
x_{jil}=\begin{pmatrix}
z_1: \qquad y_{11}\rightarrow\begin{pmatrix}
x_{111}\\
x_{112}
\end{pmatrix} &  y_{12}\rightarrow\begin{pmatrix}
x_{121}\\
x_{122}
\end{pmatrix} &  y_{13}\rightarrow\begin{pmatrix}
x_{131}\\
x_{132}
\end{pmatrix} &  y_{14}\rightarrow\begin{pmatrix}
x_{141}\\
x_{142}
\end{pmatrix}\\
z_2: \qquad y_{21}\rightarrow\begin{pmatrix}
x_{211}\\
x_{212}
\end{pmatrix} &  y_{22}\rightarrow\begin{pmatrix}
x_{221}\\
x_{222}
\end{pmatrix} &  y_{23}\rightarrow\begin{pmatrix}
x_{231}\\
x_{232}
\end{pmatrix} &  y_{24}\rightarrow\begin{pmatrix}
x_{241}\\
x_{242}
\end{pmatrix}\\
z_3:\qquad y_{31}\rightarrow\begin{pmatrix}
x_{311}\\
x_{312}
\end{pmatrix} &  y_{32}\rightarrow\begin{pmatrix}
x_{321}\\
x_{322}
\end{pmatrix} &  y_{33}\rightarrow\begin{pmatrix}
x_{331}\\
x_{332}
\end{pmatrix} &  y_{34}\rightarrow\begin{pmatrix}
x_{341}\\
x_{342}
\end{pmatrix}\\
z_4:\qquad y_{41}\rightarrow\begin{pmatrix}
x_{411}\\
x_{412}
\end{pmatrix} &  y_{42}\rightarrow\begin{pmatrix}
x_{421}\\
x_{422}
\end{pmatrix} &  y_{43}\rightarrow\begin{pmatrix}
x_{431}\\
x_{432}
\end{pmatrix} &  y_{44}\rightarrow\begin{pmatrix}
x_{441}\\
x_{442}
\end{pmatrix}\\
z_5:\qquad y_{51}\rightarrow\begin{pmatrix}
x_{511}\\
x_{512}
\end{pmatrix} &  y_{52}\rightarrow\begin{pmatrix}
x_{521}\\
x_{522}
\end{pmatrix} &  y_{53}\rightarrow\begin{pmatrix}
x_{531}\\
x_{532}
\end{pmatrix} &  y_{54}\rightarrow\begin{pmatrix}
x_{541}\\
x_{542}
\end{pmatrix}
\end{pmatrix},
\end{equation}
meaning that for the solutions of the $\ft$-parameter derived from the quintic \eqref{eq:A23}, there are eight sets of solutions for the $x$-parameter in the context of the original quintic \eqref{eq:A1}.\\\\

{\textbf{Data Availability Statement:} No Data associated in the manuscript.}


\bibliographystyle{ieeetr}
\bibliography{biblio_v1.bib}

\end{document}